\documentclass[11pt]{article}
\usepackage{amsfonts,amssymb,amsmath}
\usepackage{stmaryrd}

\def \DATE {March 29, 2009}
\def \TITLE {Anomalies in QFT and cohomologies of configuration spaces}

\newcommand{\beq}{\begin{equation}}
\newcommand{\eeq}{\end{equation}}
\newcommand{\beqa}{\begin{eqnarray}}
\newcommand{\eeqa}{\end{eqnarray}}
\newcommand{\nn}{\nonumber \\}
\def \podr {&& \hspace{-15pt}}

\def \z {{\mathrm{z}}}

\def \u {{\mathrm{u}}}
\def \x {{\mathrm{x}}}
\def \xx {\mbf{\mathrm{x}}}
\def \yy {\mbf{\mathrm{y}}}
\def \y {{\mathrm{y}}}

\def \rr {\mbf{\mathrm{r}}}
\def \qq {\mbf{\mathrm{q}}}

\def \ee {\mbf{\mathrm{e}}}

\def \Ss {\mathrsfs{S}}
\def \Dd {\mathrsfs{D}}
\def \R {{\mathbb R}}
\def \C {{\mathbb C}}
\def \Z {{\mathbb Z}}
\def \N {{\mathbb N}}

\def \spr {\cdot}
\def \id {\text{\rm id}}
\def \di {\partial}
\def \mz {\bigr\backslash\hspace{1pt}\{\Mbf{0}\}}

\def \rdf {\mathfrak{Q}}
\def \RDF {Q}

\DeclareMathAlphabet{\mathbbm}{U}{bbm}{m}{n}

\DeclareSymbolFont{ltrs}     {OT1}{pzc}{m}{it}
\DeclareSymbolFont{ltrsa}     {OMS}{cmsy}{m}{n}
\DeclareSymbolFont{ltrsA}{U}{txmia}{m}{it}
\DeclareSymbolFont{symbolsC}{U}{txsyc}{m}{n}
\DeclareSymbolFont{ltrsB}{U}{rsfs}{m}{n}

\DeclareSymbolFontAlphabet{\mfrak}{ltrsA}
\DeclareMathAlphabet{\mathpzc}{OT1}{pzc}{m}{it}
\DeclareMathAlphabet{\mathrsfs}{U}{rsfs}{m}{n}

\def \ID {\text{\it id}}

\def \Ker {\text{\it Ker}}

\newcommand{\vrestr}[2]{\!\left.\raisebox{#1}{$\,$}\!\right|_{\,
\raisebox{1pt}{\small \(#2\)}}}

\newcommand{\txfrac}[2]{\frac{\raisebox{1pt}{$#1$}}{\raisebox{-3pt}{$#2$}}}

\newcommand{\mbf}[1]{\ensuremath{\mathchoice
                    {\mbox{\boldmath$\displaystyle\mathbf{\mathit{#1}}$}}
                    {\mbox{\boldmath$\textstyle\mathbf{\mathit{#1}}$}}
                    {\mbox{\boldmath$\scriptstyle\mathbf{\mathit{#1}}$}}
                    {\mbox{\boldmath$\scriptscriptstyle\mathbf{\mathit{#1}}$}}}}

\newcommand{\Mbf}[1]{\ensuremath{\mathchoice
                    {\mbox{\boldmath$\displaystyle\mathbf{#1}$}}
                    {\mbox{\boldmath$\textstyle\mathbf{#1}$}}
                    {\mbox{\boldmath$\scriptstyle\mathbf{#1}$}}
                    {\mbox{\boldmath$\scriptscriptstyle\mathbf{#1}$}}}}

\newcounter{Theorem}
\newcounter{Definition}
\newcounter{Remark}

\def \setcntrs {\setcounter{equation}{0}\setcounter{Theorem}{0}\setcounter{Definition}{0}\setcounter{Remark}{0}}

\newcommand{\medskp}{

  \medskip

  }

\newenvironment{Theorem}[1][\bf Theorem \arabic{section}.\arabic{Theorem}]{%
        
        \refstepcounter{Theorem}\noindent\textbf{#1.}${}$\hspace{1pt}${}$\it}{}
\newenvironment{Proposition}[1][\bf Proposition \arabic{section}.\arabic{Theorem}]{%
        
        \refstepcounter{Theorem}\noindent\textbf{#1.}${}$\hspace{1pt}${}$\it}{}
\newenvironment{Lemma}[1][\bf Lemma \arabic{section}.\arabic{Theorem}]{%
        
        \refstepcounter{Theorem}\noindent\textbf{#1.}${}$\hspace{1pt}${}$\it}{}
\newenvironment{Corollary}[1][\bf Corollary \arabic{section}.\arabic{Theorem}]{%
        
        \refstepcounter{Theorem}\noindent\textbf{#1.}${}$\hspace{1pt}${}$\it}{}
\newenvironment{Remark}[1][\bf Remark \arabic{section}.\arabic{Remark}]{%
        
        \refstepcounter{Remark}\noindent\textbf{#1.}${}$\hspace{1pt}${}$}{}
\newenvironment{Definition}[1][\bf Definition~\arabic{section}.\arabic{Definition}]{%
        
        \refstepcounter{Definition}\noindent\textbf{#1.}${}$\hspace{1pt}${}$}{}

\newcounter{tmpc}
\newlength{\tmplenght}
\newlength{\tmplenghta}
\newlength{\tmplenghtb}
\newlength{\tmplenghtc}

\newenvironment{LIST}[1]{%
\setlength{\tmplenghta}{#1}
\setlength{\tmplenghtb}{#1}
\setlength{\tmplenghtc}{#1}
\advance\tmplenghtb-5pt
\advance\tmplenghtc 42pt
\setcounter{tmpc}{0}
\begin{list}{{\rm (\alph{tmpc})}}{\usecounter{tmpc}
\setlength{\leftmargin}{\tmplenghta}
\setlength{\rightmargin}{0cm}
\setlength{\itemsep}{1pt}
\setlength{\topsep}{3pt}
\setlength{\labelsep}{5pt}
\setlength{\labelwidth}{\tmplenghtb}
\setlength{\listparindent}{\tmplenghta}}
}{\end{list}}

\def\DP{\Dd'}
\def\CI{\mathcal{C}^{\infty}}
\def\SCi{\mathrsfs{N}}
\def\du{\hspace{1pt}\bullet}
\def\DSCi{\SCi\raisebox{8pt}{\hspace{0.5pt}}'}
\def\RM{\mathcal{R}}
\newcommand{\OM}[1]{\Omega^{#1}}
\newcommand{\CLO}[1]{\mathcal{Z}^{#1}}
\newcommand{\EXA}[1]{\mathcal{B}^{#1}}
\def\THETA{\mbf{\mathrm{\Theta}}}
\def\OMEGA{\mbf{\mathrm{\Omega}}}
\def\aLPHA{\mbf{\mathrm{\alpha}}}

\def\BBB{\text{\textbf{\textit{B}}}}

\newcommand{\HOM}[1]{H^{#1}}

\def\wed{^{\hspace{1pt}\wedge}}
\newcommand{\DO}[1]{\mathfrak{D}\raisebox{-3.5pt}{\hspace{-2pt}}_{#1}}
\def\emb{\text{\scriptsize $\widehat{\text{\normalsize $\iota$}}$}}
\def\scdeg{\text{\rm sc.{\hspace{1pt}}d.}}
\def\Scdeg{\text{\rm Sc.{\hspace{1pt}}d.}}

\def \Lccl {\omega}

\def \Sccl {\gamma}
\def \sccl {\Gamma}
\def \bsccl {\Mbf{\Gamma}}

\def \Cc {\Mbf{g}}
\def \Aa {\Mbf{\alpha}}
\def \Bb {\Mbf{\beta}}
\def \Rr {\Mbf{r}}
\def \Ca {\mbf{a}}
\def \Bf {\mbf{\beta}}

\def\GF{\mathpzc{k}\hspace{-1pt}}
\def\Q{\mathbb{Q}}
\def\F{F}

\def\HF{\Mbf{F}}
\def\HES{\Mbf{E}}
\def\VSP{\mathbf{V}}
\def\PA{{\mathrsfs{O}\hspace{-1pt}}}
\def\FI{\mathcal{F}}
\def\HDP{\DP}
\def\RMA{R}
\def\PRT{\mathfrak{P}}
\def\VP{\HF}
\def\SRMA{{\mathop{\RMA}\limits^{\text{\tiny $\bullet$}}}}
\def\Prma{\mathcal{P}}
\def\PRMA{P}
\def\devdeg{\text{\rm dev.{\hspace{1pt}}d.}}

\def\NORM{\text{\it n.\hspace{1pt}f.}\hspace{1pt}}

\def\Rdf{Q}

\def\RIE{\mathcal{M}}
\def\TMPSP{\mathrsfs{V}}

\def\CONE{\mathcal{U}}
\usepackage{ulem}
\def\CO{\mathcal{C}}
\newenvironment{Proof}[1][\it Proof]{\noindent\textit{#1.}${}$\hspace{7pt}${}$}{\nolinebreak$\quad$\nolinebreak$\Box$}
\def\LADIAG{\widehat{\Delta}}
\def\SLccl{{\mathop{\Lccl}\limits^{\text{\tiny $\bullet$}}}}
\newcommand{\EMPH}[1]{{\it #1}}
\newcommand{\EMPHTH}[1]{{\rm #1}}
\def\SCDEG{\text{\it deg}}
\def\Ll{\ell}
\def\MM{m}
\def\KK{k}
\def\RDf{Q}
\def\spwedge{\mathop{\wedge}\limits^{\circ}}

\def\PAE{\hspace{2pt}\widetilde{\hspace{-2pt}\PA\hspace{2pt}}\hspace{-2pt}}

\def\LL{L}
\def\TMPSPP{\mathrsfs{W}}
\newcommand{\SOM}[1]{\Omega_{#1}}
\newcommand{\SOMD}[2]{\bigl(\Omega_{#2}\bigr)_{#1}}
\def\sppro{\bullet}

\title{Anomalies in Quantum Field Theory and Cohomologies of Configuration Spaces}

\author{Nikolay M. Nikolov}

\date{\DATE}

\begin{document}

\maketitle

\thispagestyle{empty}

\vspace{-0.8cm}

\begin{center}
\scriptsize
Institute for Nuclear Research and Nuclear Energy, \\
Tsarigradsko Chaussee 72, BG-1784 Sofia, Bulgaria \\
mitov@inrne.bas.bg
\end{center}

\vspace{0cm}

\begin{abstract}
In this paper we study systematically
the Euclidean renormalization in configuration spaces.
We investigate also the deviation from
commutativity of the renormalization
and the action of all linear partial differential operators.
This deviation is the source of
the
anomalies in quantum field theory,
including the renormalization group action.
It
also
determines a Hochschild $1$--cocycle
and the renormalization ambiguity corresponds
to a \textit{nonlinear} subset in the cohomology class
of this renormalization cocycle.
We show that the related cohomology spaces
can be reduced to de Rham cohomologies of the so called
``(ordered) configuration spaces''.
We find cohomological differential equations
that determine the renormalization cocycles
up to the renormalization freedom.
This analysis is a first step towards a new approach
for computing renormalization group actions.
It can be also naturally extended to manifolds as well as
to the case of causal perturbation theory.
\end{abstract}%

\tableofcontents

\section{Introduction}\label{se1}
\setcntrs

The methods of the abstract algebra
have been
successfully applied
for several decades to
two dimensional conformal Quantum Field Theory (QFT).
As a result, many nontrivial models
have been
found with a very rich and explicit structure.
A question naturally arises
to what extent
it is possible to apply these algebraic methods
to more general QFT and, for instance, to perturbative QFT.
Let us motivate in more details this expectation.
The two dimensional conformal field theories can be described
by
a purely algebraic structure
called vertex algebra (see e.g., \cite{K}).
This structure has an analog in higher space-time dimensions and
there it characterizes
in a purely algebraic way the class of so called globally conformal invariant
models of quantum fields (\cite{N06}, \cite{NT01}).
Since the perturbative QFT can be considered as a deformation theory of
vertex algebras then
one may expect
that the perturbation theory can be
purely algebraically developed.
On the other hand, more than
half a century experience
of perturbative QFT suggests that
the usage of some transcendental methods are necessary
even if
we are perturbing
free massless fields.
The present work, in particular,
arose from an attempt
to understand the exact place and kind of
the transcendental methods that are needed to derive
the renormalization group action.
We also consider the question what transcendental numbers appear in
the perturbative expansion of the beta function.
We conjecture,
based on
further investigations, which will be published
elsewhere,
that these transcendental numbers are multiple zeta values in any
perturbative QFT on even space-time dimensions.
Let us point also out that recently there are very
intensive investigations of integrability in supersymmetric gauge theories.
These expectations are related to the anomalous dimensions and hence,
having an algebraic approach to renormalization group
will be very beneficial for these studies.

\subsection{Motivation from perturbative QFT}\label{se1.1}

Perturbation theory in QFT is one of the
technically most difficult subjects in the contemporary theoretical physics.
This is, first and foremost, due to the appearance of complicated integrals in
higher orders, as well as, to the complexity of the accompanying renormalization.
For the realistic QFT models there are practically no numerical results
for arbitrary orders in perturbation theory.
There are also a few methods that allow to perform calculations to all orders.
Without pretending to give justice to various approaches to the subject
we just point out the general analysis of perturbative renormalization
in recent work of Connes-Kreimer (see e.g., \cite{K98,CK00, CK01}).

The present work is a first step to a new approach for determining the action of
the renormalization group in perturbative QFT (i.e., for calculating beta functions).
More generally, our analysis is also applicable to any anomalies in QFT.
It offers in addition a geometric insight to the problem.
The general idea of the method is to perform a cohomological analysis
of the renormalization ambiguity and to use it to determine the renormalization group action.
Furthermore, we separate the problem from the particular models of perturbative QFT,
i.e. we consider all possible theories and even more general situations.
It is this generality that makes the geometric interpretation possible.
It is also important that we do not confine our treatment to the one parameter
action of the renormalization group but consider all linear partial differential operators.
This is done in order to restrict as much as possible the related cohomology owing to the
general properties of the algebra of all differential operators.
Our geometric view favors the study of renormalization in ``coordinate space''.
This approach has been originally developed by Bogolubov, and
Epstein and Glaser \cite{EG73} on Minkowski space and
recently applied to more general pseudo-Riemann manifolds
(see e.g., \cite{BF00, HW}).
It is also called causal perturbation theory.
This approach has a counterpart in \textit{Euclidean} QFT
(\cite{Stora}, \cite{GB}),
which is in some respects simpler.
We choose to work here within this Euclidean framework, and even on
$\R^D$,
but our analysis can be extended to manifolds, as well as to the case of the
causal perturbation theory on pseudo-Riemann manifolds.
Our choice was motivated by the fact that
the geometric structures appearing in the analysis are much more transparent
in the Euclidean approach.

For the purpose of the present work we also systematically develop
the theory of Euclidean renormalization
on configuration spaces (Sect.~\ref{se2}).
This is done in the spirit of the Epstein--Glaser approach
in the causal perturbative QFT but in contrast
in the Euclidean case we deal only with the Green functions
and there are no time--ordered (or retarded) products of fields.
This makes different the axiomatic assumptions that are needed
in order to achieve the ``universal renormalization theorem''.
The latter statement is the fact that the
change of renormalization uniquely induces a change
of the coupling constants as formal power series.
(This is true for any theory, but only for the so called
renormalizable theories
a finite number of coupling constants remains
under an arbitrary change of renormalization.)
Let us point out that
to the best of our knowledge
there is no systematic
treatment
in the literature
of Euclidean renormalization on configuration spaces.
We
split
this theory in two parts:
the first of them is model independent and
we introduce there a general concept of renormalization maps
as acting on algebras whose elements
can be \textit{bare} Feynman amplitudes of any theory.
The remaining part of this study
will treat the application of the renormalization maps
to particular models of perturbative QFT.
We intend to
consider it in a future work.

We shall briefly explain the place of our analysis
within
the perturbative QFT.

In perturbative Euclidean QFT one computes
the Green functions as formal power series of the type:
\beqa\label{E1}
\podr
G_N \bigl(\z_1,\dots,\z_N; \Cc; \{\RMA_n\}\bigr)
\nn
\podr = \, \mathop{\sum}\limits_{\Rr \, \geqslant \, 0} \,
\mathop{\int}\limits_{\R^{D|\Rr|}} d^D \x_1 \cdots d^D \x_{|\Rr|}
\ \frac{\Cc^{\Rr}}{\Rr!} \,
\RMA_{|\Rr|} \, \mathcal{A}_{\Rr} \bigl(\x_1,\dots,\x_{|\Rr|};\z_1,\dots,\z_N\bigr)
\,.
\eeqa
Here: $\x_k=(x^1_k,\dots,x^D_k)$ and $\z_k$ $\in$ $\R^D$;
$\Rr$ $=$ $(r_1,\dots,r_s)$, $|\Rr|$ $=$ $\sum_j r_j$ and $\Rr!$ $=$ $\prod_j r_j!$
are multiindex notations;
$\Cc$ $=$ $(g_1,\dots,g_s)$ is a system of coupling constants;\footnote{%
or functions, with included space cutoff}
$\mathcal{A}_{\Rr} \bigl(\x_1,\dots,\x_{|\Rr|};\z_1,\dots,\z_N\bigr)$
are Feynman amplitudes (sums over Feynman graphs)
with $|\Rr|$ internal and $N$ external vertices;
$\{\RMA_n\}$ is a system of renormalization maps to be defined later.
The important point for us in Eq.~(\ref{E1})
is that after smearing the external points $\z_k$
of the Feynman amplitudes
$\mathcal{A}_{\Rr} \bigl(\x_1,\dots,\x_{|\Rr|};\z_1,\dots,\z_N\bigr)$
by test functions they become finite sums of products of type:
\beq\label{E2}
\left(\raisebox{14pt}{\hspace{-2pt}}\right.
\mathop{\prod}\limits_{1 \, \leqslant \, j \, < \, k \, \leqslant \, n}
\!\!
G_{jk} (\x_j-\x_k)
\left.\raisebox{14pt}{\hspace{-2pt}}\right) \! \left(\raisebox{14pt}{\hspace{-2pt}}\right.
\mathop{\prod}\limits_{m \, = \, 1}^n
\!
F_m (\x_m)
\left.\raisebox{14pt}{\hspace{-2pt}}\right)\,,
\eeq
where $G_{jk} (\x_j-\x_k)$ are ``propagators''
and $F_m (\x_m)$ are smooth functions on $\R^D$,
which arise from the smearing of the external propagators
($F(\x)$ $=$ $\int G(\x-\y)$ $f(\y)$ $d^D \y$).
Since the propagators $G_{jk} (\x_j-\x_k)$
are regular functions for $\x_j \neq \x_k$,
the integrands of type (\ref{E2}) are well defined,
regular functions on the subspace of all pairwise distinct arguments
$(\x_1,\dots,\x_n) \in \bigl(\R^D\bigr)^{\times\hspace{1pt}n}( \, \cong$ $\R^{Dn})$.
The latter subspace of $\R^{Dn}$
is also called an \textit{ordered configuration space}
over $\R^D$ and is denoted by $F_n \bigl(\R^D\bigr)$.
The configuration spaces are generally introduced for arbitrary manifold (or set) $X$ by:
$$
F_n \bigl( X \bigr)
\, = \,
\Bigl\{\bigl(\x_1,\dots,\x_n\bigr) \in X^{\times n} \, : \, \x_j \, \neq \, \x_k \ \ \text{if} \ \ j \, \neq \, k\Bigr\} \,
$$
and they are well studied (see e.g., \cite{FH}).
Hence, the renormalization maps $\RMA_n$ in (\ref{E1}) are introduced in order to make
the integrals well defined.
In particular, $\RMA_n$ should extend smooth functions on configuration spaces
to distributions over the whole space.\footnote{%
Let us point that we do not consider here the infrared problem,
i.e., an additional extension related to the integration over an infinite volume
(but it can be treated by the same method).
This is because we shall be mainly interested in the renormalization group action (see below),
which in QFT can be extracted only by its ultraviolet renormalization
(or, its short distance properties).}
The system $\{\RMA_n\}$ have to satisfy certain natural properties
that we shall consider in Sect.~\ref{se2}.

A perturbative QFT is said to be renormalizable if after changing the system
of renormalization maps, $\{\RMA_n\}$ $\to$ $\{\RMA_n'\}$,
there exits a unique formal power series
$$
\Aa \bigl(\Cc\bigr) \, = \,
\mathop{\sum}\limits_{\Rr \, \geqslant \, 0} \, \Ca_{\Rr} \, \frac{\Cc^{\Rr}}{\Rr!}
$$
such that
$$
G_N \bigl(\z_1,\dots,\z_N; \Cc; \{\RMA_n'\}\bigr)
\, = \,
G_N \bigl(\z_1,\dots,\z_N; \Aa \bigl(\Cc); \{\RMA_n\}\bigr)
$$
in the sense of formal power series.
(More generally, one can consider all possible interactions each switched
with its own coupling constant.
Then the change of renormalization uniquely induces an action
on the infinite set of coupling constants in the above sense
of formal power series.
This is the statement we called above ``universal renormalization theorem''.)

In particular, one can change the renormalization maps by a dilation (i.e., changing the ``scale''):
$$
\RMA_n^{\lambda} \, := \, d_{\lambda} \circ \RMA_n \circ d_{\lambda}^{-1} \,
$$
(where $\bigl(d_{\lambda} f\bigr)(\x_1,\x_2,\dots) := f\bigl(\lambda^{-1}\x_1,\lambda^{-1}\x_2,\dots)$
for $\lambda \in \R^+$).
As a result, this generates the action of renormalization group ($R^+ \ni \lambda$)
on the coupling constants
$$
\{\RMA_n^{\lambda}\} \quad \to \quad
\Aa_{\lambda} \bigl(\Cc\bigr) \,,
$$
whose generator is the so called \textit{beta function}
$$
\Bf \bigl(\Cc\bigr) \, = \,
\lambda \, \frac{d}{d\lambda} \ \Aa_{\lambda} \bigl(\Cc\bigr) \vrestr{12pt}{\lambda \, = \, 1} \,.
$$
Clearly, to compute the formal power series of $\Bb \bigl(\Cc\bigr)$
one has to know the ``commutators''
$$
(\xx \cdot \di_{\xx}) \circ \RMA_n - \RMA_n \circ (\xx \cdot \di_{\xx}) \,,
$$
where $\xx \cdot \di_{\xx}$ $:=$ $\mathop{\sum}\limits_{k,\mu} \, x^{\mu}_k \, \frac{\di}{\di x^{\mu}_k}$.
We shall generalize this task and will look for the commutators
$$
c_n[A] \, = \, A \circ \RMA_n - \RMA_n \circ A
$$
for all linear partial differential operators $A$.
It turns out that $c_n [A]$ is a certain Hochschild cocycle
and changing the renormalization corresponds to adding a coboundary
(see Sect.~\ref{se4}).
So, we would like to find some cohomological equations that would determine $c_n[A]$.

From our analysis in Sect.~\ref{se2},
it follows that the renormalization ambiguity allows us to achieve
$c_n \bigl[f\bigr] = 0$ for smooth functions $f$ (i.e., differential operators of zeroth order).
This allows us to extend our methods also for manifolds without even any metric structure on them.
Moreover, the remaining nontrivial part of the cocycle $c_n \bigl[\di_{x_k^{\mu}}\bigr]$
($\di_{x_k^{\mu}}$ $:=$ $\frac{\di}{\di x_k^{\mu}}$)
can be characterized by certain cohomological equations (Eqs.~(\ref{eq4.6ne2}) and (\ref{eq6.9ne1})).
We prove in Theorem~\ref{Th4.1qq1} that
the cohomological ambiguity in the solutions of these equations
exactly corresponds to the renormalization freedom.
Then we reduce the related cohomologies to
de Rham cohomologies of configuration spaces.

So, in the subsequent section we shall introduce
the precise notion of renormalization maps.
Then, we analyze the remaining nontrivial cohomological properties
of the renormalization maps
and their reduction to de Rham cohomologies.
The essential material of Sect.~\ref{se2} that
is needed
for Sect.~\ref{se4} is contained
in Sects.~\ref{se2.0}--\ref{se2.2}.

\medskip

\noindent
\textit{Some
common
notations.}\
$\N$ $=$ $\{1,$ $2,$ $\dots\}$,\ $\N_0$ $=$ $\{0,$ $1,$ $2\dots\}$,\
$\Z$ $=$ $\{0,$ $\pm 1,$ $\pm 2,$ $\dots\}$;\
$\Q$, $\R$ and $\C$ are the fields of rational, real and complex numbers,
respectively.
For a finite set $S$, $|S|$ stands for the number of its elements.
The vectors in Euclidean ``space--time''
$\R^D$ ($D$ standing for the space--time dimension)
are denoted by $\x$ $=$ $(x^1,\dots,x^D)$, $\y$, $\dots$.
Some times, we shall also deal with $\R^N$ for other $N \in \N$
(for instance, $\bigl(\R^D\bigr)^{\times n}$ $\equiv$ $\R^{Dn}$)
and then we shall denote its elements by
$\xx$ $=$ $(x^1,\dots,x^N)$, $\yy$, $\dots$.
\textit{Multiindex notations}.
For
$\rr$ $=$ $(r_1,\dots,r_N)$ $\in$ $\N_0^N$:
$|\rr|$ $=$ $\sum_j r_j$,
$\rr!$ $=$ $\prod_j r_j!$,
$\xx^{\rr}$ $=$ $\prod_j \bigl(x^j\bigr)^{r_j}$
and
$\Mbf{\di}_{\xx}^{\rr}$ $=$ $\prod_j \di_{x^j}^{r_j}$
$(\, =$
$\prod_j \Bigl(\frac{\di}{\di x^j}\Bigr)\raisebox{12pt}{\hspace{-1pt}}^{r_j})$.

\section{Theory of renormalization maps}\label{se2}
\setcntrs

\subsection{Preliminary notions}\label{se2.0}

In this subsection we shall introduce for every $n=2,3,\dots$
an algebra $\PA_n$, which
should be thought of
as the algebra of
``$n$--point Feynman diagrams''.
All these algebras are built by the algebra
$\PA \equiv \PA_2$,
which is ``the algebra of propagators''.
So, we shall fix only
the propagator structure
of the theory
but not
the vertex structure.
In fact, even the propagators will be fixed only
as a type of functions
but not as an explicit system of functions
(say, scalar propagators, spinor propagators etc.).
For instance, if we
perturb
massless fields then
the algebra $\PA$ is just the algebra of rational functions $G(\x)$
on the Euclidean space-time $\R^D \ni \x$,
which denominators are only powers of the interval $\x^2$
$(:=$ $(x^1)^2$ $+$ $\dots$ $+$ $(x^D)^2)$.
(This algebra is denoted by $\Q [\x,1/\x^2]$
if the coefficients are rational numbers.)
We shall introduce later
the renormalization maps as certain linear maps acting
on the algebras $\PA_n$.

Since we wish to take into account what transcendental methods and numbers
are needed to describe renormalization we shall fix some ground field
$\GF$ $\subseteq$ $\R$,
which will be assumed to be the field $\Q$
of rational numbers
if not stated otherwise.
The vector spaces and associative algebras will be usually assumed over $\GF$
(except for some standard spaces as the distributions spaces which will
be considered on $\R$).

All the algebras we shall use
will be \textbf{differential algebras}.
A differential algebra with $N$--derivatives is
a \textit{commutative} associative algebra
endowed with linear operators
$t_j$ and $\di_{t_j}$ for $j$ $=$ $1,$ $\dots,$ $N$
such that, they satisfy
the Heisenberg commutation relations:
$$
t_jt_k - t_kt_j \, = \,
\di_{t_j}\di_{t_k} - \di_{t_k}\di_{t_j} \, = \, 0
,\quad
t_j \di_{t_k} - \di_{t_k}t_j \, = \, \delta_{jk} \,,
$$
$\di_{t_j}$ satisfy the Leibnitz rule, and $t_j$ commute with
the multiplication by the elements of the algebra:
$$
\di_{t_j} (ab) \, = \, \di_{t_j} (a) \, b + a \, \di_{t_j} b
,\quad
t_j (ab) \, = \, a \, t_j b\,.
$$
An example of differential algebra is the $\R$--algebra
$\CI \bigl(\R^{Dn}\bigr)$
of smooth functions $f(\x_1,\dots,\x_n)$ over $\R^{Dn}$,
where we shall denote the operators $t_j$ and $\di_{t_j}$
by $x_k^{\mu}$ and $\di_{x_k^{\mu}}$ $\equiv$ $\frac{\di}{\di x_k^{\mu}}$
($k=1,\dots,n$, $\mu = 1,\dots,D$),
respectively,
but some times we shall also replace the pair of indices $(k,\mu)$ with a single letter
$\xi$ and write $x^{\xi}$ and $\di_{x^{\xi}}$.

We shall
focus
our analysis on translation invariant $n$--point functions
on the Euclidean space $\R^D$ and
for short
we denote the quotient space:
\beq\label{quo1}
\HES_n \, := \, \bigl(\R^D\bigr)^{\times n} \bigl/ \R^D
\, \equiv \, \R^{Dn} \bigl/ \R^D ,
\eeq
where the quotient is taken under the action of $\R^D$ by translations:
$\bigl(\x_1,$ $\dots,$ $\x_n\bigr)$ $\mapsto$
$\bigl(\x_1$ $+$ $\u,$ $\dots,$ $\x_n$ $+$ $\u\bigr)$.
Thus, we have an isomorphism
\beq\label{isomm}
\HES_n \, \cong \, \R^{D(n-1)}
\, : \,
\bigl[\x_1,\dots,\x_n\bigr] \mapsto \bigl(\x_1-\x_n,\dots,\x_{n-1}-\x_n\bigr) \,,
\eeq
where $\bigl[\x_1,\dots,\x_n\bigr]$ stands for a class
$\bigl(\x_1,\dots,\x_n\bigr)$ mod $\R^D$.
Recall that $\F_n \bigl(\R^D\bigr)$
stands for the configuration space over $\R^D$
and denote
$$
\HF_n \, := \, \F_n \bigl(\R^D\bigr) \bigl/ \R^D
$$
where
quotient is taken again under the above action of
$\R^D$ on $\R^{Dn}$.
We have again an isomorphism
\beq\label{isom2}
\HF_n \, \cong \, \F_{n-1} \bigl(\R^D \mz\bigr)
\, : \,
\bigl[\x_1,\dots,\x_n\bigr] \mapsto \bigl(\x_1-\x_n,\dots,\x_{n-1}-\x_n\bigr) \,.
\eeq
We consider the above isomorphisms (\ref{isomm}) and (\ref{isom2})
as identifications.
For every finite nonempty subset $S \subset \N$ we similarly denote:
\beqa
&
\HES_S \, := \,
\bigl(\R^D\bigr)^S \bigl/ \R^D ,
& \nn &
\HF_S \, := \,
\bigl\{
(\x_j)_{j \in S} \in \bigl(\R^D\bigr)^S
: \x_j \neq \x_k \text{ for } j \neq k
\bigr\} \bigl/ \R^D
\ ( \, \equiv \F_S \bigl(\R^D\bigr) \bigl/ \R^D)
\,.
& \nonumber \eeqa

Now
we assume that we are given a differential subalgebra (over the ground field $\GF$):
$$
\PA \equiv \PA_2 \subseteq \CI \bigl(\R^D\mz\bigr) \,.
$$
of the algebra of smooth functions on $\R^D\mz$ $(\equiv$ $\HF_2)$.
(As we mentioned, one should
think of the algebra
$\PA$ as an algebra of ``propagators''.)
We define several embeddings:
$$
\emb_{jk} : \PA \hookrightarrow \CI \bigl(\HF_n\bigr) :
G(\x) \mapsto G(\x_j-\x_k)
, \quad
1 \leqslant j < k \leqslant n
,
$$
where (as well as further)
\textbf{we shall identify the functions belonging to
$\CI \bigl(\HF_n\bigr)$ with translation invariant functions over
$F_n\bigl(\R^D\bigr)$} (as in Eq.~(\ref{isom2})).
Then for every
$S \subseteq \{1,\dots,n\}$ with $|S| \geqslant 2$ we set
\beqa
\PA_S \, := \podr \text{\parbox[t]{200pt}{the subalgebra of $\CI \bigl(\HF_n\bigr)$ generated by all
$\emb_{jk} \bigl(\PA\bigr)$ for $j,k \in S$, $j<k$,}}
\nn
\PA_n \, := \podr \PA_{\{1,\dots,n\}} \,.
\nonumber
\eeqa
In fact, we shall not keep $n$ fixed and we consider the natural inductive limit
$\PA_{\infty}$ $:=$
$\mathop{\bigcup}\limits_{n \, = \, 2}^{\infty} \, \PA_n$,
and so, for every finite subset $S \subset \N$ with $|S| \geqslant 2$ we consider $\PA_S$
as a subalgebra of $\PA_{\infty}$.
Note that the algebra $\PA_S$ is linearly spanned by functions of the form
\beq\label{privod}
G \, = \,
\mathop{\prod}\limits_{\mathop{}\limits_{j \, < \, k}^{j,k \, \in \, S}}
\emb_{jk} G_{jk} \, \equiv \,
\mathop{\prod}\limits_{\mathop{}\limits_{j \, < \, k}^{j,k \, \in \, S}}
G_{jk} \bigl(\x_j-\x_k\bigr)
, \quad G_{jk} \, \in \, \PA .
\eeq

Basic examples for the algebra $\PA$ are the algebras
$\GF [\x,$ $1/\x^2]$ and $\GF [\x,$ $1/\x^2,$ $\log \, \x^2]$.
In the first case the corresponding algebras $\PA_n$ are:
\beq\label{PAN}
\PA_n \, = \,
\GF \bigl[\x_1,\dots,\x_{n-1}\bigr] \hspace{-2pt}
\left[\raisebox{18pt}{\hspace{-3pt}}\right.
\left(\raisebox{16pt}{\hspace{-2pt}}\right.
\mathop{\prod}\limits_{k \, = \, 1}^{n-1} \x_k^2
\left.\raisebox{16pt}{\hspace{-3pt}}\right)^{-1}
\left(\raisebox{16pt}{\hspace{-2pt}}\right.
\mathop{\prod}\limits_{1 \, \leqslant \, j \, < \, k \, \leqslant \, n-1}
(\x_j-\x_k)^2
\left.\raisebox{16pt}{\hspace{-3pt}}\right)^{-1}
\left.\raisebox{18pt}{\hspace{-3pt}}\right] \,
\eeq
(under the identification (\ref{isom2})).
These are the algebras that we need if we
perturb
massless free fields.
From the point of the algebraic geometry the algebra $\PA_n$
coincides exactly with the ring of regular functions on the
affine manifold that is complement of union of the quadrics
$\x_k^2$ $=$ $0$ and $(\x_j-\x_k)^2$ $=$ $0$.

We introduce similar notations for the distributions spaces:
\beqa
&
\HDP_n \, := \,
\DP \bigl(\HES_n\bigr)
,
& \nn &
\HDP_S \, := \, \bigl\{u \in \HDP_{\infty} :
\text{$u$ depends at most on $\x_j$ for all $j \in S$}\bigr\}
\equiv \DP \bigl(\HES_S\bigr),
& \nn &
\HDP_{S,0} \, := \, \bigl\{u \in \HDP_S : supp \, u \subseteq
\{0\} \subset \HES_S
\bigr\} .
\nonumber &
\eeqa

We note that for every inclusion $S' \subseteq S$
we have natural embeddings
$\PA_{S'}$ $\hookrightarrow$ $\PA_S$ and
$\DP_{S'}$ $\hookrightarrow$ $\DP_S$.

\subsection{Filtrations}\label{se2.2nw}

A very important notion in the Epstein--Glaser approach
to the renormalization is
the Steinmann
\textit{scaling degree} (\cite[Ch. 5]{S71}).
It corresponds to the degree of divergence in the other approaches
and introduces filtrations on our function spaces.

First, for a distribution $u \in \DP \bigl(\R^N \mz \bigr)$
the scaling degree is defined as:
$$
\scdeg \, u \, := \,
\mathop{\inf} \, \bigl\{\lambda \in \R
\, : \,
\mathop{\text{w-\!}\lim}\limits_{\varepsilon \downarrow 0} \,
\varepsilon^{\lambda} \, u (\varepsilon \xx) \, = \, 0 \bigr\} \,
$$
($\mathop{\text{w-\!}\lim}$ standing for the weak limit).
If $u \in \DP \bigl(\R^N\bigr)$ then $\scdeg \, u$
is defined similarly but note that:
\beq\label{note1}
\scdeg \, \Bigl( u \vrestr{10pt}{\R^N \mz} \Bigr)
\, \leqslant \,
\scdeg \, u
\eeq
(for instance, take $u(\x) = \delta (\x)$).
There is a
theorem (\cite[Lemma 5.1]{S71}) stating that
\textit{every distribution belonging
to $\DP \bigl(\R^N\bigr)$ has a finite scaling degree}.
Furthermore,
\textit{a necessary and sufficient condition for
a distribution on $\R^N \mz$
to posses an extension over the whole space $\R^N$
is the finiteness of its scaling degree}
(cf. Lemma \ref{Lm2.4} and the construction after it).
Let us also point out the inequalities
\beq\label{ineqq}
\scdeg \, x^{\xi} u(\xx) \, \leqslant \, -1 + \scdeg \, u
,\quad
\scdeg \, \di_{x^{\xi}} u(\xx) \, \leqslant \, 1 + \scdeg \, u
\eeq
($u \in \DP \bigl(\R^N \mz\bigr)$, $\xi=1,\dots,N$).

Thus, we obtain \textit{increasing filtrations} on all
$\DP \bigl(\R^N\bigr)$, $N \in \N$
(and hence, on each $\HDP_S$):
\beqa
\label{FI1}
& \hspace{-10pt}
\FI_{\Ll}\hspace{1pt} \DP \bigl(\R^N\bigr) \, := \,
\bigl\{u \in \DP \bigl(\R^N\bigr) : \scdeg \, u \, \leqslant \, \Ll \bigr\}
, \quad \Ll \, \in \, \R,
& \\ & \hspace{-10pt}
\FI_{\Ll} \DP \bigl(\R^N\bigr) \, \subseteq \,
\FI_{\Ll'} \DP \bigl(\R^N\bigr) \quad \text{for} \quad
\Ll \leqslant \Ll'
\, , \quad
\DP \bigl(\R^N\bigr)
\, = \,
\mathop{\bigcup}\limits_{\Ll \, \in \, \R} \,
\FI_{\Ll}\hspace{1pt} \DP \bigl(\R^N\bigr) \,.
& \nonumber
\eeqa

We introduce the distribution spaces:
$$
\DP_{temp} \bigl(\R^N \mz\bigr) \, := \,
\bigl\{u \in \DP \bigl(\R^N \mz\bigr) :
\scdeg \, u \, < \, \infty \bigr\} \,.
$$
Then on $\DP_{temp} \bigl(\R^N \mz\bigr)$ we have again a filtration:
$$
\FI_{\Ll}\hspace{1pt} \DP_{temp} \bigl(\R^N \mz\bigr)
\, := \, \bigl\{u \in \DP_{temp} \bigl(\R^N \mz\bigr) :
\scdeg \, u \, \leqslant \, \Ll \bigr\}\,
\quad (\Ll \, \in \, \R).
$$
According to the result mentioned above:
\textit{a distribution on $\R^N \mz$
belongs to the space
$\DP_{temp} \bigl(\R^N \mz\bigr)$
iff it is a restriction of a distribution on $\R^N$}.

We also introduce filtrations on the spaces $\PA_S$
(for finite subsets $S \subset \N$ with $|S| \geqslant 2$).
Starting with $\PA \equiv \PA_2 \subseteq \CI \bigl(\R^D \mz\bigr)$
we introduce first another, stronger notion of scaling degree on
$\CO \bigl(\R^N \mz\bigr)$ ($N \in \N$):
for $G$ $\in$ $\CI \bigl(\R^N \mz\bigr)$
we set
\beqa
\Scdeg \, G :=
\mathop{\inf} \, \bigl\{\lambda \in \R
: \podr
\exists \ C_{\lambda}, R \, > \, 0 \text{ such that }
\bigl| \Mbf{\di}_{\xx}^{\rr} G (\xx) \bigr|
\, < \,
C_{\lambda,\rr} \, |\xx|^{-\lambda-|\rr|}
\nn \podr
\text{for all }
\rr \in \N_0^N \text{ and }
|\xx| \, < \, R
\bigr\}
\nonumber
\eeqa
($|\xx|$ $:=$ $\sqrt{\xx^2}$).
Note that for every $G \in \CI \bigl(\R^N \mz\bigr)$ we have
\beqa\label{Scdeg}
&
\scdeg \, G \, \leqslant \, \Scdeg \, G ,
& \nn &
\Scdeg \, x^{\xi} G(\xx) \, \leqslant \, -1 + \Scdeg \, G
,\quad
\scdeg \, \di_{x^{\xi}} G(\xx) \, \leqslant \, 1 + \Scdeg \, G,
& \quad
\eeqa
where in the first inequality we view
$G$ also as a distribution over $\R^N \mz$.
Then we assume that:
$$
\PA \, \subset \, \CI_{temp} \bigl(\R^D \mz\bigr) \, := \,
\bigl\{G \in \CI \bigl(\R^D \mz\bigr) :
\Scdeg \, G \, < \, \infty \bigr\} \,
$$
and hence, we obtain an increasing filtration on $\PA$:
\beqa
&
\FI_{\Ll} \PA \, := \, \bigl\{G \in \PA :
\Scdeg \, u \, \leqslant \, \Ll \bigr\}
\quad (\Ll \, \in \, \R)
\,,
& \nn &
\FI_{\Ll} \PA \, \subseteq \, \FI_{\Ll'} \PA \quad \text{for} \quad
\Ll \leqslant \Ll'
\, , \quad
\PA \, = \,
\mathop{\bigcup}\limits_{\Ll \, \in \, \R} \, \FI_{\Ll} \PA \,.
& \nonumber
\eeqa

Now
on every $\PA_S$ with $|S| \geqslant 3$
we introduce a filtration by the ``power counting procedure''.
For a general element $G \in \PA_S$:
\beq\label{genel}
G \, = \,
\mathop{\sum}\limits_{\alpha} G_{\alpha}
,\quad
G_{\alpha} \, = \, \mathop{\prod}
\limits_{\mathop{}\limits_{j \, < \, k}^{j,k \, \in \, S}}
\emb_{jk} G_{jk}^{\alpha}
, \quad
G_{jk}^{\alpha} \in \PA \,
\eeq
we set a \textbf{degree of divergence}
$$
\devdeg \, G \, := \,
\mathop{\inf}
\limits_{\mathop{}
	\limits^{\text{all possible rep--}}_{\text{resentations (\ref{genel})}}}
\left\{\raisebox{18pt}{\hspace{-2pt}}\right.
\mathop{\max}\limits_{\alpha}
\left\{\raisebox{16pt}{\hspace{-2pt}}\right.
\mathop{\sum}
\limits_{\mathop{}\limits_{j \, < \, k}^{j,k \, \in \, S}}
\Scdeg \, G_{jk}^{\alpha}
\left.\raisebox{16pt}{\hspace{-2pt}}\right\}
\left.\raisebox{18pt}{\hspace{-4pt}}\right\} \,,
$$
Thus, we obtain an increasing filtration:
\beq\label{FI2}
\FI_{\Ll}\hspace{1pt} \PA_S \, := \,
\bigl\{G \in \PA_S : \devdeg \, G \, \leqslant \, \Ll \bigr\}
\quad (\Ll \, \in \, \R),
\eeq
$$
\FI_{\Ll} \PA_S \, \subset \,
\FI_{\Ll'} \PA_S \quad \text{for} \quad
\Ll \leqslant \Ll'
\, , \quad
\PA_S
\, = \,
\mathop{\bigcup}\limits_{\Ll \, \in \, \R} \,
\FI_{\Ll}\hspace{1pt} \PA_S \,.
$$

In the sequel we shall need the following statement

\medskp

\begin{Lemma}\label{Scdeg-lm}
$(a)$
Let $u \in \DP_{temp} \bigl(R^N \mz\bigr)$ and
$F \in \CI_{temp} \bigl(\R^N \mz\bigr)$, then we have:
$\scdeg \, (Fu)$ $\leqslant$ $\Scdeg \, F$ $+$ $\scdeg \, u$.

$(b)$ Let $u_k \in \DP_{temp} \bigl(R^{N_k} \mz\bigr)$
for $k=1,\dots,m$.
Then $u$ $=$ $u_1 \otimes$ $\cdots$ $\otimes u_n$
$\in$ $\DP_{temp} \bigl(R^N$ $\mz\bigr)$,
$N$ $=$ $N_1$ $+$ $\cdots$ $+$ $N_m$
and $\scdeg \, u$ $=$ $\scdeg \, u_1$ $+$ $\cdots$ $+$ $\scdeg \, u_m$.
\end{Lemma}%

\medskp

\noindent
\textit{Sketch of the proof.}
$(a)$ By the Banach-Steinhaus theorem~(\cite{Rud})
it follows that:
$\scdeg \, u$ $<$ $\lambda$
\textit{iff}
for every compact $K \subset \R^N \mz$
there exist $L=L(\lambda) \in \N_0$,
a test functions norm $\|$$\cdot$$\|_{K,L}$,
$$
\|f\|_{K,L} \, := \,
\mathop{\sup}
\limits_{\xx \, \in \, K,\, |\rr| \, \leqslant \, L} \,
\bigl|\Mbf{\di}_{\xx}^{\rr} f (\xx)\bigr|
,
$$
and a constant $C_{K,\lambda} > 0$
such that for every
$f \in \Dd (\R^N)$ with $\text{\it supp} \, f \subseteq K$
and $\varepsilon$ $\in$ $(0,1)$ we have:
$$
\Bigl|u \bigl[f \bigl(\varepsilon^{-1} \xx \bigr)\bigr] \Bigr|
\, \leqslant \,
C_{K,\lambda} \, \bigl\|f\bigr\|_{K,L} \, \varepsilon^{N-\lambda} \,.
$$
Let $\Scdeg \, F + \scdeg \, u$ $<$ $\lambda$.
There exist $\lambda_1,$ $\lambda_2$ such that
$\Scdeg \, F$ $<$ $\lambda_1$,
$\scdeg \, u$ $<$ $\lambda_2$ and
$\lambda$ $=$ $\lambda_1+\lambda_2$.
Hence,
\beqa
\Bigl|(Fu) \bigl[f \bigl(\varepsilon^{-1} \xx \bigr)\bigr] \Bigr|
\, \leqslant \podr
C_{K,\lambda_2} \,
\bigl\|F(\varepsilon \xx) \, f (\xx)\bigr\|_{K,L_2} \,
\varepsilon^{N-\lambda_2}
\nn
\, \leqslant \podr
C' \, \bigl\|f (\xx)\bigr\|_{K,L_2}
\, \varepsilon^{N-\lambda_1-\lambda_2}
\,,
\nonumber
\eeqa
where $L_2=L(\lambda_2)$. It follows that
$\scdeg \, (Fu)$ $<$ $\lambda$.

$(b)$ We have: $\scdeg \, u$ $<$ $\lambda$ iff for every compact
$K$ $\subset$ $\R^N$ we have inequality
$\Bigl|u \bigl[f_1 \bigl(\varepsilon^{-1} \xx_1 \bigr)$
$\cdots$ $f_m \bigl(\varepsilon^{-1} \xx_m \bigr)\bigr] \Bigr|$
$\leqslant$ $C$
$\bigl\|f_1\bigr\|_{K,L_1}$ $\cdots$
$\bigl\|f_m\bigr\|_{K,L_m}$
$\varepsilon^{N-\lambda}$.
Then by the kernel and the Banach-Steinhaus theorems
we obtain an inequality
$\Bigl|u \bigl[f \bigl(\varepsilon^{-1} \xx \bigr) \Bigr|$
$\leqslant$ $C'$
$\bigl\|f\bigr\|_{K,L}$ $\varepsilon^{N-\lambda}$
every $f$ $\in$ $\Dd \bigl(\R^N$ $\mz\bigr)$.
$\quad\Box$

\subsection{Preliminary definition of renormalization maps}\label{se2.1}

We proceed to give a preliminary
definition of renormalization maps.
It will be just
a list of desirable properties for them.

A system renormalization maps is a collection of linear maps:
\beq\label{RMA}
\RMA_S : \PA_S \to \DP_S
\quad (\RMA_{\{1,\dots,n\}} =: \RMA_n) \,,
\eeq
indexed by all finite subsets $S \subset \N$ with at least two elements.
They are assumed to satisfy the properties (r1)--(r4) listed below.
Let us stress that we do not assume any nontrivial continuity of
the maps $\RMA_S$~(\ref{RMA})
and we consider them just as linear maps.

\medskip

($r1$)
For every bijection $\sigma : S \cong S'$ we require:
$$
\sigma^* \circ \RMA_{S'} \, = \, \RMA_S \hspace{1pt} \circ \sigma^*
,\quad
$$
where $(\sigma^* F) \bigl(\x_{j_1},\dots,\x_{j_n}\bigr)$
$:=$ $F \bigl(\x_{\sigma (j_1)},\dots,\x_{\sigma (j_n)}\bigr)$
for a function or distribution $F$.

\medskip

Thus, by ($r1$) all $\RMA_S$ are characterized by $\RMA_n$ for
$n=|S|=2,3,\dots$.

\medskip

($r2$)
Every $\RMA_S$ preserves the filtrations:
$\RMA_S \, \FI_{\Ll}\hspace{1pt} \PA_S \subseteq \FI_{\Ll}\hspace{1pt} \DP_S$
($\Ll \in \R$).
In other words, we require for $|S| \geqslant 3$:
\beq\label{r2}
\scdeg \, \RMA_S \, G \, \leqslant \, \devdeg \, G \,,
\eeq
which is also true for $|S|=2$ if we set then
$\devdeg \, G$ $\equiv$ $\Scdeg \, G$.

\medskip

($r3$)
For every polynomial $f \in \GF \bigl[\HES_n\bigr]$
and $G \in \PA_n$
we have:
$$
\RMA_n f G \, = \, f \, \RMA_n G .
$$

\medskip

All the above conditions are of linear type with respect to $\{\RMA_n\}$,
but the last one is ``nonlinear''.
To state it we need more notations.
Let $\PRT = \{S_1,\dots,S_k\}$ be
an
\textit{$S$--partition},
i.e., $S$ $=$ $S_1$ $\dot{\cup}$ $\cdots$ $\dot{\cup}$ $S_k$
for nonempty $S_1,$ $\dots,$ $S_k$
(in the case $S$ $=$ $\{1,$ $\dots,$ $n\}$ we shall say
\textit{$n$--partition}).
The $S$--partition $\PRT$
can be characterized also as an equivalence relation on $S$
with equivalence classes $S_1,\dots,S_k$.
This relation we denote by $\sim_{\PRT}$.
Then, by the construction of the algebra $\PA_S$
it is linearly spanned by elements of a form:
\beq\label{GS}
G_S \, = \,
G_{\PRT} \, \cdot \,
\mathop{\prod}\limits_{S' \, \in \, \PRT} G_{S'} ,
\eeq
where $G_{S'} \in \PA_{S'}$ for $S' \in \PRT$ and
$G_{\PRT} \in \PA_{\PRT}$,
\beq\label{E2.1}
\PA_{\PRT} \, := \, \text{\parbox[t]{200pt}{the subalgebra of $\CI \bigl(\HF_n\bigr)$ generated by all
$\emb_{jk} \bigl(\PA\bigr)$ for $j,k \in S$, $j\nsim_{\PRT} k$.}}
\eeq
In the case when $\PRT$ contains $S'$ with $|S'|=1$ we shall assume
in Eq.~(\ref{GS}) that $G_{S'}$ $:=$ $1$ and set also
$$
\PA_{S'} \, = \, \PA_1 \, := \, \GF \quad (|S'|=1) .
$$
Another extreme case is when $|\PRT| = 1$ (i.e., $\PRT=\{S\}$)
and then we shall again assume
$G_{\PRT}$ $:=$ $1$ and set $\PA_{\PRT}$ $:=$ $\GF$.
Finally, we define the following open subsets of $\HES_n$
\beq\label{VP}
\VP_{\PRT} \, = \,
\bigl\{ \bigl[\x_1,\dots,\x_n\bigr]
\, \in \, \HES_n
\, : \, \x_j \, \neq \, \x_k \text{ if } j \, \nsim_{\PRT} \, k \bigr\} \,
\eeq
(this is for the case of
an
$n$--partition $\PRT$ and similarly one introduces
$\VP_{\PRT}$ for arbitrary $S$--partition).
Note that for the case $|\PRT|=1$
Eq.~(\ref{E2.1}) becomes an identity and $\VP_{\PRT} = \emptyset$.
Partitions containing at least two elements are called \textbf{proper}.
Let us also point out the
similarity
between the definition of $\VP_{\PRT}$ and
the definition of the configuration spaces $\HF_n$.
In fact, $\VP_{\{\{1\},\dots,\{n\}\}}$ $\equiv$ $\HF_n$.
Because of this
similarity
we have chosen one and the same letter
for the notations.
In the case of algebras $\PA_n$~(\ref{PAN})
the algebra $\PA_{\PRT}$ contains exactly those the elements of $\PA_n$,
which are regular functions on $\VP_{\PRT}$
(similarly, $\PA_n$ contains regular functions on $\HF_n$).

\medskip

($r4$)
For every
\textit{proper}
$S$--partition $\PRT$
we have:
\beq\label{r4}
\RMA_S G_S \vrestr{12pt}{\VP_{\PRT}} \, = \,
G_{\PRT} \, \cdot \,
\mathop{\prod}\limits_{S' \, \in \, \PRT} \RMA_{S'} G_{S'} \,
\eeq
where $G_S$ is of the form (\ref{GS}).

\medskip

In the extreme case of ($r4$) when
$\PRT$ contains $S'$ with $|S'|=1$ we set in addition
to the above convention $\PA_1 := \GF$ that:
$$
\RMA_1 \, := \, \id.
$$
Then we obtain as a consequence that
$$
\RMA_n G \vrestr{12pt}{\HF_n} \, = \, G
$$
for all $G \in \PA_n$
(since $\VP_{\{\{1\},\dots,\{n\}\}}$ $=$ $\HF_n$).

Another corollary of ($r4$) is that if we have
a system of linear maps $\RMA_{S'}$ (\ref{RMA})
that are defined
for all finite
$S' \subset \N$ with $2 \leqslant |S'| \leqslant n-1$
and satisfy ($r1$)--($r4$)
then we have linear maps for all $S$ with $|S|=n$:
$$
\SRMA_S : \PA_S \to \DP \bigl(\HES_S \mz\bigr)
$$
uniquely determined by the condition that
$\SRMA_S \, G_S \vrestr{10pt}{\VP_{\PRT}}$
is equal to the right hand side of Eq.~(\ref{r4})
for every proper $S$--partition $\PRT$.
This follows from the fact that $\VP_{\PRT}$ form an open covering of
$\HES_S \mz$:
\beq\label{ocov}
\HES_S \mz \, = \, \mathop{\bigcup}
\limits_{\mathop{}\limits^{\PRT \text{ is a proper}}_{\text{$S$--partition}}}
\VP_{\PRT} \,.
\eeq
Furthermore, if the linear maps $\RMA_{S'}$ ($|S'| \leqslant n-1$)
are part of a complete system
of renormalization maps then
\beq\label{EX}
\RMA_S G \vrestr{12pt}{\HES_S \mz} \, = \,
\SRMA_S G
\eeq
for all $G \in \PA_S$.
We set again
$$
\SRMA_n \, := \, \SRMA_{\{1,\dots,n\}} \,,
$$
and clearly, all $\SRMA_S$ with $|S|=n$ are isomorphic to $\SRMA_n$.

\medskp

\begin{Lemma}\label{Lm2.1}
The image of the linear map $\SRMA_n$
is contained
in the space
$\DP_{temp} \bigl(\HES_n$ $\mz\bigr)$.
In fact, $\SRMA_n$ preserves the filtrations:
\beq\label{E2.2}
\SRMA_n \FI_{\Ll}\hspace{1pt} \PA_n \, \subseteq \,
\FI_{\Ll}\hspace{1pt} \DP_{temp} \bigl(\HES_n \mz\bigr)
\quad (\Ll \, \in \, \R).
\eeq%
\end{Lemma}%

\medskp

It is enough to \textit{prove} Eq.~(\ref{E2.2}).
We do this
by induction in $n=2,3,\dots$.
For $n=2$ Eq.~(\ref{E2.2}) follows by definition:
here $\SRMA_2 \, G = \RMA_2 \, G \vrestr{10pt}{\R^D \mz}$ $=$ $G$
for $G \in \PA_2$,
i.e., $\SRMA_2$ is the inclusion
$\PA_2$ $\hookrightarrow$ $\DP \bigl(\HES_2 \mz\bigr)$,
and then we apply
the first of Eqs.~(\ref{Scdeg}).
For $n > 2$ we have to prove that the inequality
$\scdeg \, \SRMA_n G$ $\leqslant$ $\devdeg \, G$
(i.e., Eq.~(\ref{r2}))
holds
for every $G \in \PA_n$.

To this end we first note that the notions of scaling degrees
$\scdeg$ (resp., $\Scdeg$) can be also introduced for distributions
(resp., smooth functions) over open cones $\CONE$ $\subseteq$ $\R^N$ $\mz$.
Then Lemma~\ref{Scdeg-lm} $(a)$
holds
also for
$u \in \DP_{temp} \bigl(\CONE\bigr)$ and
$F \in \CI_{temp} \bigl(\CONE\bigr)$.
Note that $\VP_{\PRT}$ are open cones in $\HES_n \mz$
for proper $n$--partitions $\PRT$ and
$$
\scdeg \, u \, = \, \mathop{\max}
\limits_{\mathop{}\limits^{\PRT \text{ is a proper}}_{\text{$S$--partition}}}
\scdeg \, \Bigl(u\vrestr{12pt}{\VP_{\PRT}}\Bigr) \,
$$
(this can be proven by using a partition of unity, which is
subordinate to the open covering $\bigl\{\VP_{\PRT}\bigr\}$).

Thus, let us fix an arbitrary proper $n$--partition $\PRT$ and
set $S$ $:=$ $\{1,\dots,n\}$.
By the construction of the filtration on $\PA_n$,
the function $G$ has a representation of the form~(\ref{genel}),
such that for all $\alpha$:
\(
\mathop{\sum}
\limits_{1 \, \leqslant \, j \, < \, k \, \leqslant \, n}
\Scdeg \, G_{jk}^{\alpha}
\)
$\leqslant$ $\devdeg \, G$.
Then every $G_{\alpha}$ in (\ref{genel}) has a representation
of a type (\ref{GS}):
$G_{\alpha}$ $=$
$G_{\PRT}^{\alpha}$ $\cdot$
$\mathop{\prod}\limits_{S' \, \in \, \PRT} G_{S'}^{\alpha}$.
Note that $G_{\PRT}^{\alpha}\vrestr{10pt}{\VP_{\PRT}}$ $\in$
$\CI_{temp} \bigl(\VP_{\PRT}\bigr)$ and
$\Scdeg$ $\Bigl(G_{\PRT}^{\alpha}\vrestr{10pt}{\VP_{\PRT}}\Bigr)$ $\leqslant$
\(
\mathop{\sum}\limits_{\mathop{}\limits^{j,k \, \in \, S}_{j\, \nsim_{\PRT} \, k}}
\Scdeg \, G_{jk}^{\alpha}
\).
Then we have
$\Scdeg \, \Bigl(G_{\PRT}^{\alpha}\vrestr{10pt}{\VP_{\PRT}}\Bigr)$ $+$
$\mathop{\sum}\limits_{S' \, \in \, \PRT} \devdeg \, G_{S'}^{\alpha}$
$\leqslant$ $\devdeg \, G$ for every $\alpha$.
Applying the inductive assumption we obtain that
\\
$\Scdeg \, \Bigl(G_{\PRT}^{\alpha}\vrestr{10pt}{\VP_{\PRT}}\Bigr)$ $+$
\(\mathop{\sum}\limits_{S' \, \in \, \PRT}
\scdeg \, \RMA_{S'} \, G_{S'}^{\alpha}\)
$\leqslant$ $\devdeg \, G$.
Then by Lemma~\ref{Scdeg-lm} $(a)$ and $(b)$
(over $\VP_{\PRT}$) and conclude that
$\scdeg \, \Bigl(\SRMA_n G \vrestr{10pt}{\VP_{\PRT}}\Bigr)$
$\leqslant$ $\devdeg \, G$.
Since this is true for every proper $n$--partition $\PRT$
it follows that $\scdeg \, \SRMA_n G$ $\leqslant$ $\devdeg \, G$.

This completes the proof of Lemma~\ref{Lm2.1}.

\medskp

Thus, we shall consider the maps $\SRMA_n$ as linear maps
\beq\label{SRMA}
\SRMA_n : \PA_n \to \DP_{temp} \bigl(\HES_n \mz\bigr) \,.
\eeq

\medskp

\begin{Remark}\label{Rm2.1}
One can enhance the condition $(r1)$ to a stronger condition:
$$
(\emb)^* \circ \RMA_{S'} \, = \, \RMA_S \circ (\emb)^*
$$
for every \textit{injection} $\emb : S' \to S$.
In particular,
\(
\RMA_S \vrestr{10pt}{\PA_{S'}} \, = \, \RMA_{S'}
\)
if $S' \subseteq S$, the restriction being taken under the inclusion
$\PA_{S'} \subseteq \PA_S$.
This is quite natural assumption but since we shall not use it here
we do not impose it.
(See also Remarks \ref{Rm2.2} and \ref{Rm2.3}.)
\end{Remark}

\subsection[Primary renormalization maps]{Primary renormalization maps
and definition of renormalization maps}\label{se2.2}

Equation (\ref{EX}) suggests to built the renormalization maps $\RMA_n$
as compo\-si\-tions
of the recursively defined linear maps $\SRMA_n$ and extension maps
$\DP_{temp} \bigl(\HES_n$ $\mz\bigr)$ $\to$ $\DP \bigl(\HES_n\bigr)$.
These maps we call \textit{primary renormalization maps} and
using them
we shall define $\RMA_n$.

\medskip

Thus, a \textbf{system of primary renormalization maps}
is a collection of linear maps
\beq\label{PRM}
\Prma_N : \DP_{temp} \bigl(\R^N \mz\bigr) \to \DP \bigl(\R^N\bigr)
\eeq
labelled by
integers $N=1,2,\dots$
and satisfying properties $(p1)$--$(p5)$ listed below.
As in the case of the renormalization maps $\RMA_n$,
we emphasize that no continuity is assumed for the maps $\Prma_N$
either.

\medskip

($p1$) Extension property:
$$
\bigl(\Prma_N \, u\bigr) \vrestr{12pt}{\R^N \mz} \, = \, u
\quad
(u \in \DP_{temp} \bigl(\R^N \mz\bigr)).
$$

\medskip

($p2$)
Preservation of filtrations:
$$
\Prma_N \,
\FI_{\Ll}\hspace{1pt} \DP_{temp} \bigl(\R^N \mz\bigr)
\, \subseteq \,
\FI_{\Ll}\hspace{1pt} \DP \bigl(\R^N\bigr)
\quad
(N=1,2,\dots,\ \Ll \, \in \, \R).
$$

\medskip

($p3$) Orthogonal invariance:
$$
O^* \circ \Prma_N
\circ (O^*)^{-1}
\, = \,
\Prma_N
,
$$
for every orthogonal transformation $O$ of $\R^N$,
where
$(O^*F)\bigl(\xx\bigr)$ $:=$ $F(O\xx)$
for a function (distribution) $F (\xx)$ over $\R^N$.

\medskip

($p4$)
For every polynomial $f \in \GF \bigl[\R^N\bigr]$
and
$u \in \DP_{temp} \bigl(\R^N \mz\bigr)$:
$$
\Prma_N \, fu \, = \, f \Prma_N \, u.
$$

\medskip

($p5$) If
$u (\xx) \in \DP \bigl(\R^M\bigr)$ with
$supp \, u_1 \subseteq \{\Mbf{0}\}$
and $v (\yy) \in \DP \bigl(\R^N \mz\bigr)$,
where $M,N \geqslant 1$, then:
$$
\Prma_{M+N} \bigl(u \otimes v\bigr) \, = \,
u \otimes \Prma_N \bigl(v\bigr)
$$
($(u \otimes v)(\xx,\yy) := u(\xx) \, v(\yy)$).

\medskip

This completes the definition of primary renormalization maps.

Let us point out the following stronger version of $(p5)$,
which under condition $(p3)$ is equivalent to $(p5)$:

\medskip

($p5'$)
For every (orthogonal) decomposition
$\R^{M+N} = \VSP \oplus \VSP'$,
$\VSP \cong \R^M$ and $\VSP' \cong \R^N$
with $M,N \geqslant 1$:
if $u (\xx) \in \DP \bigl(\VSP\bigr)$ with
$supp \, u_1 \subseteq \{\Mbf{0}\}$
and $v (\yy) \in \DP \bigl(\VSP' \mz\bigr)$ then
\(
\Prma_{M+N} \bigl(u \otimes v\bigr) =
u \otimes \Prma_N \bigl(v\bigr).
\)

\medskip

We shall need only the renormalization maps over
$\R^{D(n-1)}$ $\cong$ $\HES_n$, where the identification (\ref{isomm})
is assumed and
we shall
also lift, by the Euclidean invariance, these maps
from $\HES_n$ to all $\HES_S$ for finite subsets $S \subset \N$.
Let us denote the resulting linear maps by:
\beq\label{PRMA}
\PRMA_S \ ( \, :\cong \, \Prma_{D(n-1)} \, ) \, :
\DP_{temp} \bigl(\HES_S \mz\bigr) \to \DP \bigl(\HES_S\bigr)
\,, \quad
\PRMA_n \, := \, \PRMA_{\{1,\dots,n\}}.
\eeq

Now
the construction and together, the definition of renormalization maps
is based on the following theorem:

\medskp

\begin{Theorem}\label{Th2.1}
Let we be given by a system of primary renormalization maps
$\bigl\{\PRMA_n\bigr\}\raisebox{8pt}{\hspace{0pt}}_{n \, = \, 2}^{\infty}$
and define recursively:
\beqa\label{DFPRMA1}
\RMA_2 \, := \podr \PRMA_2 ,\quad
\\ \label{DFPRMA2}
\RMA_n \, := \podr \Prma_n \circ \SRMA_n \quad \text{for $n>2$}.
\eeqa
Here:
having defined $\RMA_n$
by the induction
we then define $\RMA_S$ for all
$S \subset \N$ with $|S|=n$
by using the permutation symmetry and
the orthogonal invariance
implied by
$(p1)$ and $(p3)$;
and finally, we define $\SRMA_n$ provided that $\RMA_S$ satisfy
$(r1)$--$(r4)$ for $|S|<n$.
In this way we obtain a system of linear maps $\{\RMA_S\}_S$
satisfying $(r1)$--$(r4)$.
The so defined $\RMA_n$ (or, $\RMA_S$) we shall call
\textbf{renormalization maps}.
\end{Theorem}%

\medskp

\begin{Proof}
We use an induction in $n=2,3,\dots$.
The most nontrivial part in the proof
is contained in Lemma~\ref{Lm2.1} that ensures property $(r2)$.
Property $(r3)$ is a consequence of $(p4)$.
Condition $(r4)$ is satisfied due to the construction of $\SRMA_n$.
The permutation symmetry required in $(r1)$ follows from $(p1)$
and the restriction property in $(r1)$ follows by induction.
\end{Proof}%

\medskp

\begin{Remark}\label{Rm2.2}
If we had imposed for the renormalization maps instead
of condition $(r1)$ its stronger version in Remark \ref{Rm2.1},
then we would also need a stronger version for condition $(p3)$.
Namely, for every partial isometry
$O$ $:$ $\R^N$ $\to$ $\R^M$ we should impose
$$
O^* \circ \Prma_N
\, = \,
\Prma_M \circ O^*
,
$$
where for a distribution $F (\xx)$ over $\R^M$,
$(O^*F)\bigl(\xx\bigr)$ $:=$ $F(O\xx)$ is a distribution
over $\R^N$.
\end{Remark}

\subsection{Construction of primary renormalization maps}\label{se2.3}

In this subsection we shall prove the existence of primary renormalization maps.
As a more technical section it can be skipped on the first reading.

\medskip

\begin{Theorem}\label{Th2.3}
There exists a system of primary renormalization maps.
\end{Theorem}%

\medskp

Part of this theorem is based on the old results of
Epstein--Glaser and Steinmann on renormalization
and the new element here is mainly to achieve properties
$(p4)$ and $(p5)$.
Thus, we begin by
stating a known result:

\medskp

\begin{Lemma}\label{Lm2.4}
For every
$N=1,2,\dots$ there exists a unique linear map
$$
\Prma_{N,0} : \FI_N \DP \bigl(\R^N \mz\bigr) \to
\FI_N \DP \bigl(\R^N\bigr)
$$
such that $u = \Prma_{N,0} u\vrestr{10pt}{\R^N \mz}$
for every $u \in \FI_N \DP \bigl(\R^N \mz\bigr)$.
It has also the property:
$\scdeg \, \Prma_{N,0} u =\scdeg \, u$.
\end{Lemma}%

\medskp

The \textit{proof} of Lemma~\ref{Lm2.4} can be found in
\cite[Theorem 2]{BF99}.

Continuing with the \textit{proof of Theorem~\ref{Th2.3}}
we first define linear maps $\Prma_N'$ that fulfill
just properties ($p1$) and ($p2$).
(They would be extensions of the corresponding $\Prma_{N,0}$
provided by the above lemma.)

To this end we
take a test function $\vartheta (\xx) \in \Dd \bigl(\R^N\bigr)$
that is equal to $1$ in a neighborhood of $0$
and introduce for
test functions $f (\xx) \in \Dd \bigl(\R^N\bigr)$
the \textit{truncated, first order} Taylor remainder
$$
\mathop{\sum}\limits_{\xi \, = \, 1}^N
x^{\xi} \, T_{\xi} \bigl(f\bigr) (\xx)
\, = \,
f (\xx) \, - \, f (\Mbf{0}) \, \vartheta (\xx) \,,
$$
so that $T_{\xi} (f) \in \Dd \bigl(\R^N\bigr)$.
Then we set inductively for $\Ll=1,2,\dots$:
\beqa &
\Prma_{N,\Ll} :
\FI_{N+\Ll} \DP \bigl(\R^N \mz\bigr)
\to
\FI_{N+\Ll} \DP \bigl(\R^N\bigr) ,
& \nn \label{Inddef} &
\Prma_{N,\Ll} \bigl(u\bigr) \bigl[f\bigr] \, := \,
\mathop{\sum}\limits_{\xi \, = \, 1}^N
\Prma_{N,\Ll-1} \bigl(x^{\xi} \hspace{1pt} u\bigr) \bigl[T_{\xi} \bigl(f\bigr) (\xx)\bigr]
&
\eeqa
($u \in \FI_{\Ll}\hspace{1pt} \DP \bigl(\R^N \mz\bigr)$),
the right hand side being well defined due to the fact that
$x^{\xi} u \in \FI_{N+\Ll-1} \DP \bigl(\R^N \mz\bigr)$.
The so defined $\Prma_{N,\Ll}$ then satisfy
$u = \Prma_{N,\Ll} u\vrestr{10pt}{\R^N \mz}$

\medskp

\begin{Lemma}\label{LM-2}
For all $\Ll = 0,1,\dots$ we have:
\beq\label{est-1}
\scdeg \, \Prma_{N,\Ll} u \, \leqslant \, \scdeg \, u \,
\eeq
for every $u \in \FI_{N+\Ll} \DP \bigl(\R^N \mz\bigr)$.
\end{Lemma}

\medskp

\begin{Proof}
For $\Ll=0$ the statement follows by
Lemma \ref{Lm2.4}.
Assume by induction, we have proven the lemma for $\Ll-1$ and $\Ll > 0$.
We should prove that if $\scdeg \, u$ $<$ $\lambda$ then
for every compact $K \subset \R^N \mz$
there exist $L=L(\lambda) \in \N_0$,
a test functions norm $\|$$\cdot$$\|_{K,L}$,
and a constant $C_{K,\lambda}$ $>$ $0$
such that for every
$f \in \Dd (\R^N)$ with $\text{\it supp} \, f \subseteq K$
and $\varepsilon$ $\in$ $(0,1)$ we have:
\beq\label{tmp-eq-1}
\Bigl|\Prma_{N,\Ll} (u) \bigl[f_{\varepsilon}\bigr] \Bigr|
\, \leqslant \,
C_{K,\lambda} \, \bigl\|f\bigr\|_{K,L} \, \varepsilon^{N-\lambda} \,,
\eeq
where $f_{\varepsilon} (\x)$ $:=$ $f \bigl(\varepsilon^{-1} \xx \bigr)$
(as in Lemma \ref{Scdeg-lm}).
We prove Eq.~(\ref{tmp-eq-1}) separately for the case
of test functions $f$ such that $f(0)$ $=$ $0$ and the case when
$f$ $=$ $\vartheta$.

If $f(0)=0$ then
\(
\bigl|\Prma_{N,\Ll} (u) \bigl[f_{\varepsilon}\bigr] \bigr|
\leqslant
\mathop{\sum}\limits_{\xi \, = \, 1}^N
\bigl|\Prma_{N,\Ll} \bigl(x^{\xi}u\bigr)
\bigl[T_{\xi}f_{\varepsilon}\bigr] \bigr|
\).
Since $T_{\xi} f_{\varepsilon}$ $=$
$\varepsilon^{-1}\bigl(T_{\xi}f\bigr)_{\varepsilon}$
and
$\scdeg$ $\Prma_{N,\Ll-1} \bigl(x^{\xi} u\bigr)$
$\leqslant$ $\scdeg$ $x^{\xi} u$ $\leqslant$
$-1$ $+$ $\scdeg$ $u$ $\leqslant$ $-1+\lambda$
then using the inductive assumption
we obtain the estimate (\ref{tmp-eq-1}).

For $f=\vartheta$ we have
\(
\bigl|\Prma_{N,\Ll} (u) \bigl[\vartheta_{\varepsilon}\bigr] \bigr|
=
\bigl|u
\bigl[\vartheta-\vartheta_{\varepsilon}\bigr] \bigr|
\).
Then setting $\varepsilon=2^{-r}$ we get an estimate
$\bigl|\Prma_{N,\Ll} (u) \bigl[\vartheta_{\varepsilon}\bigr] \bigr|$
$\leqslant$
$\mathop{\sum}\limits_{s \, = \, 0}^{r-1}$
$\bigl|u \bigl[\vartheta_{2^{-s}}-\vartheta_{2^{-s-1}}\bigr] \bigr|$
$=$
$\mathop{\sum}\limits_{s \, = \, 0}^{r-1}$
\(\Bigl|u \Bigl[\bigl(\vartheta-\vartheta_{2^{-1}}
\bigr)\raisebox{10pt}{\hspace{0pt}}_{_{2^{-s}}}\Bigr] \Bigr|\)
$\leqslant$ $C'$
$\mathop{\sum}\limits_{s \, = \, 0}^{r-1}$
$2^{-s(N-\lambda)}$
$\leqslant$
$C''$ $\varepsilon^{N-\lambda}$ for some positive constants $C',C''$
provided that $N\neq\lambda$, which is not an essential restriction.
This completes the proof of the lemma.
\end{Proof}

\medskp

We note further that:
$$
\Prma_{N,\Ll+1} \vrestr{12pt}{\FI_{\Ll}\hspace{1pt} \DP \bigl(\R^N \mz\bigr)}
\, = \,
\Prma_{N,\Ll}
\quad \text{for all} \quad \Ll \, = \, 1,2,\dots,
$$
since
$$
\Prma_{N,\Ll+1} \, u
\, - \,
\Prma_{N,\Ll} \, u \, = \,
-\bigl(\Prma_{N,\Ll} \, u\bigr) [\vartheta] \, \delta (\xx) \,,
$$
for $u \in \FI_{N+\Ll} \DP \bigl(\R^N \mz\bigr)$,
but on the other hand,
$\bigl(\Prma_{N,\Ll} \, u\bigr) [\vartheta]$ $=$ $0$ for $\Ll > 0$,
because $T_{\xi} \bigl(\vartheta\bigr)$ $\equiv$ $0$.
Still, we have an
inconsistency:
$$
\Prma_{N,1} \vrestr{12pt}{\FI_N \DP \bigl(\R^N \mz\bigr)}
\, - \,
\Prma_{N,0} \, = \,
\alpha_0 \bigl(u\bigr) \, \delta (\xx),
$$
where $\alpha_0$ is the linear functional:
$$
\alpha_0 :
u \, \mapsto \, -\bigl(\Prma_{N,0}\, u\bigr) [\vartheta]
:
\FI_N \DP \bigl(\R^N \mz\bigr)
\to
\R.
$$
Hence, to correct this we have to modify $\Prma_{N,1}$
in the following way:
$$
\Prma_{N,1}' \, u \, := \, \Prma_{N,1} \, u - \alpha (u) \, \delta (\xx)
$$
where
$\alpha : \FI_{N+1} \DP \bigl(\R^N \mz\bigr) \to \R$
is a linear functional that is a continuation of $\alpha_0$
(such a continuation always exists).
Note that $\Prma_{N,1}'$ satisfies the inequality (\ref{tmp-eq-1}).
Then introducing inductively linear maps
$\Prma_{N,\Ll}'$ for $\Ll=1,2,\dots$, again by Eq.~(\ref{Inddef}),
we obtain a consistent system of linear maps,
all satisfying Eq. (\ref{tmp-eq-1}) according to Lemma \ref{LM-2}.
Thus, we can define a linear map
$$
\Prma_N' : \DP_{temp} \bigl(\R^N \mz\bigr) \to \DP \bigl(\R^N\bigr)
$$
by setting
$$
\Prma_N' \vrestr{12pt}{\FI_{N+\Ll} \DP \bigl(\R^N \mz\bigr)}
\, := \, \Prma_{N,\Ll}'
$$
for $\Ll > 0$.
This map then satisfies the properties $(p1)$ and $(p2)$
as well as,
$$
\Prma_N' \vrestr{12pt}{\FI_{N} \DP \bigl(\R^N \mz\bigr)}
\, = \, \Prma_{N,0} \,.
$$

Next
we consider the problem
of fulfilling conditions ($p3$), ($p4$) and ($p5$).
Note first that if we find a system of linear maps $\Prma_N$ satisfying
($p1$), ($p2$), ($p4$) and ($p5'$) then by
averaging over
the compact group $O(N)$:
$$
\int_{O(N)} O^* \circ \Prma_N \circ (O^*)^{-1} d \mu (O)
$$
we will obtain a system of linear maps satisfying all conditions
($p1$)--($p5$).

\medskp

\begin{Lemma}\label{Lm2.4ne1}
There exists a linear map~$\Prma_N''$
that satisfies properties $(p1)$, $(p2)$ and $(p4)$.
\end{Lemma}%

\medskp

\noindent
\textit{Proof.}
We have to fulfill in addition to $(p1)$ and $(p2)$ the equalities
\beq\label{eq2.9ne1}
\Prma_N'' \bigl(x^{\xi} \hspace{1pt} u\bigr)
\, = \,
x^{\xi} \, \Prma_N'' \bigl(u\bigr)
\eeq
for every $\xi=1,\dots,N$
$($where $\xx = \bigl(x^1,\dots,x^{N}\bigr))$.
By the above considerations there exists a linear map
$\Prma_N'$ $:$ $\DP_{temp} \bigl(\R^N$ $\mz\bigr)$
$\to$ $\DP \bigl(\R^N\bigr)$
satisfying properties ($p1$) and ($p2$).
If $\Prma_N''$ $:$ $\DP_{temp} \bigl(\R^N$ $\mz\bigr)$
$\to$ $\DP \bigl(\R^N\bigr)$ is another map that satisfies
($p1$) and ($p2$) we set
\beqa\label{eq2.15ne21}
\rdf := \Prma_N' - \Prma_N'' : \podr
\DP_{temp} \bigl(\R^N \mz\bigr)
\to
\DP_{\R^N\hspace{-1pt},\hspace{1pt}0} \,,
\\ \label{eq2.16ne21}
\hspace{-30pt}
c_{\xi} := x^{\xi} \circ \Prma_N' - \Prma_N' \circ x^{\xi} : \podr
\DP_{temp} \bigl(\R^N \mz\bigr)
\to
\DP_{\R^N\hspace{-1pt},\hspace{1pt}0}
\nonumber
\eeqa
($\xi = 1,\dots,N$),
where $\DP_{\R^N\hspace{-1pt},\hspace{1pt}0}$ stands for the space of distributions
on $\R^M$ supported at zero.
Then Eq.~(\ref{eq2.9ne1}) is equivalent to
\beq\label{eq2.17ne21}
c_{\xi} \, = \, x^{\xi} \circ \rdf - \rdf \circ x^{\xi} \,.
\eeq
Thus, the problem is to find a linear map $\rdf$~(\ref{eq2.15ne21})
such that it preserves the filtrations and
(\ref{eq2.17ne21}) is satisfied.

To this end we expand $\rdf$ and $c_{\xi}$
in delta functions and their derivatives:
\beqa\label{eq3.11uuu}
\hspace{-30pt}
\rdf =
\!\podr \mathop{\sum}\limits_{\rr \, \in \, \N_0^N} \ \frac{1}{\rr!} \
\delta^{(\rr)} (\xx)
\, \RDF_{\rr}
\, , \quad
\RDF_{\rr} : \DP_{temp} \bigl(\R^N \mz\bigr)
\to \R,
\\ \label{eq2.19ne21}
\hspace{-30pt}
c_{\xi} =
\!\podr \mathop{\sum}\limits_{\rr \, \in \, \N_0^N} \ \frac{1}{\rr!} \
\delta^{(\rr)} (\xx)
\, C_{\xi,\rr}
\, , \quad
C_{\xi,\rr} : \DP_{temp} \bigl(\R^N \mz\bigr)
\to \R
\eeqa
(recall the multiindex notations:
$\xx =(x^1,\dots,x^N)$ $\in$ $\R^N$;
$\rr$ $=$ $(r_1,\dots,r_N)$ $\in$ $\N_0^N$,
$|\rr|$ $=$ $\sum_j r_j$,
$\rr!$ $=$ $\prod_j r_j!$,
$\xx^{\rr}$ $=$ $\prod_j \bigl(x^j\bigr)^{r_j}$
and
$\delta^{(\rr)} (\xx)$ $=$ $\Mbf{\di}^{\rr} \delta(\xx)$).
The condition that $\rdf$ preserves the filtrations is equivalent to
\beq\label{eq3.12}
\RDF_{\rr} \, u \, = \, 0
\quad \text{if} \quad
|\rr| \, > \, \scdeg \, u - N \,.
\eeq
Equation (\ref{eq2.17ne21}) holds iff
\beq\label{eq3.13}
\RDF_{\rr+\ee_{\xi}} \bigl[u\bigr] \, = \,
- \, C_{\xi,\rr} \bigl[u\bigr]
-
\RDF_{\rr} \bigl[x^{\xi} \, u\bigr] \,
\eeq
for all $\rr \in \N_0^N$,
where
$\ee_{\xi}$ is the $\xi$th basic vector in $\R^N$
(this follows from the representations
(\ref{eq2.19ne21}) and (\ref{eq3.11uuu}), and the formula
$x^{\xi} \, \delta^{(\rr)} (\xx)$ $=$
$-r_{\xi}$ $\delta^{(\rr - \ee_{\xi})} (\xx)$).
So, we have to find a collection of linear functionals $\RDF_{\rr}$,
which satisfy Eqs. (\ref{eq3.12}) and (\ref{eq3.13}).

The linear maps $c_{\xi}$ satisfy an ``integrability'' relation
$$
c_{\xi} \circ x^{\eta} - x^{\xi} \circ c_{\eta} \, = \,
c_{\eta} \circ x^{\xi} - x^{\eta} \circ c_{\xi} \,,
$$
which implies
\beq\label{eq3.15n}
C_{\xi,\rr+\ee_{\eta}} \bigl[u\bigr] - C_{\xi,\rr} \bigl[x^{\eta} \hspace{1pt} u\bigr]
\, = \,
C_{\eta,\rr+\ee_\xi} \bigl[u\bigr] - C_{\eta,\rr} \bigl[x^{\xi} \hspace{1pt} u\bigr] \,.
\eeq
By the fact that $\Prma_N'$ preserves the filtrations and Eq.~(\ref{ineqq})
we obtain:
\beq\label{eq2.24ne21}
C_{\xi,\rr} \, u \, = \, 0 \,
\quad \text{if} \quad
|\rr| > \scdeg \, u - 1 - N \,.
\eeq

Then let us set
\beq\label{eq3.14}
\RDF_{\rr} \, u
\, := \,
\mathop{\sum}\limits_{\xi \, = \, 1}^N \,
\mathop{\sum}\limits_{s \, = \, 1}^{r_{\xi}} \,
(-1)^{|\qq(\xi,s)|} \,
C_{\xi,\rr-\qq(\xi,s)} \bigl[\xx^{\qq(\xi,s)-\ee_{\xi}} \hspace{1pt} u\bigr] \,,
\eeq
where
$\qq(\xi,s) := s \, \ee_{\xi} + \mathop{\textstyle \sum}\limits_{\eta \, = \, 1}^{\xi-1} r_{\eta} \, \ee_{\eta}$
(writing a sum $\Bigl(\mathop{\sum}\limits_{j \, = \, a}^b \cdots \Bigr)$ with $a,b \in \Z$
we set it zero if $a>b$).
Note that the so defined $\RDF_{\rr}$ satisfy condition (\ref{eq3.12})
since Eq.~(\ref{eq2.24ne21}) implies that
$C_{\xi,\rr-\qq(\xi,s)} \bigl[\xx^{\qq(\xi,s)-\ee_{\xi}} \hspace{1pt} u\bigr] = 0$%
~if
$$
|\rr-\qq(\xi,s)| \, > \, \scdeg \, \bigl( \xx^{\qq(\xi,s)-\ee_{\xi}} \, u \bigr) - 1 - N
\quad \Longleftarrow \quad
|\rr| \, > \, \scdeg \, u - N\,
$$
Equation~(\ref{eq3.13}) is also satisfied, because of (\ref{eq3.15n}).
Thus, $\RDF_{\rr}$~(\ref{eq3.14})
determine a linear map $\rdf$ such that Eq. (\ref{eq2.17ne21}) holds
and then
$\Prma_N'' := \Prma_N' - \rdf$ fulfills the conditions of the lemma.$\quad\Box$

\medskp

To complete the proof of Theorem~\ref{Th2.3} it remains to fulfill
condition $(p5)$.
In fact, we have to fulfill its stronger version $(p5')$.
To this end we modify again $\Prma_N''$
as in the proof of Lemma~\ref{Lm2.4ne1}:
\beq\label{cor1}
\Prma_N \, := \, \Prma_N'' - \rdf_N
\eeq
for
$$
\rdf_N \, = \,
\mathop{\sum}\limits_{\rr \, \in \, \N_0^N} \ \frac{1}{\rr!} \
\delta^{(\rr)} (\xx)
\, \RDf_{N,\rr}
\, , \quad
\RDf_{N,\rr} : \DP_{temp} \bigl(\R^N \mz\bigr)
\to \R
$$
such that
\beqa\label{eq3.12xx}
&
\RDf_{N,\rr} \, u \, = \, 0
\quad \text{if} \quad
|\rr| \, > \, \scdeg \, u - N \,,
& \\ \label{eq3.13xx} &
\RDf_{N,\rr+\ee_{\xi}} \bigl[u\bigr] \, = \,
-
\RDf_{N,\rr} \bigl[x^{\xi} \, u\bigr] \,
&
\eeqa
the first of which ensures
that $\Prma_N$ preserve the filtrations
and the second, that $\Prma_N$ commute with $x^{\xi}$
(cf. Eqs.~(\ref{eq3.12}) and (\ref{eq3.13})).
Solving Eq.~(\ref{eq3.13xx}) we obtain
$$
\RDf_{N,\rr} \, = \, (-1)^{|\rr|} \RDf_{N,0} \circ \xx^{\rr} \,.
$$
i.e., $\rdf_N$ is determined just by one linear functional
$\RDf_{N,0}$ $:$ $\DP_{temp} \bigl(\R^N \mz\bigr)$ $\to$ $\R$
(this fact will play also a crucial role for
the reduction of our cohomological analysis in Sect.~\ref{se4}
to ordinary de Rham cohomologies).

Now
we shall define $\RDf_{N,0}'$, inductively in $N=1,2,\dots$,
so that $\Prma_N$
satisfy
($p5'$).

\medskp

\begin{Lemma}\label{Lm2.5}
It is always possible to fulfill the condition $(p5)$
(without prime!)
by a suitable correction $\rdf_{M+N}$ to $\Prma_{M+N}''$,
which is determined by a linear functional $\RDf_{N+M}$.
Furthermore, the restriction $\RDf_{M+N,\R^M}$ of $\RDf_{M+N}$
on the subspace:
$$
\TMPSP \, ( \, \equiv \, \TMPSP_{M+N} \bigl(\R^M\bigr)) := \,
\DP_{\R^M\hspace{-1pt},\hspace{1pt}0} \otimes \DP_{temp} \bigl(\R^N \mz\bigr)
\, \subset \,
\DP_{temp} \bigl(\R^{M+N} \mz\bigr)
$$
($\DP_{\R^M\hspace{-1pt},\hspace{1pt}0}$ stands for the space of distributions
on $\R^M$ supported at zero),
is uniquely determined.
\end{Lemma}%

\medskp

\noindent
\textit{Proof.}
We first define a linear map
\beqa
\rdf_{M+N,\R^M} \podr : \, \TMPSP \to \DP_{\R^{M+N}\hspace{-1pt},\hspace{1pt}0}
,\quad
\nn
\rdf_{M+N,\R^M} \, (w \otimes u) \podr := \,
\Prma_{N+M}'' \, (w \otimes u) - w \otimes \Prma_N'' \, u \,
\nonumber \eeqa
for $w \in \DP_{\R^M\hspace{-1pt},\hspace{1pt}0}$,
$u \in \DP \bigl(\R^N \mz\bigr)$.
Thus, $\rdf_{M+N,\R^M}$ is uniquely determined and
commutes with the multiplication
by the coordinates $x^{\xi}$ ($\xi=1,\dots,M+N$).
Hence, $\rdf_{M+N,\R^M}$ is determined, as above,
by a linear functional $\RDf_{M+N,\R^M}$ on $\TMPSP$
and such a functional is unique.
To construct the full map $\rdf_{N+M}$ we just
need to extend the linear functional $\RDf_{M+N,\R^M}$
from $\TMPSP$ to $\DP_{temp} \bigl(\R^{M+N} \mz\bigr)$
in such a way that the
grading condition~(\ref{eq3.12xx})
is satisfied.
This is a simple linear algebra problem and it is always possible.$\quad\Box$

\medskp

To complete the proof of the possibility to fulfill $(p5')$
(by induction in $N=1,2,\dots$)
we note first that for $N=1$ the condition is trivial.
Then we assume that $(p5')$ is satisfied for $1,\dots,N-1$.
By Lemma~(\ref{Lm2.5}) for every orthogonal decomposition
$\R^N$ $=$ $\VSP$ $+$ $\VSP'$ we have a uniquely defined
linear functional
$$
\RDf_{N,\VSP} :  \TMPSP_N (\VSP) \to \R
,\quad
\TMPSP_N (\VSP) \, := \,
\DP_{\VSP,0} \otimes \DP \bigl(\VSP' \mz\bigr) ,
$$
where $\DP_{\VSP,0}$ stands for the space of distributions
on $\VSP$ supported at zero.
Since
$\TMPSP_{\VSP_1 \cap \VSP_2}$ $=$
$\TMPSP_{\VSP_1}$ $\cap$ $\TMPSP_{\VSP_2}$
and on the other hand, by the uniqueness of $\RDf_{N,\VSP}$ we have:
$\RDf_{N,\VSP_1 \cap \VSP_2}$ $=$
$\RDf_{N,\VSP_1} \vrestr{10pt}{\VSP_1 \cap \VSP_2}$,
then the collection of linear functionals
$\bigl\{\RDf_{N,\VSP}\bigr\}_{\VSP \, \subseteq \, \R^N}$
is consistent.
Hence, there exists (possibly nonunique)
extension of all these functionals to a single lineal functional
$\RDf_N$ $:$ $\DP_{temp} \bigl(\R^N$ $\mz\bigr)$ $\to$ $\R$
so that
$\RDf_N \vrestr{10pt}{\TMPSP_N (\VSP)}$ $=$
$\RDf_{N,\VSP}$.
Using this functional to correct $\Prma_N''$ as above
we shall fulfill the condition $(p5')$ by the construction.

This completes the proof of the existence of primary renormalization maps.

\medskp

\begin{Remark}\label{Rm2.3}
If we impose the stronger conditions of Remarks~\ref{Rm2.1}
and \ref{Rm2.2} then we can use a similar construction for $\Prma_N$
as above:
we should introduce the subspaces $\TMPSPP_{\VSP}$ $:=$
$\bigl(\Pi_{\VSP}\bigr)^*$ $\DP \bigl(\VSP \mz\bigr)$
of $\DP \bigl(\R^N \mz\bigr)$,
where $\Pi_{\VSP} : \R^N \to \VSP$ is the orthogonal projection on
$\VSP$ $\subset$ $\R^N$,
and extend the recursively defined linear maps $\Prma_M$
($1\leqslant M<N$) consistently
from every $\TMPSPP_{\VSP}$
to the whole space $\DP \bigl(\R^N \mz\bigr)$.
\end{Remark}

\subsection{Change of renormalization maps}\label{se3.2}

In this section we give a formula for change of renormalization maps.
This formula is the main tool for the proof of
the ``universal renormalization theorem'' that
provides the way
in which the Green functions of an arbitrary perturbative QFT change
under the change of renormalization.
We shall not consider the latter theorem in the present paper
but in a future work.

\medskp

\begin{Theorem}\label{ChRen}
Let $\{\PRMA_n\}_{n \, = \, 2}^{\infty}$
and $\{\PRMA_n'\}_{n \, = \, 2}^{\infty}$
be two systems of primary renormalization maps (\ref{PRMA}),
which define the systems
$\{\RMA_n\}_{n \, = \, 2}^{\infty}$
and $\{\RMA_n'\}_{n \, = \, 2}^{\infty}$
of renormalization maps, respectively.
Then for every finite $S \subset \N$
and $G_S \in \PA_S$ of the form~(\ref{GS})
we have:
\beq\label{ChRen-eq}
\RMA_S' \, G_S \, = \,
\mathop{\sum}
\limits_{\mathop{}\limits^{\PRT \text{ is a}}_{\text{$S$--partition}}}
\Bigl(\RMA_{S/\PRT} \otimes \id_{\DP_{\PRT,0}}\Bigr)
\circ \NORM_{\PRT}
\left(\raisebox{12pt}{\hspace{-2pt}}\right.
G_{\PRT} \,
\mathop{\prod}\limits_{S' \, \in \, \PRT}
u_{S'}
\left.\raisebox{12pt}{\hspace{-2pt}}\right),
\eeq
where
$$
u_{S'} \, = \,
\left\{\hspace{-2pt}
\begin{array}{cll}
\Rdf_{S'} \, G_{S'} & \text{if} & |S'| > 1 \\
1 & \text{if} & |S'| = 1
\end{array}
\right.
$$
and
\ \(
\Rdf_S \, = \, \bigl(\PRMA_S' - \PRMA_S\bigr) \circ
\SRMA_S{\hspace{-5pt}}' \hspace{2pt}
\, : \, \PA_S \to \DP_{S,0}
\) \
for every finite $S$ with $|S| > 1$.
\end{Theorem}%

\medskp

\noindent
\EMPH{Explanation of the notations}:
Recall that $\DP_{S,0}$ stands for the space of distributions
on $\HES_S$ supported at zero.
The linear map $\Rdf_S$ takes values in $\DP_{S,0}$
due to the difference $\PRMA_S' - \PRMA_S$
of the primary renormalization maps
and the property $(p1)$ of their definition.
For
an $S$--partition
$\PRT$ we have set
$\DP_{\PRT,0} := \mathop{\otimes}\limits_{S' \, \in \, \PRT} \DP_{S',0}$.
Thus, $\DP_{\PRT,0}$ is the space of distributions over $\HES_S$
with support on the partial diagonal
$\Delta_{\PRT}$ $:=$
$\bigl\{[\x_j]_{j \in S} \in \HES_S$ $:$
$\x_j = \x_k$ if $j \sim_{\PRT} k\bigr\}$.
Then the product
\(
G_{\PRT} \mathop{\prod}
\limits_{S' \, \in \, \PRT} \, u_{S'}
\)
takes values in the space
\(
\PA_{S/\PRT} \otimes \DP_{\PRT,0}
\)
where we have introduced the ``quotient''
$S/\PRT$ $:=$ $\{\min S'$ $:$ $S' \in \PRT\}$
$\subseteq$ $S$.
More precisely, the transformation of
\(
G_{\PRT} \mathop{\prod}
\limits_{S' \, \in \, \PRT} \, u_{S'}
\)
to the space
\(
\PA_{S/\PRT} \otimes \DP_{\PRT,0}
\)
includes a restriction of $G_{\PRT}$
to the partial diagonal
$\Delta_{\PRT}$,
with possible ``transverse'' derivatives
due to the possible derivatives of the delta functions
contained in $u_{S'}$ ($S' \in \PRT$).
The latter operation is denoted in Eq.~(\ref{ChRen-eq}) by
$\NORM_{\PRT}$ (``normal form''):
$$
\NORM_{\PRT} :
\PA_{\PRT} \otimes \DP_{\PRT,0} \to \PA_{S/\PRT} \otimes \DP_{\PRT,0} \,.
$$
We also remind the convention
$\RMA_1$ $=$ $\RMA_{\{k\}}$ $=$
$\id_{\GF} : \GF \to \GF$
(the identity map of the ground field $\GF$)
and so, the extreme case in the sum in Eq.~(\ref{ChRen-eq}) when
$\PRT = \bigl\{\{j\} : j \in S\bigr\}$
(i.e.,
$|\PRT| = |S|$),
corresponds to the term:
$\Rdf_S \, G_S$.
Note that the term in Eq.~(\ref{ChRen-eq}) corresponding to
$\PRT = \bigl\{S\bigr\}$
(i.e., $|\PRT|=1$) is $\RMA_S \, G_S$.

\medskp

\noindent
\textit{Proof of Theorem~\ref{ChRen}.}
We use an induction in $n=|S|=2,3,\dots$.
For $n=2$ Eq.~(\ref{ChRen-eq}) reduces to the equation
$\RMA_2'$ $=$ $\RMA_2$ $+$ $\Rdf_2$ and
$\Rdf_2$ $=$ $\PRMA_2'$ $-$ $\PRMA_2$.
But $\RMA_2=\PRMA_2$ and $\RMA_2'=\PRMA_2'$,
by the construction of renormalization maps,
and $\SRMA_2{\hspace{-5pt}}' \hspace{2pt}$ is just the inclusion
$\PA_2 \hookrightarrow \DP \bigl(\HES_2 \mz\bigr)$.

Now
let $n>2$ and
assume Eq.~(\ref{ChRen-eq}) is proven for
all finite subsets $S' \subset \N$ with $|S'| < n$.
We shall first prove the equality:
\beq\label{TTT-e1}
\SRMA_S{\hspace{-5pt}}' \hspace{2pt} \, G_S \, = \,
\mathop{\sum}
\limits_{\mathop{}\limits^{\PRT \text{ is a proper}}_{\text{$S$--partition}}}
\Bigl(\SRMA_{S/\PRT} \otimes \id_{\DP_{\PRT,0}}\Bigr)
\circ \NORM_{\PRT}
\left(\raisebox{12pt}{\hspace{-2pt}}\right.
G_{\PRT} \,
\mathop{\prod}\limits_{S' \, \in \, \PRT}
u_{S'}
\left.\raisebox{12pt}{\hspace{-2pt}}\right).
\eeq
Note that Eq.~(\ref{ChRen-eq}) implies (\ref{TTT-e1})
because of the fact that
$\SRMA_S$ $G_S$ $=$ $\bigl(\RMA_S$ $G_S\bigr) \vrestr{10pt}{\HES_S \mz}$.
Due to the covering property (\ref{ocov})
it is enough to prove that
\beq\label{TT-e1}
\hspace{0pt}
\text{left hand side of Eq.~(\ref{TTT-e1})} \vrestr{10pt}{\VP_{\PRT}}
=
\text{right hand side of Eq.~(\ref{TTT-e1})} \vrestr{10pt}{\VP_{\PRT}}
\hspace{-4pt}
\eeq
for every proper $S$--partition $\PRT$.

For the restriction of the right hand side we obtain
\beq\label{ChRen-eq1}
\mathop{\sum}
\limits_{\PRT' \, \leqslant \, \PRT}
\Bigl[
\Bigl(\SRMA_{S/\PRT'} \otimes \id_{\DP_{\PRT',0}}\Bigr)
\circ \NORM_{\PRT'}
\left(\raisebox{12pt}{\hspace{-2pt}}\right.
G_{\PRT'} \,
\mathop{\prod}\limits_{S' \, \in \, \PRT'}
u_{S'}
\left.\raisebox{12pt}{\hspace{-2pt}}\right)
\Bigr]
\vrestr{12pt}{\VP_{\PRT}},
\eeq
where the relation $\PRT' \leqslant \PRT$ for two $S$--partitions stands for:
\beq\label{order}
\PRT' \leqslant \PRT
\qquad \mathop{\Longleftrightarrow}\limits_{\text{def}} \qquad
j \sim_{\PRT'} k \text{ \ implies \ } j \sim_{\PRT} k \ \ (\forall j,k \in S).
\eeq
This is because the support of
$\mathop{\prod}\limits_{S' \, \in \, \PRT'} u_{S'}$
for $\PRT'$ $\nleqslant$ $\PRT$ is
disjoint from~$\VP_{\PRT}$.

The restriction of the left hand side of Eq.~(\ref{ChRen-eq})
is computed accordingly to property $(r4)$
of renormalization maps (Sect.~\ref{se2.1}).
We thus obtain that for every proper $S$--partition $\PRT$ we have:
$$
\SRMA_S{\hspace{-5pt}}' \hspace{2pt} G_S \vrestr{12pt}{\VP_{\PRT}} \, = \,
\RMA_S{\hspace{-5pt}}' \hspace{2pt} G_S \vrestr{12pt}{\VP_{\PRT}} \, = \,
\Bigl(
G_{\PRT} \, \cdot \,
\mathop{\prod}\limits_{S' \, \in \, \PRT} \RMA_{S'}' \, G_{S'}
\Bigr)\vrestr{12pt}{\VP_{\PRT}}
\,.
$$
By the inductive assumption we get that
\(\Bigl(
G_{\PRT} \, \cdot \,
\mathop{\prod}\limits_{S' \, \in \, \PRT} \RMA_{S'}' \, G_{S'}
\Bigr)
\vrestr{12pt}{\VP_{\PRT}}\) equals
$$
G_{\PRT} \, \cdot \,
\Bigl[
\mathop{\prod}\limits_{S' \, \in \, \PRT}
\mathop{\sum}
\limits_{\mathop{}\limits^{\PRT_{S'} \text{ is a}}_{\text{$S'$--partition}}}
\Bigl(\RMA_{S'/\PRT_{S'}} \otimes \id_{\DP_{\PRT_{S'},0}}\Bigr)
$$
$$
\circ \, \NORM_{\PRT_{S'}}
\left(\raisebox{12pt}{\hspace{-2pt}}\right.
G_{\PRT_{S'}} \,
\mathop{\prod}\limits_{S'' \, \in \, \PRT_{S'}}
u_{S''}
\left.\raisebox{12pt}{\hspace{-2pt}}\right)
\Bigr]
\vrestr{12pt}{\VP_{\PRT}} \,.
$$
Now
we expand the product in $S' \in \PRT$ of the sums
and combine all the $S'$--partitions $\PRT_{S'}$
into a single $S$--partition
$\PRT'$ $:=$
$\mathop{\bigcup}\limits_{S' \, \in \, \PRT} \PRT_{S'}$.
Taking also into account that
$G_{\PRT}$ $\cdot$
$\mathop{\prod}\limits_{S' \, \in \, \PRT} G_{\PRT_{S'}}$
$=$ $G_{\PRT'}$
together with the property $(r4)$
we arrive at
the expression
(\ref{ChRen-eq1}).

Having proven Eq.~(\ref{TTT-e1}) for $|S|$ $=$ $n$ we can
prove Eq.~(\ref{ChRen-eq})
by using the compositions
$\RMA_S$ $=$ $\PRMA_S \circ \SRMA_S$
and
$\RMA_S'$ $=$ $\PRMA_S' \circ \SRMA_S{\hspace{-5pt}}' \hspace{2pt}$.
This is done as follows:
$$
\RMA_S' \, G_S \, = \,
\Rdf_S \, G_S \, + \, \PRMA_S \, \SRMA_S{\hspace{-5pt}}' \hspace{2pt} \, G_S
$$
$$
= \,
\Rdf_S \, G_S
\, + \, \PRMA_S
\mathop{\sum}
\limits_{\mathop{}\limits^{\PRT \text{ is a proper}}_{\text{$S$--partition}}}
\!
\Bigl(\SRMA_{S/\PRT} \otimes \id_{\DP_{\PRT,0}}\Bigr)
\circ \NORM_{\PRT}
\left(\raisebox{12pt}{\hspace{-2pt}}\right.
G_{\PRT} \,
\mathop{\prod}\limits_{S' \, \in \, \PRT}
u_{S'}
\left.\raisebox{12pt}{\hspace{-2pt}}\right)
$$
$$
= \,
\Rdf_S \, G_S
\, + \,
\mathop{\sum}
\limits_{\mathop{}\limits^{\PRT \text{ is a proper}}_{\text{$S$--partition}}}
\!\!
\Bigl(
\bigl(\PRMA_{S/\PRT} \circ \SRMA_{S/\PRT}\bigr) \otimes \id_{\DP_{\PRT,0}}
\Bigr)
\circ \NORM_{\PRT}
\left(\raisebox{12pt}{\hspace{-2pt}}\right.
G_{\PRT} \,
\mathop{\prod}\limits_{S' \, \in \, \PRT}
u_{S'}
\left.\raisebox{12pt}{\hspace{-2pt}}\right) \,,
$$
where in the last step we have used the property $(p5)$
of primary renormalization maps (Sect.~\ref{se2.2}).
Thus,
we arrive at
the expression of the right hand side of
Eq.~(\ref{ChRen-eq}).
This completes the proof of Theorem~\ref{ChRen}.$\quad\Box$

\medskp

We see that a change of renormalization maps
$\{\RMA_S\}$ $\to$ $\{\RMA_S'\}$
is completely determined by a system of linear maps
\beq\label{RDF}
\Rdf_S : \PA_S \to \DP_{S,0} \,
\eeq
indexed by nonempty finite subsets $S$ of $\N$.
They are related to $\{\RMA_S\}$ and $\{\RMA_S'\}$ by
$$
\Rdf_S \, := \, \bigl(\PRMA_S' - \PRMA_S\bigr) \circ
\SRMA_S{\hspace{-5pt}}'
\quad \text{for $|S|>1$} \quad \text{and} \quad
\Rdf_S \, := \, 1 \quad \text{for $|S|=1$}\,.
$$
Let us point out that the primary renormalization maps
contain more information than those, which is encoded
in the renormalization maps.

The conditions that characterize the system $\{\Rdf_S\}$
(\ref{RDF})
are:

\medskip

($c1$) Permutation symmetry:
for every bijection $\sigma : S \cong S'$ we have
$$
\sigma^* \circ \Rdf_{S'} \, = \, \Rdf_S \hspace{1pt} \circ \sigma^*
\quad
$$
($(\sigma^* F) \bigl(\x_{j_1},\dots,\x_{j_n}\bigr)$
$:=$ $F \bigl(\x_{\sigma (j_1)},\dots,\x_{\sigma (j_n)}\bigr)$).

\medskip

So, by ($c1$) all $\Rdf_S$ are characterized by
$\Rdf_n$ $:=$ $\Rdf_{\{1,\dots,n\}}$ for $n=|S|=2,3,\dots$.

\medskip

($c2$) Preservation of the filtrations:
$\Rdf_S \, \FI_{\Ll}\hspace{1pt} \PA_S \subseteq \FI_{\Ll}\hspace{1pt} \DP_S$.

\medskip

($c3$)
For every polynomial $f \in \GF \bigl[\HES_n\bigr]$
and $G \in \PA_n$
we have:
$$
\Rdf_n f G \, = \, f \, \Rdf_n G .
$$

\medskip

There is a converse statement:

\medskp

\begin{Proposition}\label{APr}
For every system $\{\Rdf_S\}$
satisfying the above conditions $(c1)$--$(c3)$
and a given system of renormalization maps $\{\RMA_S\}$
there exists a system of primary renormalization maps
$\{\PRMA_S'\}$, which determines a system of renormalization maps
$\{\RMA_S'\}$ so that the maps $\{\Rdf_S\}$ correspond to the change
$\{\RMA_S\}$ $\to$ $\{\RMA_S'\}$.
\end{Proposition}

\medskp

Since we shall not use here the above proposition we shall
only sketch its \textit{proof}
(there is also	a similarity with the proof of Theorem~\ref{Th4.1qq1}
in the next section).
We construct the primary renormalization maps $\PRMA_n$
by induction in $n=2,3,\dots$.
For $n=2$, $\SRMA_2$ is the embedding
$\PA_2$ $\hookrightarrow$ $\DP \bigl(\HES_2\mz\bigr)$ and hence,
$\PRMA_2'$ $=$ $\PRMA_2$ $+$
\(\Rdf_2 \circ \bigl(\SRMA_2{\hspace{-5pt}}'\hspace{3pt}\bigr)^{-1}\).
Having constructed $\PRMA_m'$ for $m=2,\dots,n-1$ we define $\PRMA_n'$
as $\PRMA_n$ $+$ $\widetilde{\Rdf}_n$, where
the linear map $\widetilde{\Rdf}_n$ is an extension of
$\Rdf_n \circ \bigl(\SRMA_n{\hspace{-5pt}}'\hspace{3pt}\bigr)^{-1}$
from the subspace
$\SRMA_S{\hspace{-5pt}}'\hspace{3pt} \bigl(\PA_n\bigr)$ $\subset$
$\DP \bigl(\HES_N \mz\bigr)$ to the whole space.
The later extension is not arbitrary but is done in the way
used in the proof of Theorem~\ref{Th4.1qq1} for the construction
of the analogous map $\widetilde{\RDF}_n$ there.
This completes the proof.

\medskp

One can further relate to the systems $\{\Rdf_S\}$
a certain group product:
if $\{\Rdf_S'\}$ characterizes another change of renormalization maps,
$\{\RMA_S'\}$ $\to$ $\{\RMA_S''\}$, then
a question arises what is $\{\Rdf_S''\}$
characterizing the change $\{\RMA_S\}$ $\to$ $\{\RMA_S''\}$?
The answer is analogous to Eq.~(\ref{ChRen-eq}):
\beq\label{URG}
\Rdf''_S \, G_S \, = \,
\mathop{\sum}
\limits_{\mathop{}\limits^{\PRT \text{ is a}}_{\text{$S$--partition}}}
\Bigl(\Rdf'_{S/\PRT} \otimes \id_{\DP_{\PRT,0}}\Bigr)
\circ \NORM_{\PRT}
\left(\raisebox{12pt}{\hspace{-2pt}}\right.
G_{\PRT} \,
\mathop{\prod}\limits_{S' \, \in \, \PRT}
\Rdf_{S'} \, G_{S'}
\left.\raisebox{12pt}{\hspace{-2pt}}\right).
\eeq
The method for proving the composition law (\ref{URG})
is the same as those, which is used in the proof of Theorem~\ref{ChRen}.
It is natural to expect that the set of all systems of linear maps
$\{\Rdf_S\}$ satisfying the conditions $(c1)$--$(c3)$
form a group under the above law with a unit
$\{\Rdf_1=1,$ $\Rdf_n=0$ for $n > 1\}$.
This group can be called \textbf{universal renormalization group}
and we intend to study it in a separate work.
(See also Remark~\ref{Rm3.2} at the end of Sect.~\ref{se4.2}.)

\subsection{Remark on renormalization on Riemann manifolds}\label{se2.6}

Our renormalization scheme based on the Epstein--Glaser approach
can be generalized for renormalization on Riemann manifolds.
We shall briefly sketch this construction here.
The restriction of translation invariance we have used up to now
is not crucial but only simplifies the considerations.

If $\RIE$ is a Riemann manifold
we should introduce the algebra $\PA \equiv \PA_2$
as a subalgebra of $\CI \bigl(\F_2(\RIE)\bigr)$.
We should also assume that $\PA$ is invariant under the actions
of all smooth vector fields on $\RIE \times \RIE$.
Then the algebras $\PA_n$ and $\PA_S$ for finite $S \subset \N$
are constructed in the same way as for the flat case:
$\PA_n$ is the subalgebra of $\CI \bigl(F_n (\RIE)\bigr)$
generated by all embeddings $\emb_{jk} \PA_2$ for
$1 \leqslant j < k \leqslant n$.

The scaling degree for distributions on $\RIE^{\times n}$
is introduced with respect to the total diagonal
$$
\Delta_n (\RIE) \, := \,
\{(\x,\dots,\x): \x \in \RIE\} \, \subset \, \RIE^{\times n}.
$$
Then the filtrations on the distributions spaces
and on the algebras $\PA_n$ are introduced similarly to the flat case.

Concerning
the preliminary definition of renormalization maps, conditions
$(r1)$, $(r2)$ and $(r4)$ remains the same
(the sets $\VP_{\PRT}$ are defined now as an open covering of
$\RIE^{\times n} \backslash \Delta_n (\RIE)$).
Condition $(r3)$ should be extended not only for polynomials,
but for arbitrary smooth functions on $\RIE \times \RIE$, i.e.:
$$
\RMA_n \, f(\x_j,\x_k) \, G(\x_1,\dots,\x_n) \, = \,
f(\x_j,\x_k) \, \RMA_n \, G(\x_1,\dots,\x_n)
$$
for every $f \in \CI \bigl(\RIE \times \RIE\bigr)$,
$G \in \PA_n$
and $1 \leqslant j < k \leqslant n$.

Then again we have linear maps
$$
\SRMA_n :
\PA_n
\to
\DP_{temp} \bigl(\RIE^{\times n} \backslash \Delta_n (\RIE)\bigr)
$$
determined by condition $(r4)$ from $\RMA_2$, $\dots$, $\RMA_{n-1}$.
Thus, the primary renormalization maps we need to construct
inductively $\RMA_n$ are linear maps:
$$
\PRMA_n :
\DP_{temp} \bigl(\RIE^{\times n} \backslash \Delta_n (\RIE)\bigr) \to
\DP \bigl(\RIE^{\times n}\bigr) \,.
$$
These maps should satisfy the following properties:
\begin{LIST}{31pt}
\item[$\bullet$]
$\Ss_n$--symmetry.
\item[$\bullet$]
Preservation of filtrations.
\item[$\bullet$]
For every $f \in \CI \bigl(\RIE^{\times n}\bigr)$ and
$u \in \DP_{temp} \bigl(\RIE^{\times n} \backslash \Delta_n (\RIE)\bigr)$:\
$$
\PRMA_n \, f \, u \, = \, f \, \PRMA_n \, u .
$$
\item[$\bullet$]
Let $\PRT$ be
an $n$--partition
and set $S:=\{\min S' : S' \in \PRT\}$.
Then
$$
\PRMA_n
\left[\raisebox{12pt}{\hspace{-2pt}}\right.
\left(\raisebox{12pt}{\hspace{-2pt}}\right.
\mathop{\otimes}\limits_{S' \, \in \, \PRT} u_{S'}
\left.\raisebox{12pt}{\hspace{-3pt}}\right)
\otimes
u
\left.\raisebox{12pt}{\hspace{-3pt}}\right]
\, = \,
\left(\raisebox{12pt}{\hspace{-2pt}}\right.
\mathop{\otimes}\limits_{S' \, \in \, \PRT} u_{S'}
\left.\raisebox{12pt}{\hspace{-3pt}}\right)
\otimes
\PRMA_S \, u ,
$$
where
$u \in \DP_{temp} \bigl(\RIE^S \backslash \Delta_S (\RIE)\bigr)$
and $u_{S'} \in \DP \bigl(\RIE^{S'}\bigr)$ with
$supp \, u_{S'}$ $\subseteq$
$\Delta_{S'} (\RIE)$
(the total diagonal in the cartesian power $\RIE^{S'}$)
for every $S' \in \PRT$.
\end{LIST}
We then set
$$
\RMA_2 \, = \, \PRMA_2
,\quad
\RMA_n \, = \, \PRMA_n \circ \SRMA_n \quad (n>2) \,.
$$

Apart from the above conditions on $\RMA_n$ and $\PRMA_n$
it is also natural to impose in the curved case the so called
``\textit{general covariance}''.
This assumes that
we have the linear maps $\RMA_n =: \RMA_{n (\RIE)}$ and
$\PRMA_n =: \PRMA_{n (\RIE)}$ defined for every
Riemann manifold $\RIE$ and furthermore, they depend
\textit{naturally} on $\RIE$ under isometric embeddings
$\RIE' \hookrightarrow \RIE$.

We shall not consider further here
the problem of constructing
such families of renormalization maps but we believe
that this can be done using the methods of this work
combined with the techniques used in the
causal perturbative QFT on pseudo--Riemann manifolds
(for instance, using Riemann normal coordinates).

\section{Anomalies in QFT and cohomologies of conf\-i\-guration spaces}\label{se4}
\setcntrs

When a symmetry of unrenormalized (bare) Feynman amplitudes
is broken after the renormalization
one speaks about an \textit{anomaly} in the theory.
Clearly, the source of the anomalies in perturbative QFT
is the absence of commutativity between the action
of the partial differential operators and the renormalization,
i.e., the ``commutators''
$$
c_n [A] \, = \, A \circ \RMA_n - \RMA_n \circ A \,,
$$
are generally nonzero for linear partial differential operators
$A$ on $\HES_n$.
Note that by the extension property
$\RMA_n G \vrestr{10pt}{\HF_n}$ $=$ $G$ (cf.~Sect.~\ref{se2.1})
for $G \in \PA_n$ it follows that $c_n [A]$ is a linear map
$$
c_n [A] : \PA_n \to \DP \bigl[\LADIAG_n\bigr]
,\quad \LADIAG_n \, := \, \HES_n \backslash \HF_n
$$
($\LADIAG_n$ is the so called ``large'' diagonal in $\HES_n$),
where $\DP \bigl[\LADIAG_n\bigr]$ stands for the space of distributions
on $\HES_n$ supported at $\LADIAG_n$.
It is straightforward to observe
that
$c_n [A]$ is a Hochschild $1$--cocycle for
the associative algebra of all linear partial differential operators
on $\HES_n$ (having polynomial coefficients):
\beq\label{e3.3}
A_1 \circ c_n \bigl[A_2\bigr] - c_n \bigl[A_1 \circ A_2\bigr] +
c_n \bigl[A_1\bigr] \circ A_2 \, = \, 0 \,
\eeq
(hence, $c_n [A]$ is a Hochschild $1$--cocycle
in the bimodule of all linear maps
$\PA_n$ $\to$ $\DP \bigl[\LADIAG_n\bigr]$).
It is also clear that every change of the renormalization map
$\RMA_n \to \RMA_n'$
changes the cocycle $c_n$ by a coboundary:
$$
c_n [A] - c_n'[A] \, = \, A \circ b - b \circ A \,,
$$
where $b = \RMA_n - \RMA_n'$.
Thus, we shall call the maps $c_n$ linearly depending on partial
differential operators \textit{renormalization cocycles}.

Nevertheless,
it is not so simple to take the full renormalization ambiguity into account:
as we mentioned in the previous section, condition (\textit{r}4)
is nonlinear and hence, the class
$$
\bigl\{ \{c_n\} : \{c_n\} \text{ is generated by } \{\RMA_n\} \bigr\}
$$
forms a nonlinear subset in the direct sum of the cohomology classes of
all $c_n$.
In this section we shall find a description of the above class of
renormalization cocycles.

\subsection{Cohomological equations}\label{se4.1}

By property ($r3$) (Sect.~\ref{se2.1}) of the renormalization maps we have
$c_n [x_k^{\mu}] = 0$,
where $x_k^{\mu}$ for $k$ $=$ $1,$ $\dots,$ $n-1$,
$\mu$ $=$ $1,$ $\dots,$ $D$
are the coordinates in $\HES_n$ $\cong$ $\R^{D(n-1)}$
according to the isomorphism~(\ref{isomm}).
Thus,
what remains to be determined (due to Eq.~(\ref{e3.3})) is
\beq\label{e3.5}
\Lccl_{n;\, k,\, \mu} \, := \, c_n \bigl[\di_{x_k^{\mu}}\bigr]
\, = \, \bigl[\di_{x_k^{\mu}}, \RMA_n\bigr]
\eeq
$(\di_{x_k^{\mu}} := \txfrac{\di}{\di x_k^{\mu}})$.
For short, we denote the pair of indices $(k,\mu)$ in Eq.~(\ref{e3.5})
by a single index $\xi$ (or $\eta$, $\dots$) running from $1$ to $D(n-1)$;
then the corresponding components $x_k^{\mu}$ will be denoted by $x^{\xi}$.
In what follows we shall call the above system of linear maps
$\Lccl_{n;\, \xi}$ \textbf{renormalization cocycles}.

Applying to the definition of $\Lccl_{n; \xi}$
the construction of renormalization maps by Theorem~\ref{Th2.1},
$\RMA_n$ $=$ $\PRMA_n$ $\circ$ $\SRMA_n$ ($n>2$),
we obtain a decomposition:
\beqa\label{eq4.2ne1}
\Lccl_{n;\, \xi} \, = \podr
\Sccl_{n;\, \xi} \, + \, \SLccl_{n;\, \xi}
\qquad (n>2), \qquad \Lccl_{2;\, \xi} \, \equiv \, \Sccl_{2;\, \xi}
\, , \quad
\\ \label{eq4.3ne1}
\Sccl_{n;\, \xi} \, := \podr
\bigl[\di_{x^{\xi}},\PRMA_n\bigr] \circ \SRMA_n
\qquad (n>2)
\, , \quad
\\ \label{eq4.4ne1}
\SLccl_{n;\, \xi} \, := \podr
\PRMA_n \circ \bigl[\di_{x^{\xi}},\SRMA_n\bigr]
\qquad (n>2)
\, . \quad
\nonumber
\eeqa
The linear maps $\Sccl_{n;\, \xi}$ are simpler than $\Lccl_{n;\, \xi}$
since they take values that are distributions supported at the
\textit{reduced total diagonal}, i.e., at the origin $\Mbf{0} \in \HES_n$
(this is due to condition ($p1$)):
\beq\label{e3.9}
\Sccl_{n;\, \xi} \, : \,
\PA_n \to \DP_{n,0} \,
\eeq
(recall that $\DP_{n,0}$ stands for the space distributions on $\HES_n$
supported at zero).
On the other hand, the remaining part $\SLccl_{n;\, \xi}$
of $\Lccl_{n;\, \xi}$ in Eq.~(\ref{eq4.2ne1})
is determined by the renormalization induction.
This is first because of the presence of $\SRMA_n$ and second,
due to the commutator $[\di_{x^{\xi}}, \SRMA_n]$,
which
produces
at least one delta function (with possible derivatives)
and then by property $(p5)$ $\PRMA_n$
is reduced to $\PRMA_{n'}$ with $n'<n$.
So, the only new information is contained in $\Sccl_{n;\, \xi}$.

The linear maps $\Sccl_{n;\, \xi}$ satisfy the ``differential equations'':
\beqa\label{eq4.6ne2}
\podr \!\!
\bigl[\di_{x^{\xi}},\Sccl_{2;\, \eta}\bigr]
-
\bigl[\di_{x^{\eta}},\Sccl_{2;\, \xi}\bigr] \, = \, 0 \,,
\nn
\podr \!\!
\bigl[\di_{x^{\xi}},\Sccl_{n;\, \eta}\bigr]
-
\bigl[\di_{x^{\eta}},\Sccl_{n;\, \xi}\bigr]
\\\ \podr \!\! \hspace{20pt}
= \,
- \, \bigl[\di_{x^{\xi}},\PRMA_n\bigr] \circ \bigl[\di_{x^{\eta}},\SRMA_n\bigr]
\, + \,
\bigl[\di_{x^{\eta}},\PRMA_n\bigr] \circ \bigl[\di_{x^{\xi}},\SRMA_n\bigr]
\, \quad (n > 2)
\nonumber
\eeqa
for all $\xi,\eta$ $=$ $1,\dots,D(n-1)$,
which are derived by a straightforward computation.
We shall characterize $\Sccl_{n;\, \xi}$ by these equations.
Before that, let us point out that \textit{the right hand side of (\ref{eq4.6ne2})
is determined by the renormalization induction}.
The reason is the same as above:
the values of
$\bigl[\di_{x^{\xi}},\SRMA_n\bigr]$
are distributions supported at the large diagonal $\LADIAG_n$ and then,
$\bigl[\di_{x^{\xi}},\PRMA_n\bigr]$
act on functions whose nontrivial dependence is in less than
$n-1$ relative distances (i.e., functions on $\HES_S$ with $|S|$ $<$ $n$).
Thus, we can consider (\ref{eq4.6ne2})
as equations for $\{\Sccl_{n;\, \xi}\}_{\xi}$ for fixed $n$,
whose right hand side is determined by $\{\Sccl_{n';\, \xi}\}_{\xi}$
for $n=2,\dots,n-1$.
We shall call the maps $\Sccl_{n;\, \xi}$
\textbf{primary renormalization cocycles}
without meaning of closedness with respect to some differential.

There are important additional restrictions to the solutions
of Eqs.~(\ref{eq4.6ne2}),
which are important for us.
These are the conditions
\beqa\label{e4.2}
&
\bigl[x^{\eta}, \Sccl_{n;\, \xi} \bigr] \, = \, 0 \,,
& \\ & \label{aa2}
\Sccl_{n;\, \xi} \, \FI_{\Ll}\hspace{1pt} \PA_n \, \subseteq \, \FI_{\Ll+1} \DP_{n,0}
&
\eeqa
($\xi,\eta = 1,\dots,D(n-1)$, $\Ll=0,1,\dots$),
which have to satisfy all renormalization cocycles defined by
(\ref{eq4.3ne1})
(the first one is due to Eqs.~(\ref{e3.9}) and conditions
($r3$) and ($p4$),
and the second one is due to ($r2$), ($p2$) and Eq.~(\ref{ineqq})).

\medskip

\begin{Theorem}\label{Th4.1qq1}
Let $n>2$ and we have a system of primary renormalization maps
$\PRMA_2,$ $\PRMA_3,$ $\dots,$ $\PRMA_n$
(which therefore determine renormalization maps
$\RMA_2,$ $\RMA_3,$ $\dots,$ $\RMA_n$).
Let $\{\Sccl_{n;\, \xi}\}_{\xi}$ be defined accordingly to
Eq.~(\ref{eq4.3ne1}) and
$\{\Sccl_{n;\, \xi}'\}_{\xi}$ be a solution of Eqs.~(\ref{eq4.6ne2}),
(\ref{e4.2}) and (\ref{aa2}),
which differs from $\{\Sccl_{n;\, \xi}\}_{\xi}$ by an
\EMPHTH{exact solution}, i.e., the difference
$\Sccl_{n;\, \xi}'$ $-$ $\Sccl_{n;\, \xi}$ is of a form
\beq\label{eq4.7ne2}
\Sccl'_{n;\, \xi} - \Sccl_{n;\, \xi} \, = \,
\bigl[\di_{x^{\xi}},\RDF_n\bigr]
\eeq
($\xi=1,\dots,D(n-1)$), for some linear map
$$
\RDF_n :
\PA_n
\to
\DP_{n,0} \,.
$$
Then there exists a primary renormalization map $\PRMA_n'$,
which together with $\PRMA_2,$ $\dots,$ $\PRMA_{n-1}$ determines
a system of renormalization maps
$\RMA_2,$ $\dots,$ $\RMA_{n-1}$ and $\RMA_n'$
and a primary renormalization cocycle coinciding with
$\{\Sccl_{n;\, \xi}'\}_{\xi}$.
\end{Theorem}%

\medskip

\begin{Proof}
Equations (\ref{eq4.6ne2}) have an obvious form of cohomological equations.
If $\{\Sccl_{n;\, \xi}\}_{\xi}$ is a solution of them for some fixed $n$
then $\{\Sccl_{n;\, \xi}'\}_{\xi}$ related to $\{\Sccl_{n;\, \xi}\}_{\xi}$
by Eq.~(\ref{eq4.7ne2}) is a solution too.
If $\{\Sccl_{n;\, \xi}\}_{\xi}$ satisfies conditions
(\ref{e4.2}) and (\ref{aa2})
and $\RDF_n$ satisfy also
$$
\bigl[x^{\eta}, \RDF_n \bigr]
\, = \,
0 \,,
\quad
\RDF_n \, \FI_{\Ll} \PA_n
\, \subseteq \,
\FI_{\Ll} \DP_{\R^N\hspace{-1pt},\hspace{1pt}0}
$$
then $\{\Sccl'_{n;\, \xi}\}_{\xi}$ also obey Eqs. (\ref{e4.2}) and (\ref{aa2}).

We shall show now that if the solution $\{\Sccl_{n;\, \xi}\}_{\xi}$
corresponds by Eq.~(\ref{eq4.3ne1}) to some renormalization map $\RMA_n$
then the changed solution $\{\Sccl'_{n;\, \xi}\}_{\xi}$~(\ref{eq4.7ne2})
corresponds to another renormalization map
$\RMA_n'$ $:$ $\PA_n \to \DP_n$.

To this end
let us introduce the subspace $\mathfrak{S}$ $\subset$
$\DP \bigl(\HES_n \mz\bigr)$ of distributions
that are supported at unions of linear subspaces of $\HES_n$.
Thus, $\DP \bigl[\LADIAG_n \mz\bigr]$ $\subset$ $\mathfrak{S}$.
We observe
next
that because $\SRMA_n$ is an injection
$\PA_n$ $\hookrightarrow$ $\DP_{temp} \bigl(\HES_n$ $\mz\bigr)$
then there exists a linear map
\beq\label{eq4.9ne2}
\widetilde{\RDF}_n :
\DP_{temp} \bigl(\HES_n \mz\bigr)
\to
\DP_{n,0}
\eeq
such that
$$
\RDF_n \, = \, \widetilde{\RDF}_n \circ \SRMA_n \,.
$$
Moreover, since
$\mathfrak{S}$ $\cap$
$\SRMA_n \bigl(\PA_n\bigr)$ $=$ $\{0\}$
we can additionally choose $\widetilde{\RDF}_n$ in such a way that
\beq\label{eq-qqq}
\widetilde{\RDF}_n \vrestr{12pt}{\mathfrak{S}} \, = \, 0 \,.
\eeq
In fact, $\widetilde{\RDF}_n$ can be constructed by a linear map
\beq\label{bb1}
\DP_{temp} \bigl(\HES_n \mz\bigr) \bigl/ \mathfrak{S}
\, \to \,
\DP_{n,0}
\eeq
that extends the map $\RDF_n$ under the embedding
\beq\label{bb2}
\PA_n \, \hookrightarrow \,
\DP_{temp} \bigl(\HES_n \mz\bigr) \bigl/ \mathfrak{S} \,.
\eeq
We claim also that it is possible to construct $\widetilde{\RDF}_n$
also in such a way that it additionally satisfies the following conditions:
(\textit{a}) $\widetilde{\RDF}_n$ preserves the
gradings,
$\widetilde{\RDF}_n$ $\FI_{\Ll}\hspace{1pt} \DP_{temp} \bigl(\HES_n \mz\bigr)$
$\subseteq$ $\FI_{\Ll}\hspace{1pt} \DP_{n,0}$
($\Ll=0,1,\dots$);
(\textit{b}) $\widetilde{\RDF}_n$ commutes with all $x^{\eta}$,
$\bigl[x^{\eta},$ $\RDF_n \bigr]$ $=$ $0$;
(\textit{c}) $\widetilde{\RDF}_n$ is invariant under orthogonal transformations
of $\HES_n$.
Indeed, to achieve (\textit{a}) we note that the linear map (\ref{bb2})
preserves the
gradings
and so,
the extension (\ref{bb1})
can be made to preserve them too. Regarding
 (\textit{b}), we use the arguments of Lemma~\ref{Lm2.4ne1}
in order to construct the extension (\ref{bb1})
in such a way that it also commutes with $x^{\eta}$.
Finally, (\textit{c}) can be simply achieved by the averaging operation
used in Sect.~\ref{se2.3}.

Thus, we can introduce a new primary renormalization map
$$
\PRMA_n' \, = \, \PRMA_n \, + \, \widetilde{\RDF}_n \,.
$$
and it satisfies all conditions $(p1)$--$(p5)$ by the construction.
(For instance, the condition $(p5)$ is ensured by condition (\ref{eq-qqq}).)
Due of Eq.~(\ref{eq-qqq}) we have also for $\PRMA_n'$:
$$
\bigl[\di_{x^{\xi}},\PRMA_n'\bigr] \circ \SRMA_n
\, = \,
\Sccl'_{n;\, \xi} \, - \,
\widetilde{\RDF}_n \circ \bigl[\di_{x^{\xi}},\SRMA_n\bigr]
\, = \, \Sccl'_{n;\, \xi}
\,,
$$
since the image of $\bigl[\di_{x^{\xi}},\SRMA_n\bigr]$
is contained in the space $\DP \bigl[\LADIAG_n \mz\bigr]$
$\subset$ $\mathfrak{S}$.
We obtained that $\Sccl'_{n;\, \xi}$ correspond by Eq.~(\ref{eq4.3ne1}) to another renormalization map
$\RMA_n'$ $=$ $\PRMA_n' \circ \SRMA_n$.
This completes the proof of the theorem.
\end{Proof}%

\medskp

Thus, we have seen that the linear maps $\Sccl_{n;\, \xi}$ (for fixed $n$),
corresponding to renormalization maps,
are characterized by
solutions of Eqs.~(\ref{eq4.6ne2}) modulo ``exact $1$--cocycles''
i.e., maps $c_{n;\, \xi}$ that satisfy the equations:
$$
\bigl[\di_{x^{\xi}},c_{n;\, \eta}\bigr]
-
\bigl[\di_{x^{\eta}},c_{n;\, \xi}\bigr] \, = \, 0 \,
$$
(for $\xi,\eta=1,\dots,D(n-1)$).
In the next subsection we shall reduce the corresponding
de Rham cohomologies to simpler ones.

\subsection{De Rham cohomologies of differential modules}\label{se4.2nn}

The linear maps $\Sccl_{n;\, \xi}$~(\ref{e3.9})
that determine the renormalization cocycles
$\Lccl_{n;\, \xi}$
take values that are distributions supported at $\Mbf{0} \in \HES_n$,
i.e. belonging to the space $\DP_{n,0}$.
Let us expand them in delta functions and their derivatives:
\beq\label{eq3.11}
\Sccl_{n;\, \xi} \, = \,
\mathop{\sum}\limits_{\rr \, \in \, \N_0^{N}} \ \frac{1}{\rr!} \
\delta^{(\rr)} (\xx)
\, \sccl_{n;\, \xi;\,\rr}
\, , \qquad
\sccl_{n;\, \xi;\,\rr} \, : \, \PA_n \to \C
\eeq
($N=D(n-1)$, $\delta^{(\rr)} (\xx)$ $:=$ $\di_{\xx}^{\rr} \delta (\xx)$,
$\di_{\xx}^{\rr}$ $:=$ $\prod_{\xi} \di_{x^{\xi}}^{r_{\xi}}$).
Note that the sum in (\ref{eq3.11}) becomes finite only after
applying both sides to a function.
Thus we get a characterization of $\Sccl_{n;\, \xi}$
by an infinite set of linear functionals $\sccl_{n;\, \xi;\,\rr}$
on $\PA_n$.
In fact, we shall show now that the whole information about $\Sccl_{n;\, \xi}$
\EMPH{is contained in the leading term} $\sccl_{n;\, \xi;\,0}$
of the expansion (\ref{eq3.11}).
This follows if we take into account the relation
$\bigl[x^{\eta}, \Sccl_{n;\, \xi} \bigr]$ $=$ $0$.
Combining the latter identity with (\ref{eq3.11}) we obtain
a recursive relation
$-(r_{\eta}+1)$ $\sccl_{n;\,\xi;\, \rr+\ee_{\eta}}$ $=$ $\sccl_{n;\, \xi;\,\rr} \circ x^{\eta}$,
which then implies that
\beq\label{e4.3}
\sccl_{n;\, \xi;\, \rr} \, = \, (-1)^{|\rr|} \, \sccl_{n;\, \xi;\, 0} \circ \xx^{\rr}
\eeq
($\xx^{\rr}$ $:=$ $\prod_{\xi} \bigl(x^{\xi}\bigr)^{r_{\xi}}$).
We set
\beq\label{e4.4}
\sccl_{n;\, \xi} \, := \, \sccl_{n;\, \xi;\, 0}
\eeq
for $\xi=1,\dots,D(n-1)$ and organize them as a $1$--form
\beq\label{e4.5}
\bsccl_n \, := \,
\mathop{\sum}\limits_{\xi \, = \, 1}^N \,
\sccl_{n; \xi} \ \, dx^{\xi}
\eeq
with coefficients in the dual differential module $\PA_n'$, i.e.,
$$
\bsccl_n \, \in \, \OM{1} \bigl(\PA_n'\bigr) \,.
$$
In more details, we fix the above notions in the following definitions.

\medskip

\begin{Definition}\label{Df3.1zz11}
Let us introduce the associative $\GF$--algebra
$\DO{N}$ $(N \in \N)$
of all linear partial differential operators over $\R^N$
with polynomial coefficients belonging to $\GF [\xx]$
(recall that $\GF$ is the ground field $\GF \subset \R$).
A $\Z$--\textbf{filtered}\footnote{%
in the previous section we used $\R$--filtrations but here it
will be sufficient to restrict them to $\Z$}
$\DO{N}$--module is a module $\SCi$ of $\DO{N}$,
which is endowed with an increasing filtration
$$
\SCi \, = \, \mathop{\bigcup}\limits_{\Ll \, \in \, \Z}
\FI_{\Ll}\hspace{1pt} \SCi
\,,\quad
\FI_{\Ll}\hspace{1pt} \SCi \, \subseteq \, \FI_{\Ll+1} \SCi \,,
$$
such that
for every $A \in \DO{N}$ and $u \in \SCi$ we have
\beq\label{eq3.10pp1}
\SCDEG \, A u \, \leqslant \, \SCDEG \, A \, + \, \SCDEG \, u \,,
\eeq
where
$$
\SCDEG \, u \, := \, \min \, \bigl\{\Ll : u \in \FI_{\Ll} \SCi\bigr\}
$$
and the scaling degree of a differential operator is defined by:
$$
\SCDEG
\left(\raisebox{18pt}{\hspace{-2pt}}\right.
\mathop{\sum}\limits_{\rr \, \in \, \N_0^N}
f_{\rr} (\x) \, \di^{\rr}
\left.\raisebox{18pt}{\hspace{-2pt}}\right)
\, = \,
\mathop{\max}\limits_{\rr \, \in \, \N_0^N}
\, \bigl\{ |\rr| + \scdeg \, f_{\rr} \bigr\} \,.
$$
\end{Definition}%

\medskip

The algebras $\PA_n$ and the distribution spaces $\DP$
considered in Sect.~\ref{se2.2nw} give us examples of $\Z$--filtered
$\DO{N}$--modules (for $\PA_n$: $N$ $=$ $D(n-1)$).
In particular, $\DP_{\R^N\hspace{-1pt},\hspace{1pt}0}$
becomes a $\Z$--flirted $\DO{N}$--module in which
$\FI_{\Ll} \DP_{\R^N\hspace{-1pt},\hspace{1pt}0}$ $=$ $\{0\}$
for $\Ll < N$.

Thus, $\sccl_{n;\, \xi}$ belong to the dual module of
$\PA_n$ generally defined by:

\medskip

\begin{Definition}\label{Df4.3xx4}
For a $\DO{N}$--module $\SCi$ the \textbf{dual} module is the algebraic dual space $\DSCi$ endowed with
the dual action of $\DO{N}$: for $\Phi \in \DSCi$ and $\xi=1,\dots,N$ the dual actions
of $x^{\xi}$ and $\di_{x^{\xi}}$ are
$x^{\xi} \bigl(\Phi\bigr)$ $:=$ $\Phi \circ x^{\xi}$ and
$\di_{x^{\xi}} \bigl(\Phi\bigr)$ $:=$ $-\Phi \circ \di_{x^{\xi}}$,
respectively.
\end{Definition}%

\medskip

Let us point out that the dual differential module $\SCi'$
of a $\Z$--filtered $\DO{N}$--module $\SCi$ is not naturally $\Z$--graded.
But it has a differential submodule that is $\Z$--graded.
A simple computation shows that
\beq\label{eq4.16xx6}
\SCi^{\du}
\, := \,
\mathop{\bigcup}\limits_{\Ll \, \in \, \Z} \,
\bigl(\FI_{\Ll}\hspace{1pt}\SCi\bigr)^{\perp}
\ \ ( \, \subseteq \DSCi)
\, , \quad
\bigl(\FI_{\Ll}\hspace{1pt}\SCi\bigr)^{\perp} \, := \,
\Bigl\{
\Phi \in \DSCi
:
\Phi\vrestr{10pt}{\FI_{\Ll}\hspace{1pt}\SCi} = 0
\Bigr\}
\,
\eeq%
is a $\DO{N}$--submodule of $\SCi'$ and it becomes $\Z$--filtered with an increasing
$\Z$--filtration if we set
$$
\FI_{\Ll}\hspace{1pt}\SCi^{\du} \, := \, \bigl(\FI_{-\Ll}\SCi\bigr)^{\perp} \,.
$$

The linear functionals $\sccl_{n;\, \xi}$
that characterize the primary renormalization cocycles $\Sccl_{n;\, \xi}$
are elements of $\bigl(\PA_n\bigr)^{\du}$.
We shall show this in a slightly more general situation.

Let $\SCi$ be a $\Z$--graded $\DO{N}$--module and denote
\beqa\label{eq4.26ne31}
\hspace{-5pt}
\RM_N \bigl(\SCi\bigr) :=
\Bigl\{ \podr\!\!
\phi :
\phi \text{ linearly maps } \SCi \text{ to } \DP_{\R^N\hspace{-1pt},\hspace{1pt}0},\
\bigl[x^{\xi},\phi\bigr] = 0 \ (\forall \xi = 1,\dots,N),\,
\nn \podr\!\!
\text{and }\exists \, L \in \Z \text{ such that } \phi \FI_{\Ll} \SCi \subseteq
\FI_{\Ll+L} \DP_{\R^N\hspace{-1pt},\hspace{1pt}0} \ (\forall \Ll \in \Z)
\Bigr\} \,.
\nonumber
\eeqa
(so for example, $\Sccl_{n;\, \xi}$ $\in$ $\RM_N \bigl(\PA_n\bigr)$
for $N=D(n-1)$).
Expand $\phi$ $\in$ $\RM_N \bigl(\SCi\bigr)$ in delta functions and their derivatives:
\beq\label{eq4.12xx5}
\phi \, = \, \mathop{\sum}\limits_{\rr \, \in \, \N_0^N} \ \frac{1}{\rr!} \
\delta^{(\rr)} (\xx)
\, \Phi_{\rr}
\, , \quad
\Phi_{\rr} : \SCi \to \R \,.
\eeq
Then the assignment
\beq\label{eq4.23ne41}
\phi \, \mapsto \, \Phi_0
\eeq
is injective and
$\phi$ is determined by $\Phi_0$ by the formula
\beq\label{eq4.15new}
\Phi_{\rr} \, = \, (-1)^{|\rr|} \, \Phi_0 \circ \xx^{\rr}
\,.
\eeq%
This is proven exactly as above for the case of $\Sccl_{n;\, \xi}$.
Furthermore, under the assignment (\ref{eq4.23ne41}):
\beq\label{eq4.24ne2}
\text{if} \quad \
\phi \mapsto \Phi_0
\quad \ \text{then} \quad \
\bigl[\di_{x^{\xi}},\phi\bigr] \mapsto - \Phi_0 \circ \di_{x^{\xi}} \,.
\eeq

\medskip

\begin{Proposition}\label{Pr4.3xx1}
The image of $\RM_N \bigl(\SCi\bigr)$
under the linear map~(\ref{eq4.23ne41}) is the vector space $\SCi^{\du}$.
\end{Proposition}%

\medskip

\begin{Proof}
First, let $\phi \in \RM_N \bigl(\SCi\bigr)$ and let
$\phi \FI_{\Ll} \SCi$ $\subseteq$
$\FI_{\Ll+L} \DP_{\R^N\hspace{-1pt},\hspace{1pt}0}$ $(\Ll \in \Z)$.
Then $\Phi_0$ $\in$ $\bigr(\FI_{\Ll}\SCi\bigl)^{\perp}$
if $\Ll+L < N$.

Conversely, let
$\Phi \in \SCi^{\du}$ and
define by (\ref{eq4.12xx5}) and (\ref{eq4.15new}) with $\Phi_0 = \Phi$
a linear map
$\phi : \SCi \to \DP_{\R^N\hspace{-1pt},\hspace{1pt}0}$.
Note that the sum in (\ref{eq4.12xx5})
is always finite when we apply it on an element of $\SCi$,
since $\Phi$ $\in$ $\bigl(\FI_{\Ll}\hspace{1pt}\SCi\bigr)^{\perp}$
for some $\Ll \in \Z$ and
$\xx^{\rr} \FI_{\MM}\SCi \subseteq \FI_{\MM-|\rr|}\SCi$
for every $\MM \in \Z$.
The latter also implies that if $\SCDEG \, u = \MM$
then $\scdeg \, \phi \bigl(u\bigr)$ $\leqslant$ $N+\KK$,
where $\Ll = \MM-\KK$.
Hence, $\phi \in \RM_N \bigl(\SCi\bigr)$
since the equations $\bigl[x^{\xi},\phi\bigr]=0$ ($\xi=1,\dots,N$)
follow as above.
By the construction $\phi$ is mapped on $\Phi$
via the assignment (\ref{eq4.15new}).
\end{Proof}%

\medskip

\begin{Corollary}\label{Cr4.2}
If $\Sccl_{n;\, \xi}$ are primary renormalization cocycles~(\ref{eq4.3ne1})
then the linear functionals $\sccl_{n;\, \xi}$~(\ref{e4.4})
belong to $\bigl(\PA_n\bigr)^{\du}$.
Conversely, every set of functionals
$\sccl_{n;\, \xi}$ $\in$ $\bigl(\PA_n\bigr)^{\du}$
determine by Eqs.~(\ref{e4.3}), (\ref{e4.4}) and (\ref{eq3.11})
a set of linear maps
$\Sccl_{n;\, \xi}$ $:$ $\PA_n$ $\to$ $\DP_{n,0}$
that satisfy the conditions $\bigl[x^{\eta},\Sccl_{n;\, \xi} \bigr] = 0$.
\end{Corollary}%

\medskip

Thus, we see that the cohomology behind
the cohomological equations (\ref{eq4.6ne2})
is exactly the de Rham cohomology of the differential module
$\bigl(\PA_n\bigr)^{\du}$.
Let us recall its general definition.

\medskip

\begin{Definition}\label{Dfx1}
The \textit{de Rham complex} for an arbitrary
$\DO{N}$--module $\SCi$ is defined as
the complex:
\beq\label{eq4.20ww6}
\{0\}
\mathop{\longrightarrow}\limits^{d} \,
\OM{0} \bigl(\SCi\bigr)
\, \mathop{\longrightarrow}\limits^{d} \,
\OM{1} \bigl(\SCi\bigr)
\, \mathop{\longrightarrow}\limits^{d} \,\, \cdots \,\, \mathop{\longrightarrow}\limits^{d} \,
\OM{N} \bigl(\SCi\bigr)
\, \mathop{\longrightarrow}\limits^{d} \, \{0\}\,,
\eeq
where
$$
\OM{0} \bigl(\SCi\bigr) \, \equiv \, \SCi
\, , \quad
\OM{m} \bigl(\SCi\bigr)
\, := \,
\Lambda^m \bigl(\R^N\bigr) \otimes
\SCi \,,
$$
and $\Lambda^m \bigl(\R^N\bigr)$ stands for the $m$th antisymmetric power of $\R^N$.
Thus, the elements of $\OM{m} \bigl(\SCi\bigr)$
are represented by sequences $\THETA$ $=$ $\bigl(\Theta_{\xi_1,\dots,\xi_m}\bigr)$ with
coefficients
$\Theta_{\xi_1,\dots,\xi_m}$ $\in$
$\SCi$
for $\xi_1,\dots,\xi_m = 1,\dots,N$, which are antisymmetric,
$\Theta_{\xi_1,\dots,\xi_m}$ $=$
$(-1)^{\text{sgn} \, \sigma}$ $\Theta_{\xi_{\sigma_1},\dots,\xi_{\sigma_m}}$.
The differential of $\Theta_{\xi_1,\dots,\xi_m}$ is
\beq\label{eq4.21ww5}
\bigl(d \THETA\bigr)_{\xi_1,\dots,\xi_{m+1}}
\, = \,
\mathop{\sum}\limits_{\ell \, = \, 1}^m \,
(-1)^{\ell+1} \,
\di_{x^{\xi_{\ell}}} \,
\Theta_{\xi_1,\dots,\widehat{\xi_{\ell}},\dots,\xi_{m+1}} \,.
\eeq
Denote by $\HOM{m} \bigl(\SCi\bigr)$, for $m=0,\dots,N$, the cohomology group of the complex (\ref{eq4.20ww6}):
\beqa\label{eq4.22ww5}
&& \hspace{50pt}
\HOM{m} \bigl(\SCi\bigr) \, := \,
\CLO{m} \bigl(\SCi\bigr)
{\hspace{1pt}}\Bigl/{\hspace{1pt}}
\EXA{m} \bigl(\SCi\bigr)
\, , \quad
\nn &&
\CLO{m} \bigl(\SCi\bigr) \, := \,
\Ker \ d \vrestr{12pt}{\OM{m} \bigl(\SCi\bigr)}
\, , \quad
\EXA{m} \bigl(\SCi\bigr) \, := \,
d \bigl(\OM{m-1} \bigl(\SCi\bigr)\bigr) \,.
\qquad
\nonumber
\eeqa
For a $Z$--filtered $\DO{N}$--module $\SCi$ we also set:
\beqa\label{xxt}
\FI_{\Ll}\hspace{1pt}\OM{m} \bigl(\SCi\bigr) \, := \podr
\Lambda^m \bigl(\R^N\bigr) \otimes \FI_{\Ll}\hspace{1pt}\SCi
\, , \quad
\nn
\FI_{\Ll}\hspace{1pt}\CLO{m} \bigl(\SCi\bigr) \, := \podr
\CLO{m} \bigl(\SCi\bigr) \cap
\FI_{\Ll}\hspace{1pt}\OM{m} \bigl(\SCi\bigr)
\, , \quad
\nn
\FI_{\Ll}\hspace{1pt}\EXA{m} \bigl(\SCi\bigr) \, := \podr
\EXA{m} \bigl(\SCi\bigr) \cap
\FI_{\Ll}\hspace{1pt}\OM{m} \bigl(\SCi\bigr) \,
\nonumber
\eeqa%
for $\Ll \in \Z$.
\end{Definition}%

\medskip

Applying the above abstract results
to our primary renormalization cocycles $\{\Sccl_{n;\,\xi}\}_{\xi}$
we can claim that they are characterized by a cohomology class in
$\HOM{1} \bigl(\PA_n^{\hspace{1pt}\du}\bigr)$
($\PA_n^{\hspace{1pt}\du}$ $\equiv$ $\bigl(\PA_n\bigr)^{\!\!\du}$).
We can further simplify the description of this cohomology class
due to the following result.

\medskip

\begin{Theorem}\label{ThXX}
In the case of the algebra $\PA_n$
given by space of rational functions~(\ref{PAN})
there exists a natural isomorphism:
$$
\HOM{m}\Bigl( \PA_n^{\hspace{1pt}\du} \Bigr)
\, \cong \,
\Bigl(\HOM{N-m} \bigl( \PA_n \bigr)\Bigr)\raisebox{12pt}{\hspace{-1pt}}'
\,
$$
for $0 \leqslant m \leqslant N$.
\end{Theorem}%

\medskp

Let us point out that there is a
result \cite[Ch.~1, Theorems 5.5 and 6.1]{Bj}
stating that
\textit{all the de Rham cohomologies
$\HOM{m} \bigl( \PA_n \bigr)$ are finite dimensional for the case
when $\PA_n$ are given by (\ref{PAN})}
and hence, $\HOM{m} \bigl( \PA_n^{\hspace{1pt}\du} \bigr)$
are finite dimensional too.

We conclude this subsection with the \textit{proof} of Theorem~\ref{ThXX}.
Since it will not be used in the remaining part of this paper
the reader can skip it in
the first reading.

First we start with an analog of the Poincar\'e lemma in the case of
de Rham cohomologies of $\DO{N}$--modules.
It uses partial inevitability of the Euler operator (vector field) on $\R^N$,
$$
\xx \spr \di_{\xx} \, = \, \mathop{\sum}\limits_{\xi \, = \, 1}^N \, x^{\xi} \, \di_{x^{\xi}} \,.
$$
Due to Eq.~(\ref{eq3.10pp1}) this operator keeps invariant every
$\FI_{\Ll} \SCi$.

\medskip

\begin{Lemma}\label{Pr5.1qq1}
Let $\SCi$ be a $\Z$--filtered $\DO{N}$--module,
which is such that for some $\Ll \in \Z$
the operator $k+\xx \spr \di_{\xx}$
it is invertible on $\FI_{\Ll}\hspace{1pt}\SCi$ for every $k \in \N_0$.
Then every closed form with coefficients in $\FI_{\Ll}\hspace{1pt}\SCi$
is exact, i.e., if $\THETA$ $\in$
$\FI_{\Ll}\hspace{1pt}\OM{m} \bigl(\SCi\bigr)$
then $d \hspace{1pt} \THETA$ $=$ $0$
implies that $\THETA = d \BBB$
for some $\BBB$ $\in$ $\OM{m-1} \bigl(\SCi\bigr)$.
\end{Lemma}%

\medskip

\begin{Proof}
The proposition can be proven by using the $\DO{N}$--analog
of the Po\-i\-n\-c\-a\-r\'e homotopy operator
$$
\bigl(K \hspace{1pt} \THETA\bigr)_{\xi_1,\dots,\xi_{m-1}} \, = \,
\bigl(m-1+\xx \spr \di_{\xx}\bigr)^{-1} \,
\mathop{\sum}\limits_{\xi \, = \, 1}^N \,
x^{\xi} \, \Theta_{\xi,\xi_1,\dots,\xi_{m-1}}
$$
where $\THETA = \bigl(\Theta_{\xi_1,\dots,\xi_m}\bigr)$ $\in$
$\FI_{\Ll}\hspace{1pt}\OM{m} \bigl(\SCi\bigr)$ $=$
$\Lambda^m \bigl(\R^N\bigr) \otimes \FI_{\Ll}\hspace{1pt}\SCi$.
Since the operators $k+\xx \spr \di_{\xx}$ commute with
the derivatives $\di_{x^{\xi}}$
then the operators $\bigl(k+\xx \spr \di_{\xx}\bigr)^{-1}$
restricted to $\FI_{\Ll-1}\SCi$ and $\FI_{\Ll}\SCi$
also commute with $\di_{x^{\xi}}$
$:$ $\FI_{\Ll-1}\SCi$ $\to$ $\FI_{\Ll}\SCi$.
Hence, we derive the identity $Kd + dK = \id$
from which the proposition follows.
\end{Proof}%

\medskip

We continue with the \textit{proof of Theorem~\ref{ThXX}}.
There is a natural pairing
$$
\OM{m} \bigl(\SCi^{\du}\bigr)
\otimes
\OM{N-m} \bigl(\SCi\bigr)
\, \to \, \R
\, : \, \bigl(\OMEGA,\, \aLPHA \bigr) \, \mapsto \, \OMEGA\wed \bigl[\aLPHA\bigr]
\,
$$
for every $m = 0,\dots,N$,
where $\OMEGA\wed \bigl[\aLPHA\bigr]$ means the action of $\OMEGA$ as a linear functional
under the external product $\wedge$.
Precisely, for $\OMEGA$ $=$ $a \otimes \Phi$ $\in$ $\Lambda^m \bigl(\R^N\bigr)$ $\otimes$ $\SCi^{\du}$
and $\aLPHA$ $=$ $b \otimes f$ $\in$ $\Lambda^{N-m} \bigl(\R^N\bigr)$ $\otimes$ $\SCi$
we set:
$$
\OMEGA\wed \bigl[\aLPHA\bigr] \, := \,
\bigl(a \wedge b \bigr) \, \Phi \bigl[f\bigr] \,,
$$
where $a \wedge b$ is considered as an element of $\R$ $\cong$ $\Lambda^N \bigl(\R^N\bigr)$.
Then we have
\beq\label{eq4.29hh1}
\bigl(d \hspace{1pt} \OMEGA\bigr)\wed \bigl[\aLPHA\bigr] \, = \,
(-1)^{m+1} \, \OMEGA\wed \bigl[d \hspace{1pt} \aLPHA\bigr] \,
\eeq
for $\OMEGA$ $\in$ $\OM{m} \bigl(\SCi\bigr)$ and $\aLPHA$ $\in$ $\OM{N-m-1} \bigl(\SCi\bigr)$,
according to Eq.~(\ref{eq4.21ww5}).
We denote
$$
\OMEGA\wed \, : \, \OM{N-m} \bigl(\SCi\bigr) \, \to \, \R \, : \, \aLPHA \, \mapsto \, \OMEGA\wed \bigl[\aLPHA\bigr] \,.
$$
Note now that if $\OMEGA$ is closed then
$\OMEGA\wed \bigl[d\hspace{1pt}\aLPHA'\bigr] = 0$ for every $\aLPHA'$ $\in$ $\OM{N-m-1} \bigl(\SCi\bigr)$
and if $\OMEGA$ is exact then
$\OMEGA\wed \bigl[\aLPHA\bigr] = 0$ for every closed $\aLPHA$ $\in$ $\OM{N-m} \bigl(\SCi\bigr)$.
Hence, for closed $\OMEGA$ and $\aLPHA$ the number $\OMEGA\wed \bigl[\aLPHA\bigr]$
depends only on the cohomology classes of $\OMEGA$ and $\aLPHA$
and thus we obtain a natural linear map
\beq\label{eq4.30hh1}
\HOM{m} \Bigl(\SCi^{\du}\Bigr) \, \to \,
\Bigl( \HOM{N-m} \bigl(\SCi\bigr) \Bigr)\raisebox{12pt}{\hspace{-1pt}}'
\,.
\eeq

\medskip

\begin{Lemma}\label{Pr5.2qq1}
For a $\DO{N}$--module $\SCi$ satisfying the conditions of
Lemma~\ref{Pr5.1qq1}
the above natural map (\ref{eq4.30hh1}) is an isomorphism.
\end{Lemma}%

\medskip

\begin{Proof}
First we prove that (\ref{eq4.30hh1}) is surjective.
Let us have a linear functional
\beq\label{eq5.7jj6}
\CLO{N-m} \bigl(\SCi\bigr){\hspace{-1pt}}\Bigl/{\hspace{0pt}}\EXA{N-m} \bigl(\SCi\bigr)
\ ( = \HOM{N-m} \bigl(\SCi\bigr) )
\ \mathop{\longrightarrow}\limits^{\Omega'} \ \R \,.
\eeq
Then our task is to extend $\Omega'$ to a linear functional
\beq\label{eq5.8jj6}
\OM{N-m} \bigl(\SCi\bigr){\hspace{-1pt}}\Bigl/{\hspace{0pt}}\EXA{N-m} \bigl(\SCi\bigr)
\, \mathop{\longrightarrow}\limits^{\Omega''} \, \R
\eeq
such that $\Omega'' \circ \pi = \OMEGA\wed$ for an element $\OMEGA \in \OM{m} \bigl(\SCi^{\du}\bigr)$,
where $\pi$ is the natural projection
$\OM{N-m} \bigl(\SCi\bigr)$ $\to$
$\OM{N-m} \bigl(\SCi\bigr){\hspace{-1pt}}\Bigl/{\hspace{0pt}}\EXA{N-m} \bigl(\SCi\bigr)$.
It is always possible to extend $\Omega'$~(\ref{eq5.7jj6}) to some linear functional $\Omega''$~(\ref{eq5.8jj6})
and we point out also that every linear functional
$\Theta : \OM{N-m} \bigl(\SCi\bigr) \to \R$
is of a form $\THETA\wed$ for a unique $\THETA \in \OM{m} \bigl(\DSCi\bigr)$.
So, we have an element $\OMEGA \in \OM{m} \bigl(\DSCi\bigr)$ such that $\OMEGA\wed = \Omega'' \circ \pi$
and it remains only to achieve $\OMEGA \in \OM{m} \bigl(\SCi^{\du}\bigr)$.
The latter requires to impose further conditions on the extension $\Omega''$~(\ref{eq5.8jj6}).
We require that $\Omega''$ is zero on
$$
\pi \bigl( \FI_{\Ll}\hspace{1pt}\OM{N-m} \bigl(\SCi\bigr)\bigr) \, \equiv \,
\FI_{\Ll}\hspace{1pt}\OM{N-m} \bigl(\SCi\bigr){\hspace{-1pt}}\Bigl/{\hspace{0pt}}\EXA{N-m} \bigl(\SCi\bigr)
$$
for some $\Ll \in \Z$.
This is always possible since
\beqa
\podr
\Bigl(\FI_{\Ll}\hspace{1pt}\OM{N-m} \bigl(\SCi\bigr){\hspace{-1pt}}\Bigl/{\hspace{0pt}}\EXA{N-m} \bigl(\SCi\bigr)\Bigr)
\cap
\Bigl(\CLO{N-m} \bigl(\SCi\bigr){\hspace{-1pt}}\Bigl/{\hspace{0pt}}\EXA{N-m} \bigl(\SCi\bigr)\Bigr)
\nn \podr \quad = \,
\FI_{\Ll}\hspace{1pt}\CLO{N-m} \bigl(\SCi\bigr){\hspace{-1pt}}\Bigl/{\hspace{0pt}}\EXA{N-m} \bigl(\SCi\bigr)
\, = \, \{0\}
\nonumber
\eeqa
for $\Ll$ chosen according to the assumptions of Proposition~\ref{Pr5.1qq1}.
In this way we have
$\OMEGA\wed\bigl[\aLPHA\bigr]=0$ if
$\aLPHA \in \FI_{\Ll}\hspace{1pt} \OM{N-m} \bigl(\SCi\bigr)$
and hence, $\OMEGA \in \OM{m} \bigl(\SCi^{\du}\bigr)$.
Thus, the linear map (\ref{eq4.30hh1}) is surjective.

To prove that the map (\ref{eq4.30hh1}) is injective
assume that $\OMEGA \in \OM{m} \bigl(\SCi^{\du}\bigr)$ is such that
$\OMEGA\wed \bigl[\aLPHA\bigr] = 0$ for all $\aLPHA \in \CLO{N-m} \bigl(\SCi\bigr)$.
We should prove that $\OMEGA = d \THETA$ for $\THETA \in \OM{m-1} \bigl(\SCi^{\du}\bigr)$.
To this end we note first that
$\OMEGA\wed\vrestr{10pt}{\FI_{\Ll'}\OM{N-m} \bigl(\SCi\bigr)}$ $=$ $0$
for some $\Ll' \in \Z$
and we set $\Ll_0$ $=$ $\min\{\Ll,\Ll'\}$, where $\Ll \in \Z$
is the integer from the assumptions of Proposition~\ref{Pr5.1qq1}.
Thus, $\OMEGA\wed\vrestr{10pt}{\FI_{\Ll_0}\OM{N-m} \bigl(\SCi\bigr)}$ $=$ $0$.
Now
consider the short exact sequence
\beqa\label{eq5.9jj6}
&&
0 \, \to \, \CLO{N-m} \bigl(\SCi\bigr){\hspace{-1pt}}\Bigl/{\hspace{0pt}}\FI_{\Ll_0-1}\CLO{N-m} \bigl(\SCi\bigr)
\, \hookrightarrow \,
\OM{N-m} \bigl(\SCi\bigr){\hspace{-1pt}}\Bigl/{\hspace{0pt}}\FI_{\Ll_0-1}\OM{N-m} \bigl(\SCi\bigr)
\nn && \hspace{50pt}
\mathop{\to}\limits^{d} \,
\EXA{N-m+1} \bigl(\SCi\bigr){\hspace{-1pt}}\Bigl/{\hspace{0pt}}\FI_{\Ll_0}\EXA{N-m+1} \bigl(\SCi\bigr) \, \to \, 0 \,
\,,
\nonumber
\eeqa
which is due to $\FI_{\Ll_0}\EXA{N-m+1} \bigl(\SCi\bigr)$
$=$ $d \, \FI_{\Ll_0-1}\OM{N-m} \bigl(\SCi\bigr)$
(Proposition~\ref{Pr5.1qq1}).
Then we obtain a linear functional
$$
\Theta' : \EXA{N-m+1} \bigl(\SCi\bigr){\hspace{-1pt}}\Bigl/{\hspace{0pt}}\FI_{\Ll_0}\EXA{N-m+1} \bigl(\SCi\bigr) \to \R
$$
such that $\OMEGA\wed = \Theta' \circ \pi' \circ d$,
where $\pi'$ is the projection
$$
\EXA{N-m+1} \bigl(\SCi\bigr) \, \mathop{\longrightarrow}\limits^{\pi'} \,
\EXA{N-m+1} \bigl(\SCi\bigr){\hspace{-1pt}}\Bigl/{\hspace{0pt}} \FI_{\Ll_0}\EXA{N-m+1} \bigl(\SCi\bigr)\,.
$$
Finally, we extend $\Theta'$ to a linear functional $\Theta''$,
$$
\Theta'' : \OM{N-m+1} \bigl(\SCi\bigr){\hspace{-1pt}}\Bigl/{\hspace{0pt}}\FI_{\Ll_0}\OM{N-m+1} \bigl(\SCi\bigr) \to \R
$$
(under the natural embedding
$$
\EXA{N-m+1} \bigl(\SCi\bigr){\hspace{-1pt}}\Bigl/{\hspace{0pt}}\FI_{\Ll_0}\EXA{N-m+1} \bigl(\SCi\bigr)
\, \hookrightarrow \,
\OM{N-m+1} \bigl(\SCi\bigr){\hspace{-1pt}}\Bigl/{\hspace{0pt}}\FI_{\Ll_0}\OM{N-m+1} \bigl(\SCi\bigr) \,)
$$
and setting $\THETA\wed := \Theta'' \circ \pi''$, where
$$
\OM{N-m+1} \bigl(\SCi\bigr) \, \mathop{\longrightarrow}\limits^{\pi''} \,
\OM{N-m+1} \bigl(\SCi\bigr){\hspace{-1pt}}\Bigl/{\hspace{0pt}} \FI_{\Ll_0}\OM{N-m+1} \bigl(\SCi\bigr)\,,
$$
we get by Eq.~(\ref{eq4.29hh1}) that
$\OMEGA$ $=$ $(-1)^{m+1} \, d \,\THETA$ and $\THETA \in \OM{m-1} \bigl(\SCi^{\du}\bigr)$
(i.e., $\OMEGA\wed\bigl[\aLPHA\bigr] = \THETA\wed\bigl[d\hspace{1pt}\aLPHA\bigr]$
for all $\aLPHA \in \OM{m-1} \bigl(\SCi\bigr)$).
Hence, $\OMEGA$ is exact and thus, the linear map~(\ref{eq4.30hh1})
is also injective.
\end{Proof}%

\medskp

To complete the proof of Theorem~\ref{ThXX}
we need to apply Lemma~\ref{Pr5.2qq1} to the
$\DO{D(n-1)}$--module $\PA_n$~(\ref{PAN}).
This is possible due to the following result.

\medskp

\begin{Lemma}\label{Pr5.3jj1}
The $\DO{D(n-1)}$--modules
$\PA_n$~(\ref{PAN})
satisfy the assum\-p\-tions of Lem\-ma~\ref{Pr5.1qq1}.
\end{Lemma}%

\medskip

\begin{Proof}
We shall prove that
all the operators
$\ell+\xx \spr \di_{\xx}$, for $\ell=0,1,2,\dots$, are invertible
on the subspace of $\OM{m} \bigl(\PA_n\bigr)$, which consists of
elements with negative scaling degree.
Indeed, if
$\THETA$ $=$ $\bigl(\Theta_{\xi_1,\dots,\xi_m} (\xx)\bigr)$
$\in$ $\OM{m} \bigl(\PA_n\bigr)$
and $\SCDEG \, \THETA < 0$
then for every $\xx$ $\notin$ $\LADIAG_n$ ($:=$ $\HES_n \backslash \HF_n$)
the function
$\lambda^{-1}$ $\Theta_{\xi_1,\dots,\xi_m} (\lambda \hspace{1pt} \xx)$
is integrable for $\lambda \in \bigl(0,1\bigr)$.
Hence, we define $\bigl(\ell+\xx \spr \di_{\xx}\bigr)^{-1} \, \THETA$ by
$$
\Bigl(\bigl(\ell+\xx \spr \di_{\xx}\bigr)^{-1} \, \THETA\Bigr)
\raisebox{-5pt}{\hspace{0pt}}_{\xi_1,\dots,\xi_m} \bigl(\xx\bigr)
\, = \,
\mathop{\int}\limits_{\hspace{-7pt}0}^{\hspace{5pt}1}
\lambda^{\ell-1} \,
\THETA_{\xi_1,\dots,\xi_m} \bigl(\lambda \hspace{1pt} \xx\bigr) \, d\hspace{1pt}\lambda
\,.
$$
The right hand side above defines an element of $\PA_n$
since the
integrand
is a polynomial in $\lambda$
with coefficients in $\PA_n$.
\end{Proof}%

\subsection{Reduction of the cohomological equations}\label{se4.2}

By the results of the previous subsections we have characterized
the primary renormalization cocycles $\{\Sccl_{n;\,\xi}\}_{\xi}$,
for every $n=2,3,\dots$ by $1$--forms
$$
\bsccl_n \, = \,
\mathop{\sum}\limits_{\xi \, = \, 1}^N
\sccl_{n;\, \xi} \, dx^{\xi} \, \in \,
\OM{1} \bigl(\PA_n^{\hspace{1pt}\du}\bigr) \,
$$
($N=D(n-1)$, $\PA_n^{\hspace{1pt}\du}$ $\equiv$ $\bigl(\PA_n\bigr)^{\!\du}$).
Then $\Sccl_{n;\,\xi}$ are constructed by the formula:
$$
\Sccl_{n;\, \xi} (G) \, = \,
\mathop{\sum}\limits_{\rr \, \in \, \N_0^{N}}
\ \frac{(-1)^{|\rr|} \, }{\rr!} \
\sccl_{n;\, \xi} \bigl(\xx^{\rr} \hspace{1pt} G\bigr) \,
\delta^{(\rr)} (\xx)
$$
($G \in \PA_n$).
On the other hand, $\{\Sccl_{n;\,\xi}\}_{\xi}$
satisfy the cohomological equations (\ref{eq4.6ne2})
and we argued in Sect.~\ref{se4.1} that their right hand sides
depend on renormalization maps of lower order.
Thus, one can expect that these equations are equivalent to some
equations for $\bsccl_n$ of a form:
\beqa\label{eq6xx}
d \hspace{1pt} \bsccl_2 \, = \podr 0,
\nn
d \hspace{1pt} \bsccl_n \, = \podr
\mathcal{F}_n \bigl[\bsccl_1,\dots,\bsccl_{n-1}\bigr]
\qquad (n>2) \,.
\eeqa

In order to derive the right hand side of (\ref{eq6xx})
in a simple explicit form we shall need some notations.
Similarly to Sect.~\ref{se2} we introduce
set--dependent notations:
\beqa &&
\Sccl_{S;\,\xi} \, := \,
\bigl[\di_{x^{\xi}},\PRMA_S\bigr] \circ \SRMA_S
\, = \,
\mathop{\sum}\limits_{\rr \, \in \, \N_0^{N}}
\ \frac{(-1)^{|\rr|} \, }{\rr!} \
\delta^{(\rr)}_S \
\sccl_{S;\, \xi} \circ \xx^{\rr}
,\quad
\nn && \hspace{45pt}
\bsccl_S \, := \,
\mathop{\sum}\limits_{\xi \, = \, 1}^N
\sccl_{S;\, \xi} \, dx^{\xi} \, \in \,
\OM{1} \bigl(\PA_S^{\hspace{1pt}\du}\bigr) \,.
\nonumber \eeqa
($N=D(|S|-1)$)
for every finite subset $S \subset \N$ with at least two elements.
In the above equations we assume that we have fixed some
(linear) coordinates on $\HES_S$, which we denote by
$x^{\xi}$ ($\xi=1,\dots,N$).

For $S'$ $\subseteq$ $S$ $(\!\mathop{\subset}\limits_{\text{\ fin.}}$ $\N\,)$
we set
$$
S/S' \, := \, (S\backslash S') \cup \{\min \, S'\}
$$
and for every two subsets $S'$, $S''$ $\subseteq$ $S$ such that
$S' \cup S''$ $=$ $S$ and $|S' \cap S''|$ $=$ $1$
(for instance $S'$ and $S''$ $=$ $S/S'$ is such a pair)
there is a canonical isomorphism
\beq\label{isom-ss1}
\HES_S \, \cong \, \HES_{S'} \oplus \HES_{S''}
: \bigl[\x_j\bigr]_{j \in S} \mapsto
\bigl(\bigl[\x_{j'}\bigr]_{j' \in S'},
\bigl[\x_{j''}\bigr]_{j'' \in S''}\bigr) \,
\eeq
(recall that $\HES_S$ $:=$ $\bigl(\R^D\bigr)^S \bigl/ \R^D$ and
its elements, denoted by $[\x_j]_{j \in S}$,
are equivalence classes of elements
$(\x_j)_{j \in S}$ $\in$ $\bigl(\R^D\bigr)^S$).
Let us consider the pair $(S',S''=S/S')$ for a nonempty $S' \subsetneq S$
and we assume that we have equipped $\HES_{S'}$ with (linear) coordinates
$\xx'$ $=$
$\bigl(x'{}^{\xi'}\bigr)$ for $\xi'=1,\dots,N'$ ($N'=D(|S'|-1)$)
and $\HES_{S/S'}$ with coordinates
$\xx''$ $=$
$\bigl(x''{}^{\xi''}\bigr)$ for $\xi''=1,\dots,N''$
($N''=D(|S/S'|-1)$ $=$ $D (|S|-|S'|)$).
Then we denote
$\bigl(\Pi'_{S'}\bigr){\hspace{-1pt}}^{\xi'}_{\xi}$
$:=$ $\frac{\di x'{}^{\xi'}}{\di x^{\xi}}$
and
$\bigl(\Pi''_{S'}\bigr){\hspace{-1pt}}^{\xi''}_{\xi}$
$:=$ $\frac{\di x''{}^{\xi''}}{\di x^{\xi}}$
($\xi=1,\dots,N$, $\xi'=1,\dots,N'$, $\xi''=1,\dots,N''$)
using the change of coordinates $\bigl(x^{\xi}\bigr)$
$\mapsto$ $\bigl(x'{}^{\xi'},$ $x''{}^{\xi''}\bigr)$
implied by Eq.~(\ref{isom-ss1}).

Using the above notations we introduce bilinear operations
(see Lemma \ref{hlm})
\beq\label{BOP}
\spwedge \, : \,
\OM{k} \bigl(\PA_{n-m+1}^{\hspace{1pt}\du}\bigr)
\otimes
\OM{\ell} \bigl(\PA_m^{\hspace{1pt}\du}\bigr)
\, \to \,
\OM{k+\ell} \bigl(\PA_n^{\hspace{1pt}\du}\bigr)
\eeq
for every $1 <  m < n$ in the following way.
Set $S:=\{1,\dots,n\}$ and for every
$\Mbf{\Theta}$ $\in$ $\OM{k} \bigl(\PA_m^{\hspace{1pt}\du}\bigr)$
and
$S' \subset S$ with $|S'|=m$
we denote by $\Mbf{\Theta}_S$ the natural lift of $\Mbf{\Theta}$
from $\OM{k} \bigl(\PA_m^{\hspace{1pt}\du}\bigr)$
to $\OM{k} \bigl(\PA_{S'}^{\hspace{1pt}\du}\bigr)$
under the bijection $\{1,\dots,m\}$ $\cong$ $S'$ $:$ $s \mapsto j_s$
for $j_1 < \cdots < j_s$.
Then for
$\Mbf{\Theta}' \in \OM{k} \bigl(\PA_m^{\hspace{1pt}\du}\bigr)$ and
$\Mbf{\Theta}'' \in \OM{\ell} \bigl(\PA_{n-m+1}^{\hspace{1pt}\du}\bigr)$
we set
\beqa\label{qqq}
\bigl(
\Mbf{\Theta}''
\spwedge \,
\Mbf{\Theta}'
\bigr)
\bigl(G_S\bigr)
\hspace{8pt}
\podr \nn
\, = \,
\mathop{\sum}
\limits_{\mathop{}\limits^{S' \, \subsetneq \, S}_{|S'| \, = \, m}}
\mathop{\sum}\limits_{\rr' \, \in \, \N_0^{N'}}
\podr
\frac{1}{\rr'!} \
\bigl(\Pi''_{S'}\bigr)^*
\Mbf{\Theta}''_{S/S'}
\Bigl(
\Mbf{\di}_{\xx{}'}^{\rr'} \hspace{1pt} G_{\PRT(S')}\vrestr{12pt}{\xx'=0}
\Bigr)
\nn \podr \wedge
\,
\bigl(\Pi'_{S'}\bigr)^*
\Mbf{\Theta}'_{S'}
\bigl(\xx{}'{}^{\rr'} \hspace{1pt} G_{S'}\bigr)
,
\eeqa
where
$G_S$ $\in$ $\PA_S$
is taken of the form (\ref{GS}),
\beq\label{GSP}
G_S \, = \, G_{\PRT(S')} \, G_{S'} ,
\eeq
for the $S$--partition
$\PRT(S')$ $:=$ $\bigl\{\{j\}$ $:$ $j$ $\in$
$S \backslash S'\bigr\}$ $\cup$ $\{S'\}$,
$\bigl(\Pi'_{S'}\bigr)^*$ and
$\bigl(\Pi''_{S'}\bigr)^*$
stand for the obvious pullbacks, e.g.,
\beqa
\bigl(\Pi'_{S'}\bigr)^* \, : &\hspace{-4pt}
\Lambda^k \bigl(\HES_{S'}\bigr) &\hspace{-4pt} \to \,
\Lambda^k \bigl(\HES_S\bigr)
\nn
\, : &\hspace{-4pt}
\lambda_{\xi_1',\dots,\xi_k'} &\hspace{-4pt} \mapsto \,
\mathop{\sum}\limits_{\xi_1',\dots,\xi_k' \, = \, 1}^{N'}
\bigl(\Pi_{S'}\bigr){\hspace{-1pt}}^{\xi_1'}_{\xi_1}
\cdots
\bigl(\Pi_{S'}\bigr){\hspace{-1pt}}^{\xi_k'}_{\xi_k} \
\lambda_{\xi_1',\dots,\xi_k'}
\nonumber
\eeqa
and finally,
$\Mbf{\Theta} \bigl(G\bigr)$ for
$\Mbf{\Theta} \in \OM{r} \bigl(\PA_s^{\hspace{1pt}\du}\bigr)$
and $G \in \PA_s$ is considered as an element of
$\Lambda^r \bigl(\HES_s\bigr)$.

\medskp

\begin{Lemma}\label{hlm}
The bilinear operation $\spwedge$ (\ref{BOP})
is well defined by Eq.~(\ref{qqq}).
\end{Lemma}

\medskp

\begin{Proof}
First we point out that the sum in the right hand side of Eq. (\ref{qqq})
is always finite since the scaling degree of
$\xx{}'{}^{\rr'} \hspace{1pt} G_{S'}$ decreases as $|\rr'|$ increases.
Next, we need to show that Eq. (\ref{qqq}) defines a linear map in
$G \in \PA_n$.
This follows from the fact that for every $S' \subsetneq S$
the assignment
$$
G_S \, \mapsto \,
\mathop{\sum}\limits_{\rr' \, \in \, \N_0^{N'}} \
\frac{1}{\rr'!} \
\Bigl(
\Mbf{\di}_{\xx{}'}^{\rr'} \hspace{1pt} G_{\PRT(S')}\vrestr{12pt}{\xx'=0}
\Bigr)
\! \cdot \!
\bigl(
\xx{}'{}^{\rr'} \hspace{1pt} G_{S'}
\bigr)
$$
extends to a linear map in $G \in \PA_n$ since this is just
the Taylor expansion of $G_{\PRT(S')}$ in $\xx'$.
Finally, it is not difficult to show that the so defined
$\Mbf{\Theta}''$ $\spwedge$ $\Mbf{\Theta}'$
indeed belongs to $\OM{k+\ell} \bigl(\PA_n^{\hspace{1pt}\du}\bigr)$,
i.e., it gives zero on all functions $G_S$
belonging to $\FI_{\LL}\PA_n$ for some $\LL \in \Z$
(cf. Eq.~\ref{eq4.16xx6})).
\end{Proof}

\medskp

The operation $\spwedge$ (\ref{BOP}) \EMPH{is not $\Z/2\Z$--symmetric}.
Under the above notations we have the following result:

\medskp

\begin{Theorem}\label{Th-4.3}
The cohomological equations (\ref{eq4.6ne2})
for $\{\Sccl_{n;\,\xi}\}_{\xi}$ are equivalent to the following equations
for $\bsccl_n$:
\beqa\label{eq6.9ne1}
d \hspace{1pt} \bsccl_2 \, = \podr 0
\, , \qquad
\nn
d \hspace{1pt} \bsccl_n \, = \podr
\mathop{\sum}\limits_{m \, = \, 2}^{n-1} \,
\bsccl_{n-m+1}
\, \spwedge \,
\bsccl_m
\qquad (n > 2)
\,.
\eeqa
\end{Theorem}%

\medskp

For the \textit{proof} of this theorem
we shall also use some notations used in Sect.~\ref{se3.2}:
every function $G_{\PRT(S')} \, u_{S'}$ $\in$
$\PA_{\PRT(S')}$ $\otimes$ $\DP_{S',0}$ can be uniquely
transformed to a function belonging to $\PA_{S/S'}$ $\otimes$ $\DP_{S',0}$,
which includes a restriction to the partial diagonal
$\bigl\{\x_{\min S'} = \x_j$ for all $j \in S'\bigr\}$
(with possible ``transverse'' derivatives
due to the possible derivatives of the delta functions
contained in $u_{S'}$).
The latter transformation we denote by
$\NORM_{S'}$
(``normal form''
corresponding to $\NORM_{\PRT(S')}$ used in Sect.~\ref{se3.2}):
$$
\NORM_{S'} :
\PA_{\PRT(S')} \otimes \DP_{S',0}
\to
\PA_{S/S'} \otimes \DP_{S',0} \,.
$$

\medskp

\begin{Lemma}\label{Lm-3.8}
Let $|S| \geqslant 3$, $N=D(|S|-1)$, $N'=D(|S'|-1)$ for $S'\subseteq S$.
For every $\xi=1,\dots,N$ and
$G_S \in \PA_S$ of the form (\ref{GSP}) we have the following identities
\beqa\label{eq6.3n51}
&& \hspace{-20pt}
\bigl[\di_{x^{\xi}}, \SRMA_S \bigr] \, G_S
\\ && =
\mathop{\sum}
\limits_{\mathop{}\limits^{S' \, \subsetneq \, S}_{|S'| \, \geqslant \, 2}} \,
\Bigl( \SRMA_{S / S'} \otimes \ID_{\DP_{S',0}} \Bigr)
\circ
\NORM_{S'}
\Biggl(
G_{\PRT(S')} \cdot
\mathop{\sum}\limits_{\xi' \, = \, 1}^{N'}
\bigl(\Pi_{S'}\bigr){\hspace{-1pt}}^{\xi'}_{\xi}
\Sccl_{S';\,\xi'} (G_{S'})
\Biggr)
.
\nonumber
\eeqa
\end{Lemma}%

\medskp

\begin{Proof}
Due to the covering property (\ref{ocov})
it is enough to prove that
\beq\label{T-e1}
\hspace{0pt}
\text{left hand side of Eq.~(\ref{eq6.3n51})} \vrestr{10pt}{\VP_{\PRT}}
=
\text{right hand side of Eq.~(\ref{eq6.3n51})} \vrestr{10pt}{\VP_{\PRT}}
\hspace{-4pt}
\eeq
for every proper $S$--partition $\PRT$.
For the restriction of the right hand side we obtain:
\beq\label{T-e2}
\mathop{\sum}
\limits_{\mathop{}\limits^{S' \, \leqslant \, \PRT}_{|S'| \, \geqslant \, 2}}
\hspace{-1pt}
\Bigl( \SRMA_{S / S'} \otimes \ID_{\DP_{S',0}} \Bigr)
\hspace{2pt}\circ\hspace{2pt}
\NORM_{S'}
\Biggl(\hspace{-1pt}
G_{\PRT(S')}
\hspace{2pt}
\cdot
\hspace{-2pt}
\mathop{\sum}\limits_{\xi' \, = \, 1}^{N'}
\bigl(\Pi_{S'}\bigr){\hspace{-1pt}}^{\xi'}_{\xi}
\Sccl_{S';\,\xi'} (G_{S'})
\hspace{-1pt}\Biggr)
\!\vrestr{18pt}{\VP_{\PRT}}
\!,
\eeq
where the notation $S' \, \leqslant \, \PRT$ stands for the relation:
$j,k$ $\in$ $S'$\ $\Rightarrow$\ $j$ $\sim_{\PRT}$ $k$.
This is because if $S' \nleqslant \PRT$ then the support of
$\Sccl_{S';\,\xi'} \, G_{S'}$ is
disjoint from~$\VP_{\PRT}$.
For the restriction of the left hand side of Eq.~(\ref{eq6.3n51})
we obtain form condition $(r4)$ that:
\beqa\label{T-e22}
&& \hspace{-20pt}
\bigl[\di_{x^{\xi}}, \SRMA_S \bigr] \, G_S \vrestr{12pt}{\VP_{\PRT}}
\\ && \, = \,
G_{\PRT} \, \cdot \,
\Biggl(
\mathop{\sum}\limits_{S' \, \in \, \PRT} \,
\mathop{\sum}\limits_{\xi' \, = \, 1}^{N'} \
\bigl(\Pi_{S'}\bigr){\hspace{-1pt}}^{\xi'}_{\xi} \,
\bigl[\di_{x'{}^{\xi'}}, \RMA_{S'} \bigr] (G_{S'})
\Biggr)
\mathop{\prod}
\limits_{\mathop{}\limits^{S'' \, \in \, \PRT}_{S'' \, \neq \, S'}}
\RMA_{S''} \, G_{S''} \,.
\nonumber \eeqa
We then use the identities
\beqa\label{T-e3} & \hspace{-30pt}
\bigl[\di_{x'{}^{\xi'}}, \RMA_{S'} \bigr] (G_{S'})
\, = \,
\Sccl_{S';\,\xi'} (G_{S'})
\quad \text{for } |S'| = 2,
& \\ \label{T-e4} & \hspace{-35pt}
\bigl[\di_{x'{}^{\xi'}}, \RMA_{S'} \bigr] (G_{S'})
=
\Sccl_{S';\,\xi'} (G_{S'})
+ \,
\PRMA_{S'} \circ \bigl[\di_{x'{}^{\xi'}}, \SRMA_{S'} \bigr] (G_{S'})
\ \ \text{for } |S'| > 2.
& \eeqa
From the first of these identities and Eq.~(\ref{T-e2})
we obtain Eq.~(\ref{T-e1}) for the case $|S|=3$.

For $|S| > 3$ we proceed by induction in $|S|$
and apply the inductive assumption to the second term in
the right hand side of Eq.~(\ref{T-e4}).
In this way, substituting the result in Eq.~(\ref{T-e22})
we arrive at
the expression (\ref{T-e2})
by using the properties of renormalization maps and in particular, $(p5)$.
Thus, we obtain again Eq.~(\ref{T-e1}), which proves the lemma.
\end{Proof}%

\medskp

Now
to complete the proof of Theorem~\ref{Th-4.3}
we need only to compose both sides of Eq.~(\ref{eq6.3n51}) with
$\bigl[\di_{x^{\eta}},\PRMA_S\bigr]$
and use property $(p5)$ of Sect.~(\ref{se2.2})
in order to reduce
$\bigl[\di_{x^{\eta}},\PRMA_S\bigr]$ to
\(\sum_{\eta'' \, = \, 1}^{N''}
\bigl(\Pi''_{S'}\bigr){\hspace{-1pt}}^{\eta''}_{\eta}\)
$\bigl[\di_{x''{}^{\eta''}},\PRMA_{S/S'}\bigr]$.
In this way we obtain:
\beqa\label{eq6.3n51n}
&& \hspace{-57pt}
\bigl[\di_{x^{\eta}}, \PRMA_S \bigr] \circ
\bigl[\di_{x^{\xi}}, \SRMA_S \bigr] \, G_S
\\ =
\mathop{\sum}
\limits_{\mathop{}\limits^{S' \, \subsetneq \, S}_{|S'| \, \geqslant \, 2}}
\podr
\Biggl(
\sum_{\eta'' \, = \, 1}^{N''}
\bigl(\Pi''_{S'}\bigr){\hspace{-1pt}}^{\eta''}_{\eta}
\Sccl_{S/S';\,\eta''}
\otimes \ID_{\DP_{S',0}} \Biggr)
\nn \podr
\circ \
\NORM_{S'}
\Biggl(
G_{\PRT(S')} \cdot
\mathop{\sum}\limits_{\xi' \, = \, 1}^{N'}
\bigl(\Pi_{S'}\bigr){\hspace{-1pt}}^{\xi'}_{\xi}
\Sccl_{S';\,\xi'} (G_{S'})
\Biggr)
\nn =
\mathop{\sum}
\limits_{\mathop{}\limits^{S' \, \subsetneq \, S}_{|S'| \, \geqslant \, 2}}
\podr
\mathop{\sum}\limits_{\xi',\eta''} \,
\bigl(\Pi''_{S'}\bigr){\hspace{-1pt}}^{\eta''}_{\eta}
\bigl(\Pi_{S'}\bigr){\hspace{-1pt}}^{\xi'}_{\xi}
\Bigl(
\Sccl_{S/S';\,\eta''}
\otimes \ID_{\DP_{S',0}} \Bigr)
\nn \podr
\circ \
\NORM_{S'}
\Bigl(
G_{\PRT(S')} \cdot
\mathop{\sum}\limits_{\rr' \, \in \, \N_0^{N'}}
\ \frac{(-1)^{|\rr'|} \, }{\rr'!} \
\sccl_{S';\,\xi'} \bigl(\xx{}'{}^{\rr'} \hspace{1pt} G_{S'}\bigr) \,
\delta^{(\rr')} (\xx')
\Bigr)
\nn =
\mathop{\sum}
\limits_{\mathop{}\limits^{S' \, \subsetneq \, S}_{|S'| \, \geqslant \, 2}}
\podr
\mathop{\sum}\limits_{\xi',\eta''} \,
\bigl(\Pi''_{S'}\bigr){\hspace{-1pt}}^{\eta''}_{\eta}
\bigl(\Pi_{S'}\bigr){\hspace{-1pt}}^{\xi'}_{\xi} \,
\mathop{\sum}\limits_{\rr' \, \in \, \N_0^{N'}}
\ \frac{1}{\rr'!} \
\sccl_{S/S';\,\eta''}
\Bigl(
\Mbf{\di}_{\xx{}'}^{\rr'} \hspace{1pt} G_{\PRT(S')}\vrestr{12pt}{\xx'=0}
\Bigr)
\nn \podr \times \,
\sccl_{S';\,\xi'} \bigl(\xx{}'{}^{\rr'} \hspace{1pt} G_{S'}\bigr)
\, \cdot \,
\delta (\xx) + \cdots
,
\nonumber
\eeqa
where in the last line the dots include terms with
derivatives of the delta function $\delta (\xx)$.
In this way we arrive to Eqs.~(\ref{eq6.9ne1})
and we note that there is a change in the sign
due to the passage in the left hand side of Eqs.~(\ref{eq4.6ne2})
from $\bigl[\di_{x^{\xi}},\Sccl_{n;\, \eta}\bigr]$
to $\di_{x^{\xi}} \sccl_{n;\, \eta}$ in accordance to Eq.~(\ref{eq4.24ne2}).
This completes the proof of Theorem \ref{Th-4.3}.

\medskp

\begin{Remark}\label{Rm3.1}
An element $\Theta \in \PA_m^{\hspace{1pt}\du}$ is called symmetric
iff $\Theta$ $\circ$ $\sigma^*$ $=$ $\Theta$ for every $\sigma \in \Ss_m$
($(\sigma^* F) \bigl(\x_{j_1},\dots,\x_{j_n}\bigr)$
$:=$ $F \bigl(\x_{\sigma (j_1)},\dots,\x_{\sigma (j_n)}\bigr)$).
An element $\Mbf{\Theta} \in \OM{k} \bigl(\PA_m^{\hspace{1pt}\du}\bigr)$
is called symmetric iff all its components $\Theta_{\xi_1,\dots,\xi_m}$ are
symmetric elements of $\PA_m^{\hspace{1pt}\du}$.
One can prove that if
$\Mbf{\Theta} \in \OM{k} \bigl(\PA_m^{\hspace{1pt}\du}\bigr)$ and
$\Mbf{\Theta}' \in \OM{k'} \bigl(\PA_{m'}^{\hspace{1pt}\du}\bigr)$
are symmetric then so is
$\Mbf{\Theta} \spwedge \Mbf{\Theta}'$.
Furthermore, the bilinear operations $\spwedge$~(\ref{BOP})
are associative on symmetric elements in the sense that:
$$
\Mbf{\Theta} \spwedge \bigl(\Mbf{\Theta}' \spwedge \Mbf{\Theta}''\bigr)
\, = \,
\bigl(\Mbf{\Theta} \spwedge \Mbf{\Theta}'\bigr) \spwedge \Mbf{\Theta}''
$$
for $\Mbf{\Theta} \in \OM{k} \bigl(\PA_m^{\hspace{1pt}\du}\bigr)$,
$\Mbf{\Theta}' \in \OM{k'} \bigl(\PA_{m'}^{\hspace{1pt}\du}\bigr)$ and
$\Mbf{\Theta}'' \in \OM{k''} \bigl(\PA_{m''}^{\hspace{1pt}\du}\bigr)$
(all being symmetric).
We also claim (without a proof) that
$\spwedge$ obeys the $\Z/2\Z$--graded Leibnitz rule:
$$
d \bigl(\Mbf{\Theta} \spwedge \Mbf{\Theta}'\bigr) \, = \,
\bigl(d\Mbf{\Theta}\bigr) \spwedge \Mbf{\Theta}'
+ (-1)^{m'}
\Mbf{\Theta} \spwedge \bigl(d\Mbf{\Theta}'\bigr) \,.
$$
In this way, we obtain the integrability condition for Eqs.~(\ref{eq6.9ne1}):
if $\bsccl_m$ are solutions of Eqs.~(\ref{eq6.9ne1}) for $m=2,\dots,n-1$
then the right hand side of the second of Eqs.~(\ref{eq6.9ne1})
is closed.
One can also make the direct sum
$$
\SOM{\PA} =
\mathop{\bigoplus}\limits_{m \, = \, 0}^{\infty}
\SOMD{m}{\PA}
, \quad
\SOMD{0}{\PA} := \GF
, \quad
\SOMD{m}{\PA} :=
\bigl(\OM{*} \bigl(\PA_m^{\hspace{1pt}\du}\bigr)\bigr)^{\text{symm}}
\quad \text{for } m > 0
$$
into
a graded associative differential noncommutative algebra in which
Eqs. (\ref{eq6.9ne1}) simply read:
$$
d \underline{\bsccl} \, = \, \underline{\bsccl} \spwedge \underline{\bsccl}
\,,
$$
where $\bigl(\underline{\bsccl}\bigr)_0$ $=$ $0$ and
$\bigl(\underline{\bsccl}\bigr)_n$ $=$ $\bsccl_{n+1}$
for $n>0$.
\end{Remark}

\medskp

\begin{Remark}\label{Rm3.2}
At the end of Sect.~\ref{se3.2} we have introduced a notion
of ``universal renormalization group'' with a product
given by Eq.~(\ref{URG}).
The elements of this group can be presented by sets
$\underline{\Rdf}$ $=$ $\bigl(\Rdf_n\bigr){}_{n \, = \, 1}^{\infty}$
and we shall lift the grading, as in the previous remark, by setting
$\bigl(\underline{\Rdf}\bigr)_n$ $:=$ $\Rdf_{n+1}$ for $n=0,1,\dots$.
Note that $\bigl(\underline{\Rdf}\bigr)_n$ for $n>0$ is a linear map
$\PA_{n+1}^{\hspace{1pt}\du}$ $\to$ $\DP_{n+1,0}$,
which commutes with the multiplication by polynomials.
Hence, expanding it in delta functions and derivatives,
as in Sect.~\ref{se4.2nn},
we can further reduce its description to a linear functional
belonging to $\PA_{n+1}^{\hspace{1pt}\du}$.
In this way,
the Lie algebra corresponding to the universal renormalization group
will be formed by sets
$\underline{\Theta}$ $=$
$\bigl(\bigl(\underline{\Theta}\bigr)_n\bigr){}_{n \, = \, 1}^{\infty}$
such that $\bigl(\Theta\bigr)_0$ $=$ $0$ and
$\bigl(\Theta\bigr)_n$ $\in$ $\PA_{n+1}^{\hspace{1pt}\du}$ are symmetric
(see the previous remark) for $n=1,2,\dots$.
The Lie algebra bracket is very close to the product $\spwedge$~(\ref{BOP})
and it is just the commutator
$$
\bigl[\underline{\Theta}',\underline{\Theta}''\bigr] \, = \,
\underline{\Theta}' \sppro \underline{\Theta}'' -
\underline{\Theta}'' \sppro \underline{\Theta}',
$$
where $\sppro$ $:$
$\PA_{n'+1}^{\hspace{1pt}\du}$ $\otimes$ $\PA_{n''+1}^{\hspace{1pt}\du}$
$\to$ $\PA_{n'+n''+1}^{\hspace{1pt}\du}$
are associative bilinear operations defined by
\beqa\label{qqq1}
\bigl(
\underline{\Theta}''
\sppro \,
\underline{\Theta}'
\bigr)_{n-1}
\bigl(G_S\bigr)
\hspace{8pt}
\podr \nn
\, = \,
\mathop{\sum}
\limits_{\emptyset \, \subsetneq \, S' \, \subsetneq \, S}
\mathop{\sum}\limits_{\rr' \, \in \, \N_0^{N'}}
\podr
\frac{1}{\rr'!} \
\underline{\Theta}''_{S/S'}
\Bigl(
\Mbf{\di}_{\xx{}'}^{\rr'} \hspace{1pt} G_{\PRT(S')}\vrestr{12pt}{\xx'=0}
\Bigr)
\,
\underline{\Theta}'_{S'}
\bigl(\xx{}'{}^{\rr'} \hspace{1pt} G_{S'}\bigr)
\hspace{30pt}
\eeqa
(we use the same type of notations like in Eq.~(\ref{qqq})).
\end{Remark}

\subsection{Concluding remarks}\label{Concl}

We intend to study the cohomological equations (\ref{eq6.9ne1})
and their solutions
in the future.
On $\PA_n$ the cohomological equations
do not completely characterize the renormalization cocycles
since, as we have pointed out, we have
\beq\label{INEQ}
\HOM{1} \bigl(\PA_n^{\hspace{1pt}\du}\bigr)
\, \cong \,
\Bigl(\HOM{D(n-1)-1} \bigl(\PA_n\bigr)\Bigr){\raisebox{10pt}{\hspace{-1pt}}}'
\, \neq \,
\{0\} \,.
\eeq
In fact, $\HOM{m} \bigl(\PA_n\bigr)$ are the cohomology groups of
the complement of union of quadrics in $\C^{D(n-1)}$:
$\x_k^2$ $=$ $0$ $=$ $\bigl(\x_j-\x_k\bigr)^2$
($j,k=1,\dots,n-1$).
Equation (\ref{INEQ})
is exactly the reason for which we need to introduce some
transcendental methods in order
to derive the renormalization cocycles.
This is because otherwise, all the solutions of the cohomological equations
would correspond to some renormalization scheme
and hence, there will be no need to extend the ground field $\GF$ ($=$ $\Q$).

Let us point out that if we had used instead of $\PA_n$ the algebra
$\CI_{temp} \bigl(\HF_n\bigr)$ then the primary renormalization cocycles
would belong to $\OM{1} \Bigl(\CI_{temp} \bigl(\HF_n\bigr)^{\du}\Bigr)$,
whose cohomology group is now isomorphic to
\(\Bigl(\HOM{D(n-1)-1} \Bigl(\CI_{temp} \bigl(\HF_n\bigr)\Bigr)\Bigr)
{\raisebox{10pt}{\hspace{-1pt}}}'\).
One can show that
$\HOM{D(n-1)-1} \Bigl(\CI_{temp} \bigl(\HF_n\bigr)\Bigr)$
is isomorphic to the usual de Rham cohomology group
$\HOM{D(n-1)-1} \bigl(\HF_n\bigr)$ of the configuration space $\HF_n$.
On the other hand, there is a theorem \cite{FH} stating that
for $n>2$: $\HOM{D(n-1)-1} \bigl(\HF_n\bigr)$ $=$ $\{0\}$
(recall that $\HF_n$ $\cong$ $\F_{n-1} \bigl(\R^D \mz\bigr)$).
Thus, in this case the cohomological equations
would completely characterize the renormalization cocycles for $n>2$.
In the preprint \cite[Sects.~2 and 3]{N07} we have considered
renormalization on $\CI_{temp} \bigl(\HF_n\bigr)$ exactly for this purpose.
The problem then is that the right hand side of
the cohomological equations (\ref{eq6.9ne1})
do not have a simple, algebraic construction.
The space $\CI_{temp} \bigl(\HF_n\bigr)$
is too large to work on it ``algebraically''.
So, the problem is to find an intermediate differential extension
$$
\PA_n \, \subseteq \, \PAE_n \, \subseteq \, \CI_{temp} \bigl(\HF_n\bigr)
$$
for every $n>2$ such that first,
it possesses an algebraic interpretation
of the cohomological equations (\ref{eq6.9ne1}) and second,
the de Rham cohomologies of degree $D(n-1)-1$ are trivial:
$$
\HOM{D(n-1)-1}\bigl(\PAE_n\bigr) \, = \,
\HOM{D(n-1)-1}\Bigl(\CI_{temp} \bigl(\HF_n\bigr)\Bigr) \, = \, \{0\} \,.
$$

In fact, one possible strategy for solving Eqs.~(\ref{eq6.9ne1}) on $\PAE_n$
would be to find a homotopy operator
$$
K_n \, : \, \OM{D(n-1)-1} \bigl(\PAE_n\bigr) \to
\OM{D(n-1)-2} \bigl(\PAE_n\bigr)
\, , \quad
K_n \circ d + d \circ K_n \, = \, \id \,,
$$
and then a solution of (\ref{eq6.9ne1}) for $n > 2$ will be
$$
\bsccl_n \, = \,
\mathcal{F}_n \circ K_n \,,
$$
where $\mathcal{F}_n$ is the right hand side of Eq.~(\ref{eq6.9ne1})
(as in (\ref{eq6xx})).
Let us note that for the case of two dimensional QFT ($D=2$)
the quadric $\x^2=0$ is reducible and the problem reduces to the case
``$D=1$'', where such a differential extension
$\PA_n$ $\subseteq$ $\PAE_n$
is provided by a \textit{polylogarithmic extension}
of the ring of rational functions $\PA_n$
(see e.g., \cite{B}).

\bigskip

\noindent
\textbf{Acknowledgments.}
I am grateful to Spencer Bloch, Francis Brown, Maxim Kontsevich,
Dirk Kreimer and Ivan Todorov
for enlightening discussions.
I am also grateful to Ivan Todorov for a critical reading of the manuscript.
Parts of this work were developed during my stay at
the Institut des Hautes \'{E}tudes Scientifiques in 2008
and also during the program ``Operator algebras and conformal field theory''
at the Erwin Schr\"{o}dinger International Institute
for Mathematical Physics.
I am grateful to these institutions for the support and hospitality.
This work was partially supported by the French--Bulgarian project Rila
under the contract Egide -- Rila N112.

\newcommand{\Bibitem}[1]{\bibitem[#1]{#1}}
\newcommand{\BIbitem}[2]{\bibitem[#1]{#2}}
\addtocontents{toc}{\protect\vspace{-8pt}}
\addtocontents{toc}{\contentsline {section}{References}{\arabic{page}}}

\end{document}